\pdfoutput=1

\documentclass[12pt, letterpaper]{article}
\usepackage{enumerate}
\usepackage{natbib}

\usepackage{amsmath, amsfonts}
\usepackage{graphicx}
\usepackage{caption}
\usepackage{subcaption}
\usepackage[labelformat=simple]{subcaption}

\usepackage{url}
\usepackage{multirow}
\usepackage{afterpage}

\usepackage{morefloats}
\usepackage{pdflscape}

\newcommand{\bmbeta}{\boldsymbol{\beta}}

\newcommand{\bmpi}{\boldsymbol{\pi}}
\newcommand{\bmxi}{\boldsymbol{\xi}}
\newcommand{\bmnu}{\boldsymbol{\nu}}
\newcommand{\bmmu}{\boldsymbol{\mu}}

\newcommand{\bmgamma}{\boldsymbol{\gamma}}
\newcommand{\bmomega}{\boldsymbol{\omega}}

\newcommand{\bmY}{\mathbf{Y}}
\newcommand{\bmZ}{\mathbf{Z}}
\newcommand{\bmX}{\mathbf{X}}
\newcommand{\bmV}{\mathbf{V}}
\newcommand{\bmW}{\mathbf{W}}
\newcommand{\bmU}{\mathbf{U}}

\newcommand{\bmb}{\mathbf{b}}

\newcommand{\bmSigma}{\boldsymbol{\Sigma}}
\newcommand{\bmI}{\mathbf{I}}
\newcommand{\bmTheta}{\boldsymbol{\Theta}}

\newcommand{\bea}{\begin{eqnarray}} 
\newcommand{\eea}{\end{eqnarray}} 
\newcommand{\beas}{\begin{eqnarray*}} 
\newcommand{\eeas}{\end{eqnarray*}} 
\newcommand{\benum}{\begin{enumerate}} 
\newcommand{\eenum}{\end{enumerate}} 
\newcommand{\bd}{\begin{description}}
\newcommand{\ed}{\end{description}}
\newcommand{\bi}{\begin{itemize}}
\newcommand{\ei}{\end{itemize}}

\newcommand{\mydots}{...}

\newcommand{\blind}{0}

\begin{document}

\def\spacingset#1{\renewcommand{\baselinestretch}%
{#1}\small\normalsize} \spacingset{1.45}

\if0\blind
{
  \title{\bf  A Bayesian hierarchical model for prediction of latent health states from multiple data sources with application to active surveillance of prostate cancer}
  \author{\hspace{-0.75cm}R. Yates Coley$^\text{a}$\thanks{
  	This research was supported by the Patrick C. Walsh Prostate Cancer Research Fund and  a Patient-Centered Outcomes Research Institute (PCORI) Award (ME-1408-20318). The statements presented in this article are solely the responsibility of the authors and do not necessarily represent the views of the Patient-Centered Outcomes Research Institute (PCORI), its Board of Governors or Methodology Committee.
    The authors gratefully acknowledge Ruth Etzioni, Tom Louis, and Gary Rosner for their helpful comments.}\hspace{.2cm}, Aaron J. Fisher$^\text{a}$, Mufaddal Mamawala$^\text{b}$\\
    \hspace{-0.75cm}H. Ballentine Carter$^\text{b}$, Kenneth J. Pienta$^\text{b}$, and Scott L. Zeger$^\text{a}$\\
	\hspace{-0.75cm}$^\text{a}$Department of Biostatistics, Johns Hopkins Bloomberg School of Public Health\\
	\hspace{-0.75cm}$^\text{b}$James Buchanan Brady Urological Institute, Johns Hopkins Medical Institutions}
\maketitle
\newpage
} \fi

\if1\blind
{
  \bigskip
  \bigskip
  \bigskip
  \begin{center}
    {\LARGE\bf  A Bayesian hierarchical model for prediction of latent health states from multiple data sources with application to active surveillance of prostate cancer}
\end{center}
  \medskip
} \fi



\newpage

\spacingset{1.45} 

\section{Introduction}

Medicine is in a period of transition. An ever-increasing amount of information is available on patients ranging from genetic and epigenetic profiles enabled by next-generation sequencing to moment-to-moment data collected by physical activity monitors. With this wealth of information comes the opportunity to provide more targeted healthcare including, for example, prediction of pre-clinical atherosclerosis \citep{McGeachie2009}, individualized cancer screening (Saini, van Hees, and Vijan, 2014)\nocite{Saini2014}, sub-typing of scleroderma (Schulam, Wigley, and Saria, 2015)\nocite{Schulam2015},  and personalized cancer treatment \citep{Hayden2009}. In order to fully realize the promise of patient-focused medicine, principled statistical methods are needed that integrate data from a variety of sources in order to provide physicians and patients with relevant syntheses to inform their decision-making. These methods must also accommodate limitations common to data generated in an observational setting including measurement error and informative missing data patterns.

An excellent example of this challenge is low-risk prostate cancer diagnosis. Tumor lethality is an aspect of an individual's health state that is not directly observable but is manifest in multiple types of measurements including biomarkers, histology of biopsied tissue, genetic markers, and family history of the disease. Individualized predictions of the latent disease state are critical to guide treatment decisions. If the tumor is potentially lethal, immediate treatment (including surgery or radiation) can be life-saving. Yet, some tumors are indolent and not life-threatening. In this case, treatment is not recommended due to the risk of lasting side effects including urinary incontinence and erectile dysfunction \citep{Chou2011b}. 

Active surveillance (AS) offers an alternative to early treatment for individuals with lower risk disease  \citep{DallEra2012}. Though AS regimes vary, the approach generally entails regular biopsies (e.g., annually) with intervention recommended upon detection of higher risk histological features, as determined by the Gleason grading system \citep{Gleason1992}. Biopsies with a Gleason score of 6 (the minimum for prostate cancer diagnosis) indicate low risk disease while a subsequent Gleason score of 7 or above is considered ``grade reclassification'' \citep{Tosoian2015}; treatment is recommended once grade reclassification is observed. Prostate-specific antigen (PSA), a blood serum biomarker of inflammation in the prostate, is also routinely measured and may be used as the basis for a biopsy recommendation. 

The success of AS programs depends on clinicians' ability to identify tumors with metastatic potential with sufficient time for curative intervention to be effective. Yet, biopsies used to characterize tumors typically sample less than one percent of the prostate tissue and so have imperfect sensitivity and specificity \citep{Epstein2012}. Existing decision support tools that predict biopsy outcomes for AS patients (including, most recently, \citet{Ankerst2015}) provide patients and physicians with valuable information to guide decisions about biopsy timing and frequency but are insufficient to directly address patients' primary concerns about their tumors' lethality. Patients and clinicians need predictions of the pathological make-up of the entire prostate to guide their decision-making.

With this application in mind, we have developed a Bayesian hierarchical model that enables prediction of an individual's underlying disease state via joint modeling of repeated PSA measurements and biopsies. Specifically, we predict a binary cancer state-- \textit{indolent} or \textit{aggressive}-- with the latter defined as a ``true" Gleason score of 7 or higher. Predictions are informed by a subset of patients for whom the true state is observed-- patients who, either before or after biopsy grade reclassification, chose to undergo prostatectomy and have post-surgery, entire-prostate Gleason score determinations. In this sense, cancer state operates as a partially-latent class in the proposed model \citep{Wu2015}. 

\begin{figure}
\begin{center}
\begin{subfigure}[b]{0.45\textwidth}
\includegraphics[width=\textwidth]{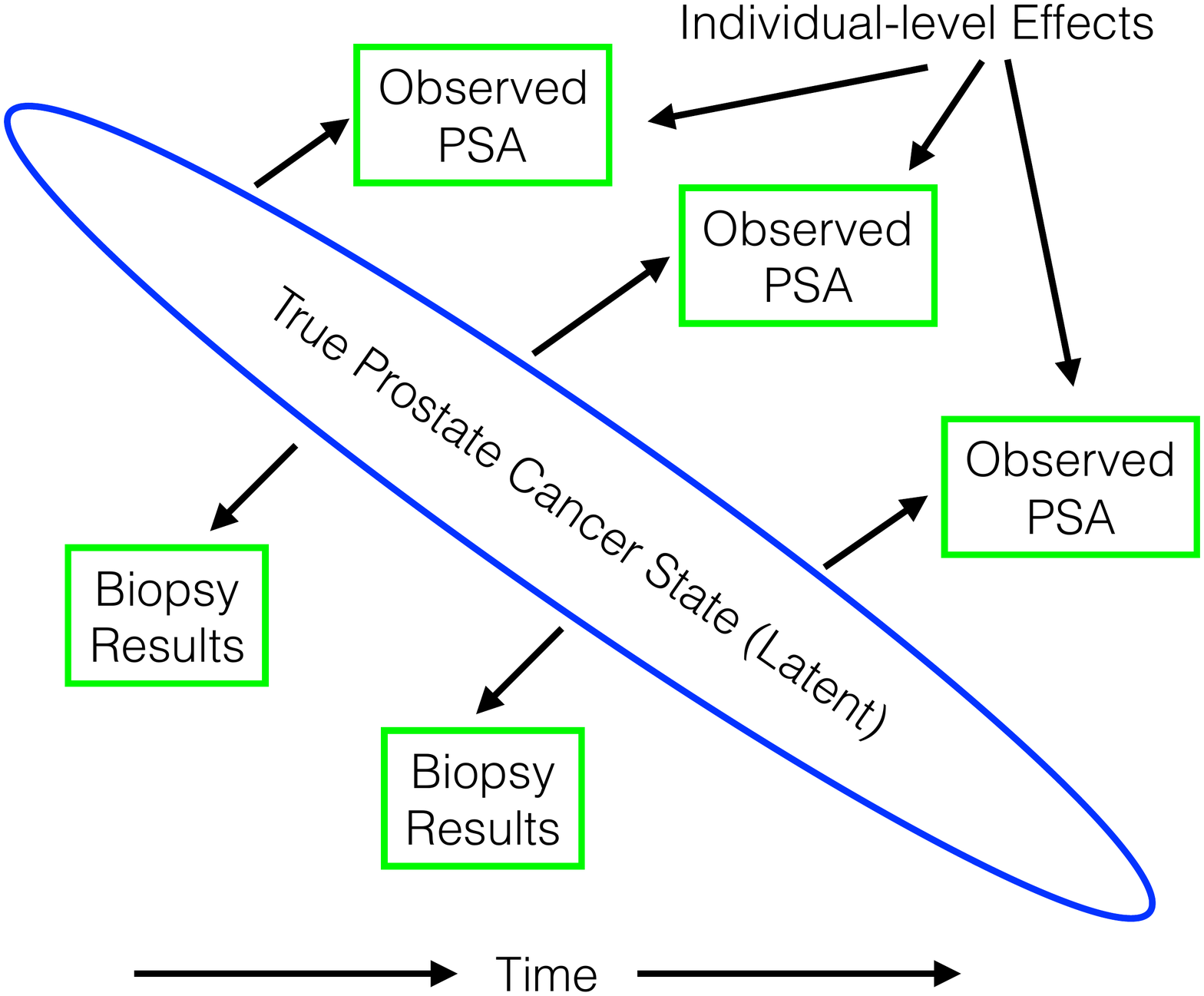}
\caption{Time-varying scenario}
\label{fig:dag1}
\end{subfigure}
\begin{subfigure}[b]{0.45\textwidth}
\includegraphics[width=\textwidth]{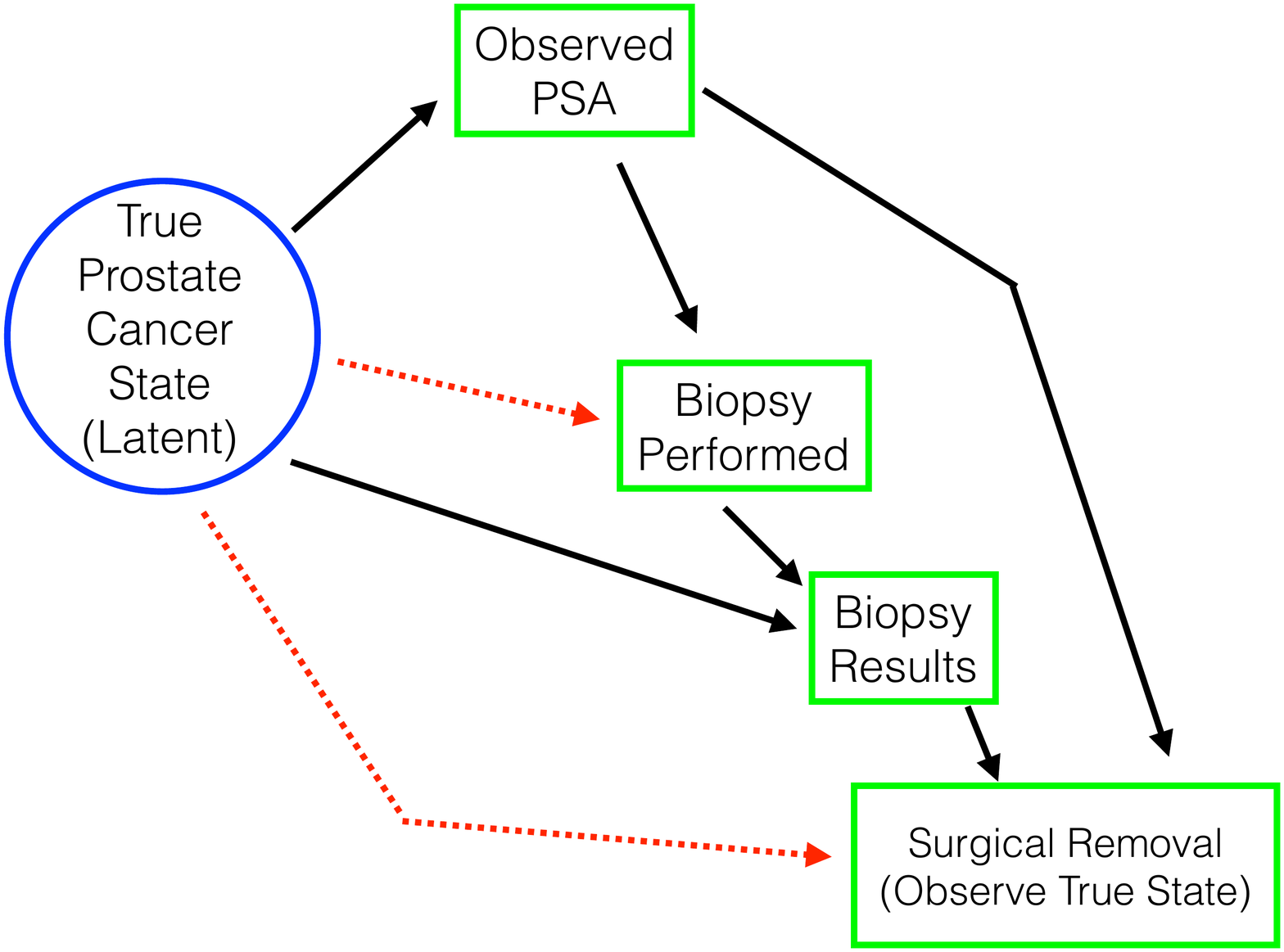}
\caption{Informative observation process scenarios}
\label{fig:dag2}
\end{subfigure}
\label{fig:dags}
\caption{DAGs describing the relationships between latent class (circled) and clinical outcomes (squared).}
\end{center}
\end{figure}

An individual's cancer state is assumed to be manifest in both the level and trajectory of PSA measurements as well as in the outcomes from repeated biopsies. These relationships are illustrated by the directed acyclic graph (DAG) in Figure \ref{fig:dag1}. In the model we are proposing, PSA measurements follow a multilevel model with mean intercept and age effects varying across latent classes. Then, repeated annual biopsies constitute a time-to-event outcome since patients exit AS after grade reclassification on biopsy. So, time until reclassification on biopsy is modeled using pooled logistic regression under the assumption that biopsy results are independent conditional on cancer state and covariates \citep{Cupples1988}. Pooled logistic regression provides survival estimates equivalent to those of a time-varying Cox model for discrete event times and conditionally independent intervals \citep{D'Agostino1990}. As indicated in Figure \ref{fig:dag1}, PSA and biopsy results are also assumed to be conditionally independent given latent class. 

The model depicted in Figure \ref{fig:dag1} is related to previous work by \citet{Lin2002}, who proposed a joint latent class model (JLCM) to analyze longitudinal PSA and time-to-diagnosis of prostate cancer, extending earlier joint models by \citet{Schluchter1992}, \citet{DeGruttola1994}, and Henderson, Diggle, and Dobson (2000)\nocite{Henderson2000}. \citet{Inoue2008} used a Bayesian approach to jointly model PSA and time-to-diagnosis at various stages of disease in order to estimate the underlying natural history process for prostate cancer initiation and progression. \citet{Proust2009} developed a dynamic extension of the JLCM to predict prostate cancer recurrence after radiation therapy.

A credible statistical solution to the active surveillance problem requires three extensions of existing latent variable models for multivariate outcomes (such as the JLCM). First, the model must accommodate measurement error inherent in monitoring disease state. In our approach, we focus prediction on a partially-observed true Gleason score, instead of relying on biopsy Gleason scores for accurate characterization of the latent health state, and model a stochastic, rather than deterministic, relationship between the two.

Second, the model must allow disease monitoring to reflect patterns of clinical practice, including discrete, possibly informative, observation times. Specifically, prostate biopsies are scheduled to occur annually, but patients may opt to forgo the procedure. Our method replaces the JLCM's usual survival model for right-censored outcomes in favor of a pooled logistic regression model for biopsy grade reclassification where the possibility of reclassification in any year is conditional on a biopsy being performed. Furthermore, it is possible that the choice to receive a biopsy depends on the true cancer state or, more generally, that unobserved confounding exists, as shown by the dotted arrow from true cancer state to the ``Biopsy Performed" node in Figure \ref{fig:dag2}. If so, biopsy results are missing not at random (MNAR), and predictions of the true state that ignore the MNAR mechanism will be biased \citep{Little2014}. In response, our approach also includes a regression model for the probability of receiving a biopsy in each interval; the occurrence of a biopsy is allowed to depend on the latent health state, as well as previous biopsy and PSA observations.

Third, the active surveillance model must allow surgical removal of the prostate and subsequent observation of the underlying cancer state to be informative of that latent state. Consider the dotted arrow from true cancer state to the ``Surgical Removal" node in Figure \ref{fig:dag2}. If, after conditioning on clinical observations, an individual's true cancer state is associated with his choice to undergo surgery, whether through direct causation or unmeasured confounding, then informative missingness is present and failure to accommodate this in the model will result in biased predictions of the cancer state. While the association between the true cancer state and a binary indicator of its observation is not identifiable, we propose to model the time until surgery (and true state observation) conditional on the latent state. Evidence of this relationship among the subset of patients with surgery and mild assumptions on the structure of the hazard function (such as an additive or smooth effect) provide identifiability. This approach shares a similar intuition with missing data models for repeated attempt designs in which the estimated association between the number of attempts needed to elicit a response and its value is used to account for outcomes suspected to be MNAR \citep{Jackson2012}. In this application, patients have the opportunity to elect surgery throughout their participation. For simplicity, we also model the time until surgery with a pooled logistic regression model.

This paper is organized as follows. In Section 2, a hierarchical model for latent class prediction is described and estimation procedures are outlined. In Section 3, we specify our model to predict latent cancer states for patients in the Johns Hopkins Active Surveillance cohort and outline a simulation study based on this application. Results are presented in Section 4. We close with a discussion.

\vspace{-14pt}

\section{Hierarchical Latent Class Model}

We propose a Bayesian hierarchical model of the underlying cancer state, measurement process, and clinical outcomes of patients enrolled in active surveillance (AS). Predictions are made by incorporating information from repeated PSA and biopsy measurements for all patients and true cancer state observations in a potentially non-random subset of the cohort. Predictions are also informed by the presence of some observations, which we refer to as an \textit{informative observation process} (IOP). In this section, we introduce notation and conditional distributions for the observed data given the latent variables and parameters, then give the likelihood function. The model is completed by specifying appropriate priors and defining the joint posterior distribution. Overall model structure is summarized in Figure \ref{fig:model-summary}.

\subsection{Latent cancer state $\eta_i$ for patient $i$, $i=1,\dots,n$}

Define individual $i$'s true cancer state, $\eta_i$, as either indolent, $\eta_i=0$, or aggressive, $\eta_i=1$, $i=1,\dots,n.$ We use the Gleason score that would be assigned if his entire prostate were to be surgically removed and analyzed to define $\eta_i=0$ if Gleason $=$ 6 and $\eta_i=1$ if Gleason $\geq$ 7. Note that this definition assumes that cancer state is constant during the time under consideration. This assumption is discussed in more detail in Section 5.

True cancer state is then modeled as a Bernoulli random variable, $\eta_i \sim Bern(\rho_i)$. We assume a shared underlying probability of aggressive cancer, $\rho_i=\rho$, for simplicity in initial presentation. We observe this true cancer state on a possibly non-random subset of patients who choose surgical removal of the prostate and, hence, $\eta_i$ is a partially-latent variable. 

\subsection{Longitudinal data $Y_{im}$ given latent class $\eta_i$, $m=1,\dots,M_i$}

Next, we consider PSA, which is influenced by the true cancer state $\eta_i$ as well as covariates including age and prostate volume. Unlike biopsies, PSA measurements are a routine part of each clinic visit so the times of observation are assumed to be independent of $\eta_i$. We use a multilevel model to estimate the linear trend (on a log scale) of an individual's PSA as he ages \citep{Gelman2006}. Patient-level coefficients, $\bmb_i$, vary about an $\eta_i$-specific mean intercept and slope ($\bmmu_{\eta_i}$). Specification follows that of a hierarchically-centered multilevel model to speed convergence of the posterior sampling algorithm (Gelfand, Sahu, and Carlin, 1995)\nocite{Gelfand1995}. Specifically, given $\bmb_i$, the log-transformed PSA for patient $i$'s $m$th visit, $Y_{im}$, is assumed equal to $\bmX_{im}\bmbeta + \bmZ_{im}\bmb_i + \epsilon_{im}$ where $\bmX_{im}$ and $\bmZ_{im}$ are covariate vectors for individual $i$ at visit $m$, $\bmbeta$ is a parameter vector of population-level coefficients, and residual $\epsilon_{im}$ is assumed to follow a Gaussian distribution with mean zero and variance $\sigma^2$. In comparison to the commonly used mixed effects model of \citet{Laird1982}, covariates in $\bmZ_{im}$ are not a subset of covariates in $X_{im}$; covariates corresponding to patient-level effects $\bmb_i$ are only included in $\bmZ_{im}$, and the $\bmb_i$ are not centered at zero. In our application, $\bmZ_{im}$ includes an intercept and age so that PSA intercepts and slopes vary across individuals. $\bmX_{im}$ includes prostate volume, and $\bmbeta$ is the population-level association between volume and log-PSA.  

Modeling of patient-level coefficients follows the recommendation of \citet{Gelman2006} who advocate the use of a scaled inverse Wishart prior on the covariance matrix. The inverse Wishart prior, which is commonly used for Bayesian estimation of multilevel models \citep{Gelfand1995}, imposes dependence between variance and correlation components of the covariance matrix. To reduce prior dependence and allow for a flat prior on the correlation between individual-level intercepts and slopes, \citet{OMalley2008} introduce a scale parameter, $\bmxi$, for the patient-level random effects:  $\bmb_i = diag(\check{\bmb_i}\bmxi^T)$. Unscaled random effects, $\check{\bmb_i}$, are assumed to follow a  latent-class specific multivariate Gaussian distribution with mean vector $\bmmu_{\eta_i}$ and covariance matrix $\Sigma_{\eta_i}$.

\subsection{Biopsy Occurrence $B_{ij}$ and Result $R_{ij}$ for patient $i$ in time interval $j$, $j=1,\dots,J_i$}
We then consider information about the true cancer state contained in the occurrence and results of prostate biopsies. Biopsy data are categorized into discrete time intervals with $(B_{ij}, R_{ij})$ denoting binary outcomes for individual $i$ in time interval $j$. $B_{ij}$ indicates whether a biopsy was performed ($B_{ij}=1$) or not ($B_{ij}=0$) and, when it was performed, $R_{ij}$ indicates if grade reclassification occurred ($R_{ij}=1$) or not ($R_{ij}=0$). $B_{ij}$ and $R_{ij}$ are defined for $j=1,\dots, J_i$,  where $J_i$ is the time interval of reclassification or censoring for patient $i$. For each time interval, we use logistic regression to model the occurrence of a biopsy and, when a biopsy was performed, its result; both outcomes are conditional on true cancer state:
\bea
\label{eq:p_bx}
\text{logit} \{ \, P(B_{ij}=1 | \eta_i, \bmU_{ij}, \bmnu) \, \}= \bmU_{ij}\bmnu_1 + \eta_i\bmnu_2 + \bmU_{ij}\eta_i\bmnu_3 \\
\label{eq:p_rc}
\text{logit}\{\, P(R_{ij}=1 | \eta_i, \bmV_{ij}, B_{ij}=1, \bmgamma) \,\}= \bmV_{ij}\bmgamma_1 + \eta_i\bmgamma_2 + \bmV_{ij}\eta_i\bmgamma_3
\eea
where $\bmU_{ij}$ and $\bmV_{ij}$ are covariate vectors including time-varying predictors and $\bmnu=(\bmnu_1, \bmnu_2, \bmnu_3)$ and $\bmgamma=(\bmgamma_1, \bmgamma_2, \bmgamma_3)$ are parameter vectors to be estimated that include the main effects of covariates $\bmU_{ij}$ or $\bmV_{ij}$, $\eta_i$, and the possible interactions $\bmU_{ij}\eta_i$ and $\bmV_{ij}\eta_i$, respectively. Since reclassification occurs at most once, Equation (\ref{eq:p_rc}) corresponds to a modified pooled logistic regression model for time-to-reclassification in which only intervals with biopsies contribute.

This model specification represents three important aspects of data generated in active surveillance:  whether a biopsy is performed may be informative of true cancer state, time-to-reclassification depends on a patient's decision to receive a biopsy, and biopsy outcomes are prone to measurement error. In this application, $U_{ij}$ and $V_{ij}$ may include age, time since diagnosis, and calendar date. Previous PSA and biopsy results may also influence the decision to get a biopsy, but they do not influence biopsy findings. 

\vspace{-8pt}

\subsection{Surgical Removal of Prostate $S_{ij}$ and its Cancer Lethality $\eta_i$}
Lastly, to allow for the possibility that surgical removal of the prostate (and subsequent observation of the true cancer state) is informative, we define $S_{ij}$ to be a binary indicator of surgery ($S_{ij}=1$) or not ($S_{ij}=0$) for individual $i$ during time interval $j$ for $j=1,\dots,J_{S_i}$, where $J_{S_i}$ is the time of surgery or other censoring for patient $i$ and $J_{S_i} \geq J_i$ for all $i$. The probability of surgery in each time interval is modeled with logistic regression and conditional on the true cancer state: $\text{logit} \{\,P(S_{ij}=1 | \eta_i, \bmW_{ij}, \bmomega)\,\} =   \bmW_{ij}\bmomega_1 + \eta_i\bmomega_2 + \bmW_{ij}\eta_i\bmomega_3 \nonumber$ where $\bmW_{ij}$ is a vector of time-varying predictors and $\bmomega=(\bmomega_1, \bmomega_2, \bmomega_3)$ is a parameter vector to be estimated. Age, time since diagnosis, calendar date, and previous PSA and biopsy results may all be considered as possible predictors of surgery.

\vspace{-8pt}

\subsection{Posterior Distribution Estimation}
Having specified models for each information source, we define the likelihood of the latent states, patient-level coefficients, and population-level parameters given the observed data as the product of the contribution of each component described above:
\bea
&&L\big( \, \textbf{parameters, patient-level latent class and coefficients}\, | \,\textbf{data}\,\big) \nonumber \\ 
&& \quad =  \prod_{i=1}^{n}  \,  \rho^{\eta_i}\,(1-\rho)^{1-\eta_i}\, f(\bmY_i | \underline{\bmX_i}, \underline{\bmZ_i}, \bmbeta, \bmxi, \check{\bmb_i}, \sigma^2) \, g(\check{\bmb_i} |  \bmmu_{\eta_i}, \Sigma_{\eta_i})  \nonumber \\
&& \qquad \qquad \prod_{j=1}^{J_i}  P(B_{ij}=1 | \eta_i, \bmU_{ij}, \bmnu) ^{B_{ij}} P(B_{ij}=0 | \eta_i, \bmU_{ij}, \bmnu)^{1-B_{ij}}  \nonumber \\
&& \qquad \qquad \qquad  \qquad \{P(R_{ij}=1 | \eta_i, \bmV_{ij}, \bmgamma) ^{R_{ij}} P(R_{ij}=0 |  \eta_i, \bmV_{ij}, \bmgamma)^{1-R_{ij}} \}^{B_{ij}}\nonumber \\
\label{eq:lik-inf}
&& \qquad \qquad  \prod_{j=1}^{J_{S_i}}  P(S_{ij}=1 | \eta_i, \bmW_{ij}, \bmomega) ^{S_{ij}} P(S_{ij}=0 |  \eta_i, \bmW_{ij}, \bmomega)^{1-S_{ij}}  
\eea
where $f$ and $g$ are multivariate normal densities for the vector of log-transformed PSAs $\bmY_i$ and unscaled patient-level effects $\check{\bmb_i}$, respectively, each with mean and covariance as defined in Section 2.2. $\underline{\bmX_i}$ denotes the matrix of covariate vectors $[\bmX_{i1},\dots, \bmX_{iM_i}]$; $\underline{\bmZ_i}$ is similarly defined. 

\begin{figure}
\begin{center}
\includegraphics[width=\textwidth]{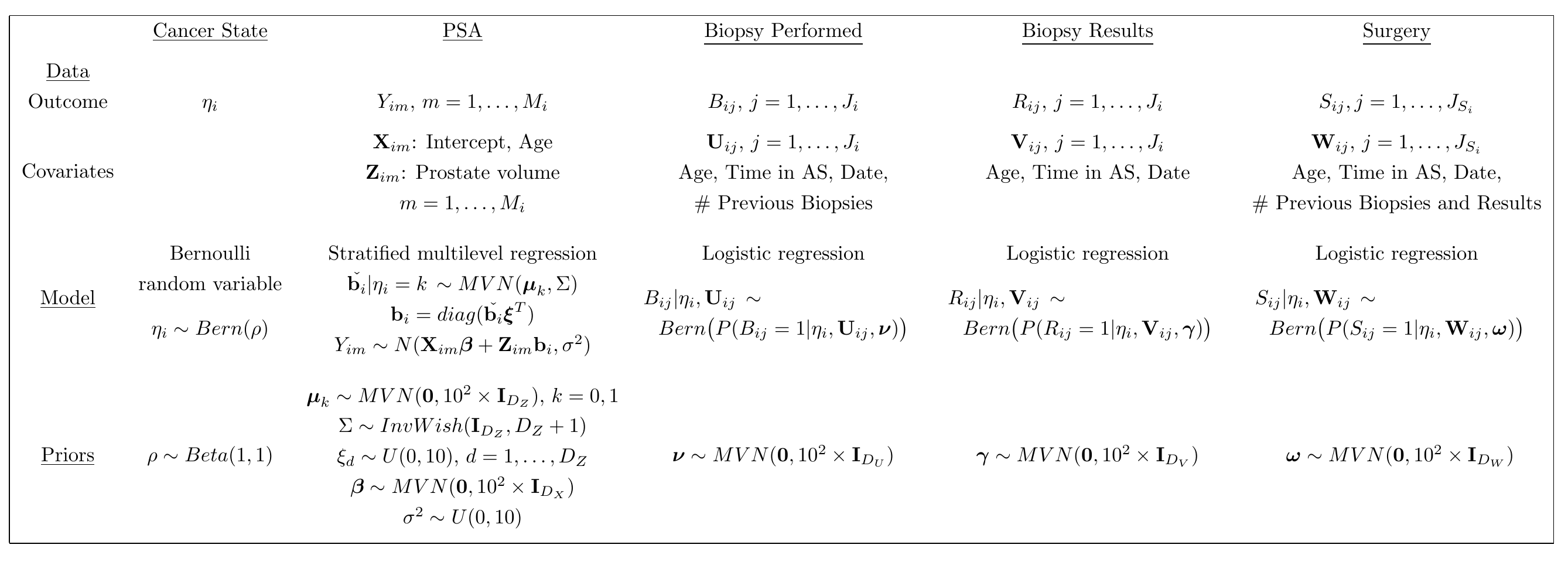}
\caption{Model summary with priors used for application to Johns Hopkins Active Surveillance data. $D_X$ is the length of vector $\bmX$ and $\bmI_{D_X}$ is the identity matrix with dimension $D_X \times D_X$. $D_Z, \, D_U, \, D_V,$ and $D_W$ and the associated identity matrices are similarly defined for covariate vectors $\bmZ,\,\bmU,\,\bmV,$ and $\bmW$.}
\label{fig:model-summary}
\end{center}
\end{figure}

We use a Bayesian approach for model estimation. The prior distributions for the latent states and patient-level coefficients have been described (Sections 2.1 and 2.2, respectively). Standard prior distributions are used for model parameters, including a beta prior on the probability of having aggressive cancer ($\rho$) and minimally informative Gaussian priors on logistic regression model coefficients, as shown in the model summary given in Figure \ref{fig:model-summary}. The joint posterior distribution of the parameters, latent states, and patient-level effects is proportional to the product of the likelihood and joint prior density of model parameters and is given explicitly in the online supplement.

For those patients without $\eta_i$ observed, a data augmentation approach is used to sample the true unobserved cancer state from its full conditional posterior at each iteration of the MCMC sampling algorithm \citep{Tanner1987}. Averaging the resulting posterior sample produces the posterior probability that a patient has a true Gleason 7 or higher prostate cancer, $P(\eta_i=1)$. 

\section{Johns Hopkins Active Surveillance Cohort}

\subsection{The Data}
From January 1995 to June 2014, the Johns Hopkins Active Surveillance (JHAS) cohort enrolled 1,298 prostate cancer patients \citep{Tosoian2015}. This study prospectively follows patients with very-low-risk or low-risk prostate cancer diagnoses (according to criteria outlined in \citet{Epstein1994}) who elect to delay curative intervention in favor of active surveillance (AS). Results of all prior PSA tests and diagnostic biopsies are collected at enrollment. As part of the surveillance regimen, PSA tests are performed every six months and biopsies are performed annually, though biopsy intervals may vary based upon patient preferences and clinician recommendations. Treatment is recommended upon biopsy grade reclassification, that is, when the Gleason score assigned on a biopsy first exceeds six. Some patients also choose to undergo treatment prior to reclassification. For patients who elect surgical removal of the prostate, the true Gleason score assigned to the entire prostate after pathologic assessment is recorded when available.

\begin{figure}
\begin{center}
\includegraphics[width=\textwidth]{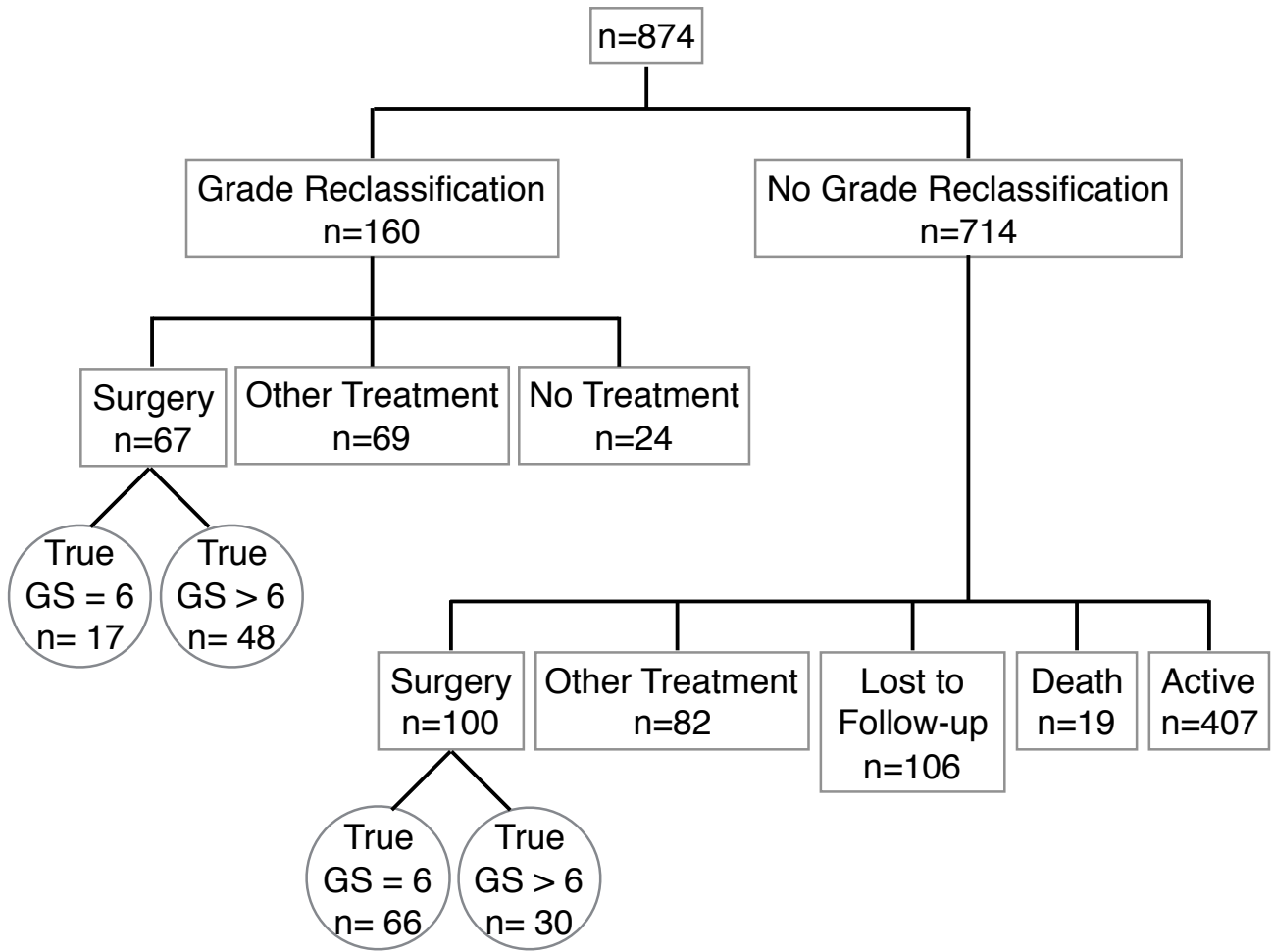}
\caption{CONSORT diagram for Johns Hopkins Active Surveillance prospective cohort patients included in this analysis. Post-surgery full prostate Gleason score (GS) observations are also given (circled). (Six patients who underwent prostatectomy did not have true GS observations available.)}
\label{fig:consort}
\end{center}
\end{figure}

A total of 874 patients who met study criteria and had at least two PSA measurements and at least one post-diagnosis biopsy as of October 1, 2014 were included in the analysis. Patient outcomes are given in Figure \ref{fig:consort}. Grade reclassification was observed in 160 patients (18$\%$ of the analysis cohort). Notably, over a quarter of patients with grade reclassification who underwent prostatectomy were downgraded after surgery (17/65) while nearly a third of patients who underwent prostatectomy in the absence of grade reclassification were upgraded (30/96). Further details on the analysis dataset are given in the online supplement. 

\subsection{Model Specification}
We applied models with and without the biopsy and surgery informative observation process (IOP) components to data from the JHAS cohort. 

PSA observations were modeled with a hierarchically-centered multilevel model, as described in Section 2.2. Patient-level coefficients for intercept and age were estimated for each patient. A shared covariance matrix was assumed for the unscaled patient-level effects, that is,  $\Sigma_0=\text{Var}(\check{\bmb_i}|\eta_i=0) = \text{Var}(\check{\bmb_i}|\eta_i=1) = \Sigma_1,$ in order to reduce model complexity. (The plausibility of this assumption was checked by examining estimated covariance matrices in the subset of patients with known cancer state.) The PSA model also included a population-level coefficient for prostate volume.

Biopsy, reclassification, and surgery observations were categorized into annual intervals, and exploratory data analysis was performed to identify predictors of each. Covariates were selected that lowered the Akaike Information Criterion (AIC) of a multivariable logistic regression model for each outcome \citep{Akaike1998} and are listed in Figure \ref{fig:model-summary}. Natural splines with up to four degrees of freedom and knots at percentiles of the predictor variable were used when doing so lowered the AIC. Additional details are given in the online supplement. 

Model parameters and their minimally informative priors are presented in the model summary given in Figure \ref{fig:model-summary}. Posterior sampling was performed with \texttt{JAGS} \citep{Plummer2015} via the \texttt{R} package \texttt{R2JAGS} \citep{R2jags2015}. Parallel chains were run to confirm model estimates converged to similar values. Cumulative quantile and trace plots were also used to monitor convergence. Analysis code, posterior sampler settings, and diagnostic plots are included in the supplementary material.


\subsection{Model Assessment}

Predictive accuracy was assessed among patients with post-surgery true Gleason score observations. Out-of-sample posterior predictions of $\eta$ were obtained for each patient by removing his true state observation from the analysis dataset and re-running the posterior sampler with an additional data augmentation step for the patient of interest. Out-of-sample predictions of $\eta_i$ were then compared to known values with receiver operating characteristic (ROC) curves \citep{Hanley1982} and calibration plots \citep{Steyerberg2010}. For the former, the area under the curve (AUC) and associated 95$\%$ bootstrapped intervals were calculated. For the latter, a plot comparing posterior predictions to observed rates of class membership was constructed by performing logistic regression of the observed true state on a natural spline representation of out-of-sample posterior predictions (degrees of freedom = 2). The mean squared error (MSE) between observed and predicted cancer state was also calculated. For comparison, posterior predictions were obtained from a logistic regression model fit with data from patients with post-surgery observations of $\eta$; covariates included age, time since diagnosis, and PSA and biopsy results. We also compared specificity of model predictions to the specificity of using final biopsy results to predict the true cancer state by fixing sensitivity at the observed true positive rate of biopsy Gleason score (dichotomized $<$ 7 or $\geq 7$).

Calibration plots were also drawn to assess model fit for outcomes observed on all patients: the occurrence of a biopsy, grade reclassification on biopsy, and the occurrence of surgery. 
Code for reproducing all plots is available in the supplementary material.


\subsection{Simulations}
We performed a simulation study to examine model performance in this application. 200 simulated datasets were generated using posterior estimates of model parameters obtained from the biopsy and surgery IOP analysis of JHAS data. For each dataset, the proposed model was estimated under four settings: unadjusted (no IOP components), biopsy IOP only, surgery IOP only, and both biopsy and surgery IOP components. Posterior predictions of the latent state were obtained for all simulated patients and compared to known (data-generating) values. For patients without surgery, posterior samples of $\eta$ were generated with a data augmentation step as a matter of course in model estimation. For patients with surgery, the posterior probability of $\eta=1$ was estimated via an importance sampling algorithm performed on the joint posterior \citep{Bishop2006}, which is less computationally intensive than the out-of-sample methods used in Section 3.3. (See technical report of \citet{Fisher2015} for further details.) Posterior predictions of $\eta$ were also compared to fitted probabilities from a logistic regression model. Code for generating data, estimating the joint posterior, and obtaining predictions is included in the supplementary material.

\vspace{-14pt}

\section{Results}

The estimated marginal probability of harboring a prostate cancer with Gleason score above 7 was 0.23 (95$\%$ CI: 0.16, 0.33) for the proposed model with biopsy and surgery IOP components, 0.20 (0.14, 0.28) with surgery IOP only, 0.31 (0.24, 0.39) with biopsy IOP only and 0.30 (0.23, 0.38) with no IOP components. Patients with $\eta=1$ were less likely to receive biopsies--leading to underestimation of $\rho$ in models without the biopsy IOP component--and more likely to elect surgery, such that $\rho$ was overestimated when not accounting for informative observation. Parameter estimates and credible intervals from all models are given in the online supplement (Appendix Tables A3-A7). 

\begin{figure}
\begin{center}
\begin{subfigure}[b]{0.49\textwidth}
\includegraphics[width=\textwidth]{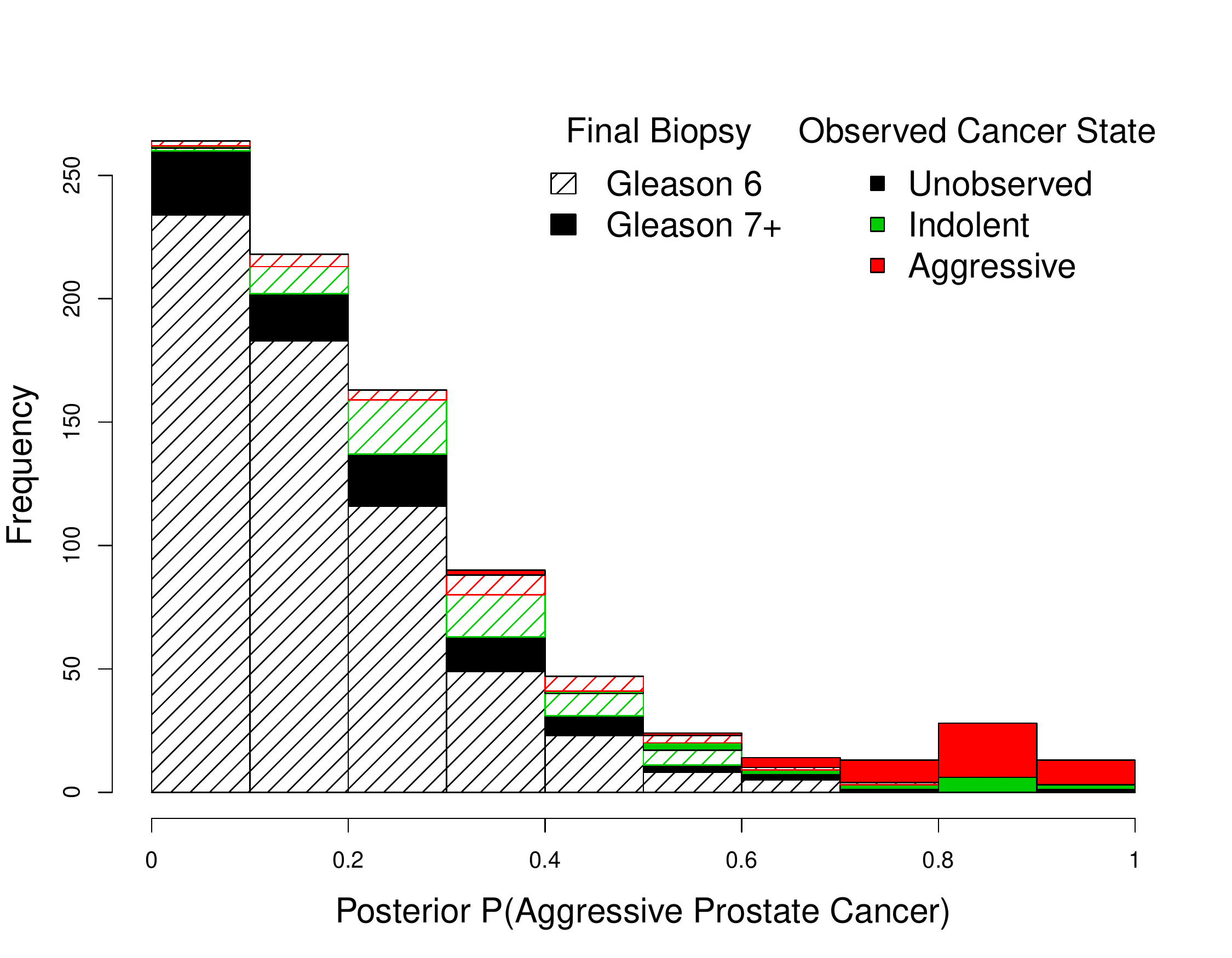}
\caption{Histogram of $B,\,S$ IOP predictions}
\label{fig:hist-eta-pred}
\end{subfigure}
\begin{subfigure}[b]{0.49\textwidth}
\includegraphics[width=\textwidth]{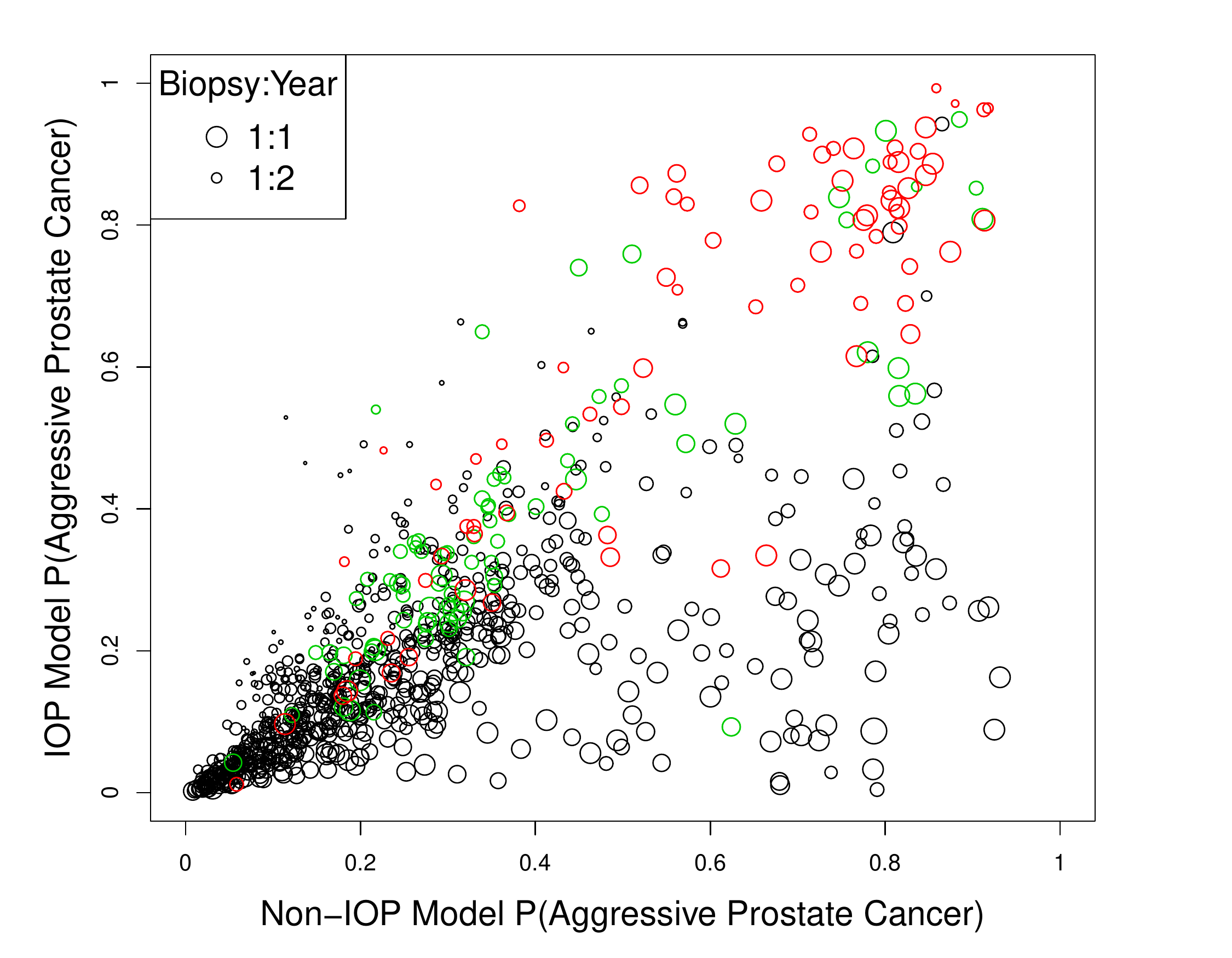}
\caption{$B,\,S$ IOP vs Non-IOP predictions}
\label{fig:scatter-eta-pred}
\end{subfigure}
\caption{Posterior predictions of true prostate cancer state. On both plots, coloring indicates whether $\eta$ was observed and, if so, its value. In the histogram (a), diagonal shading represents patients whose final biopsy was assigned a Gleason score of 6 while solid bars represent patients whose final biopsy was assigned a Gleason score of 7 or higher (i.e., grade reclassification). In the scatterplot (b), circle size indicates the frequency with which a patient received prostate biopsies; larger circles represent more frequent biopsies while smaller circles represent less frequent biopsies.}
\label{fig:post-preds-eta}
\end{center}
\end{figure}

A histogram of predictions of $\eta$ from the model with biopsy and surgery IOP components is given in Figure \ref{fig:hist-eta-pred}. Patients with posterior predictions above 60$\%$ are primarily those who both experienced grade reclassification (solid bars) and elected prostatectomy (red and green). 95$\%$ of AS patients who neither reclassified (diagonal shading) nor underwent surgery (black) have posterior predictions that are lower than 50$\%$; a majority have predictions below 20$\%$. Figure \ref{fig:scatter-eta-pred} shows a scatterplot comparing posterior probabilities of aggressive cancer, $P(\eta=1)$, between models with and without IOP components. The models produce similar posterior predictions for most patients, particularly those patients for which the non-IOP model assigns very low risk. Inclusion of biopsy and surgery IOP components decreases posterior predictions of $\eta$ most markedly for patients with frequent biopsies and no surgery (larger black circles below the x=y axis) and tends to increase posterior predictions for patients who elect surgery or have infrequent biopsies (colored or smaller black circles, respectively, above the x=y axis). These trends are further illustrated by density plots of predictions of $\eta$ stratified by reclassification and surgery given in the online supplement.

\begin{figure}
\begin{center}
\begin{subfigure}[b]{0.9\textwidth}
\includegraphics[width=\textwidth]{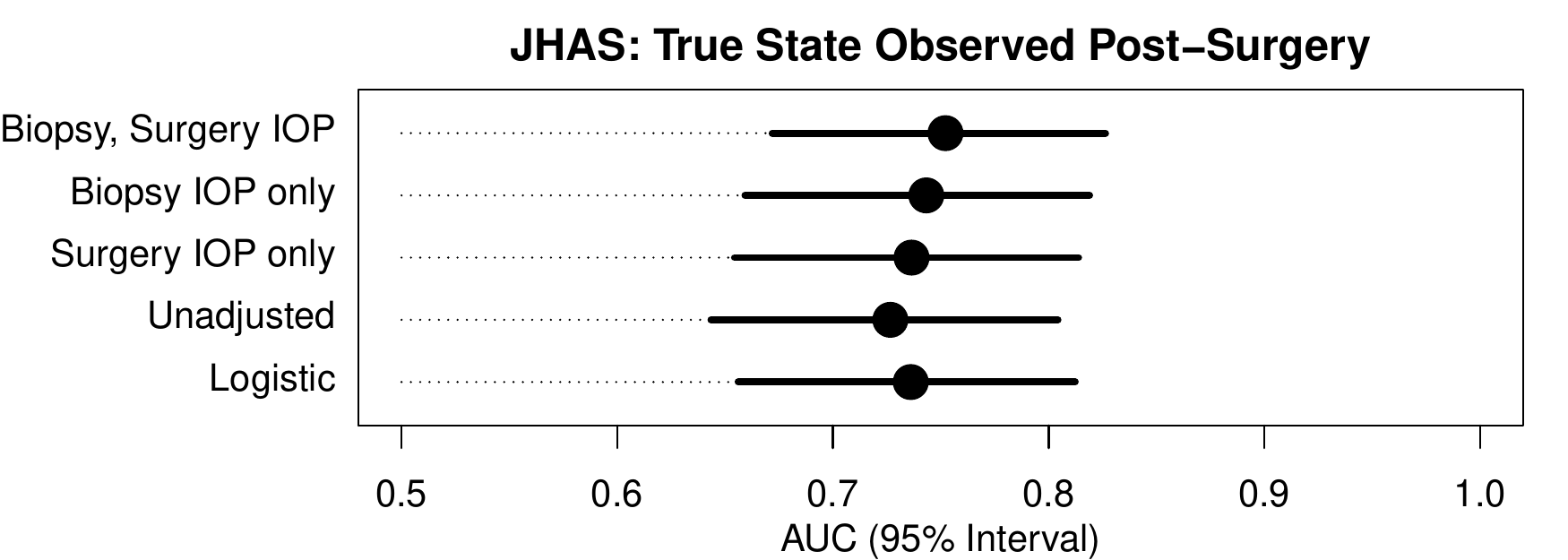}
\caption{} 
\label{fig:jhas-auc}
\end{subfigure}

\begin{subfigure}[b]{0.9\textwidth}
\vspace{20pt}
\includegraphics[width=\textwidth]{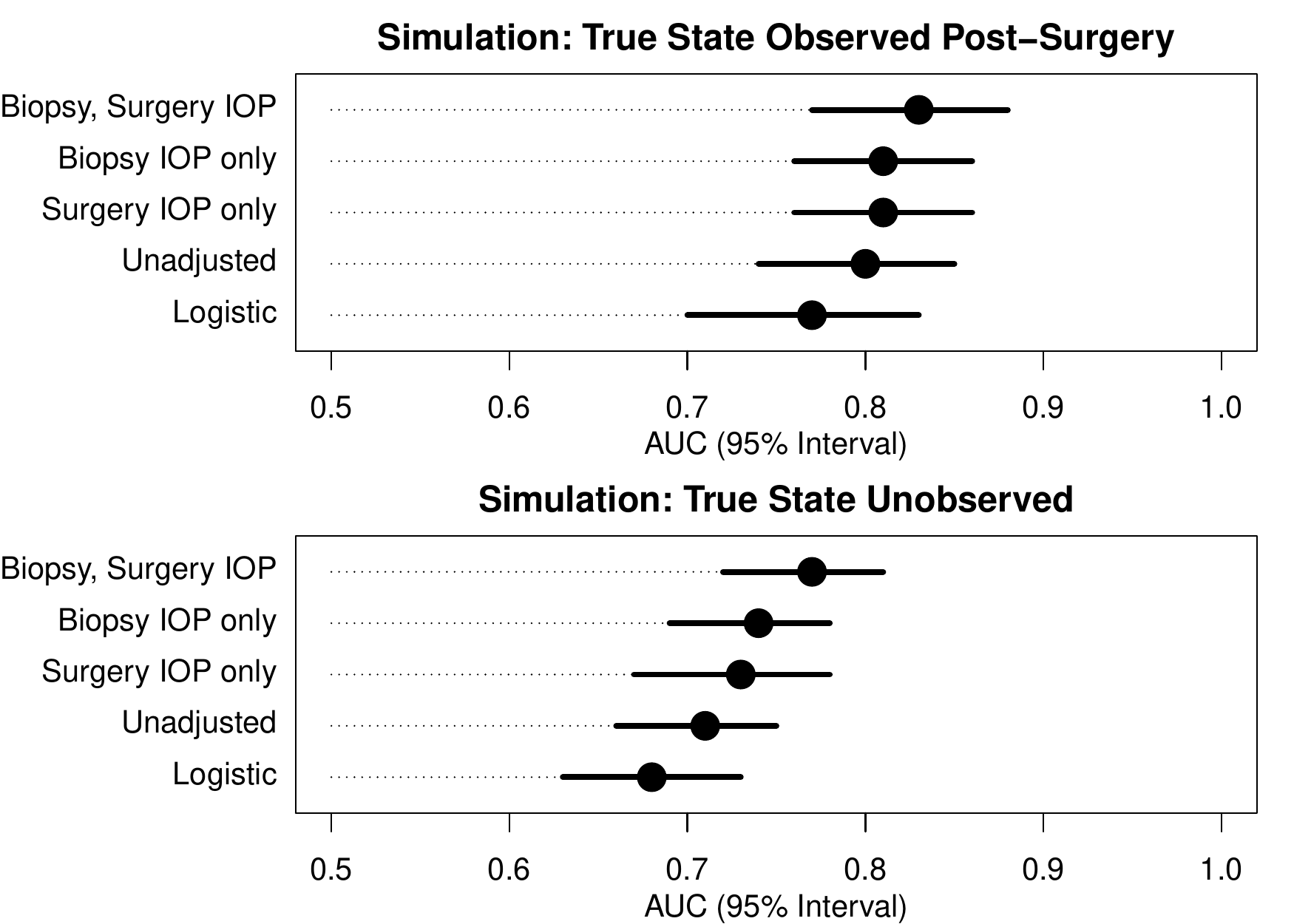}
\caption{} 
\label{fig:auc-sim}
\end{subfigure}
\caption{Estimated AUC for predictions of $\eta$ from the JHAS cohort (a) and simulation studies (b). Intervals in (a) are quantile-based 95$\%$ intervals from 10,000 bootstrap samples of patients with post-surgery observations of $\eta$. Intervals in (b) are quantile-based 95$\%$ intervals from the estimated AUC in 200 simulation studies.}
\end{center}
\end{figure}

\begin{figure}
\begin{center}
\begin{subfigure}[b]{0.475\textwidth}
\includegraphics[width=\textwidth]{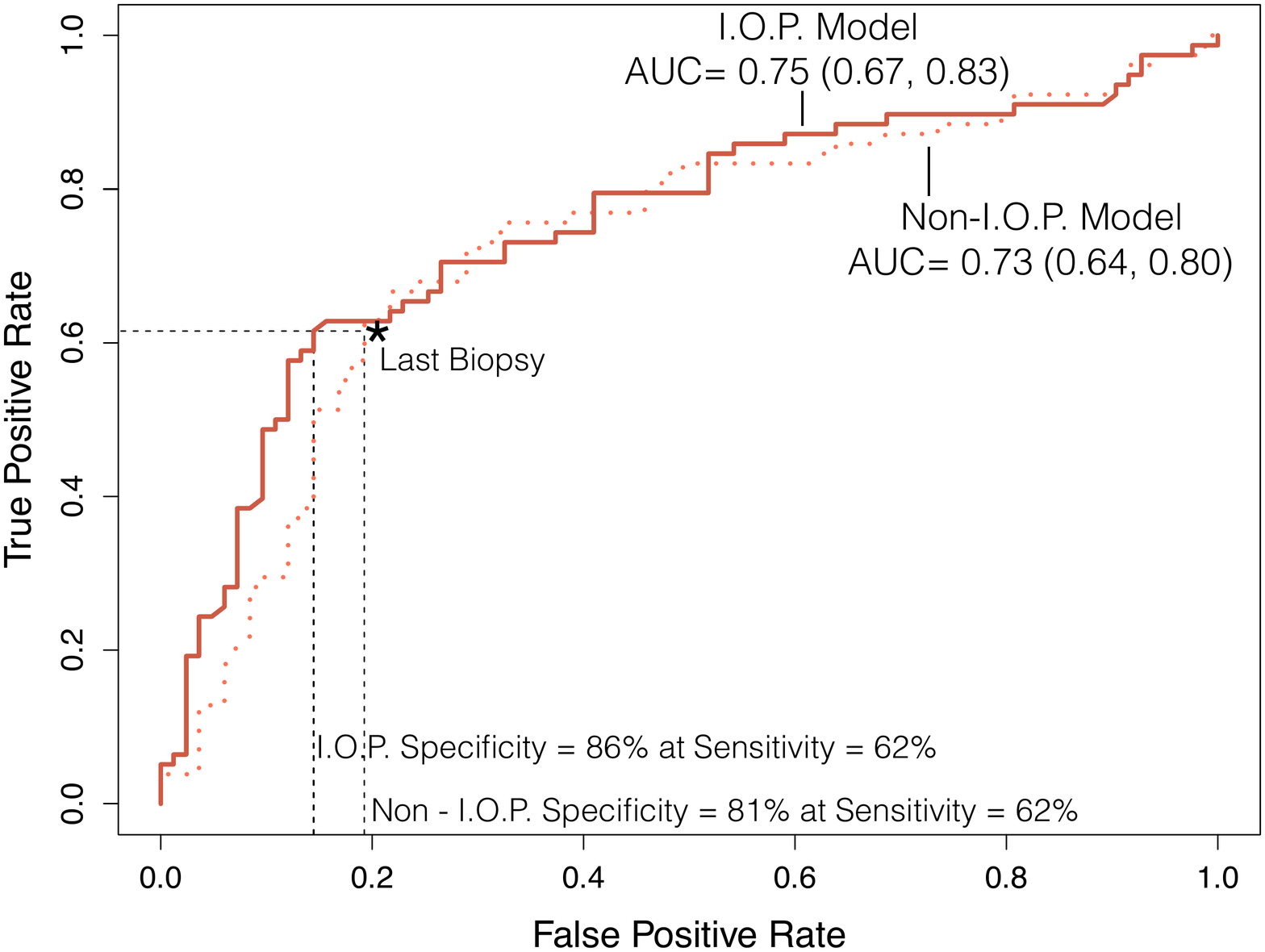}
\caption{ROC curves: $B,S$ IOP $\&$ non-IOP predictions}
\label{fig:roc}
\end{subfigure}
\begin{subfigure}[b]{0.475\textwidth}
\includegraphics[width=\textwidth]{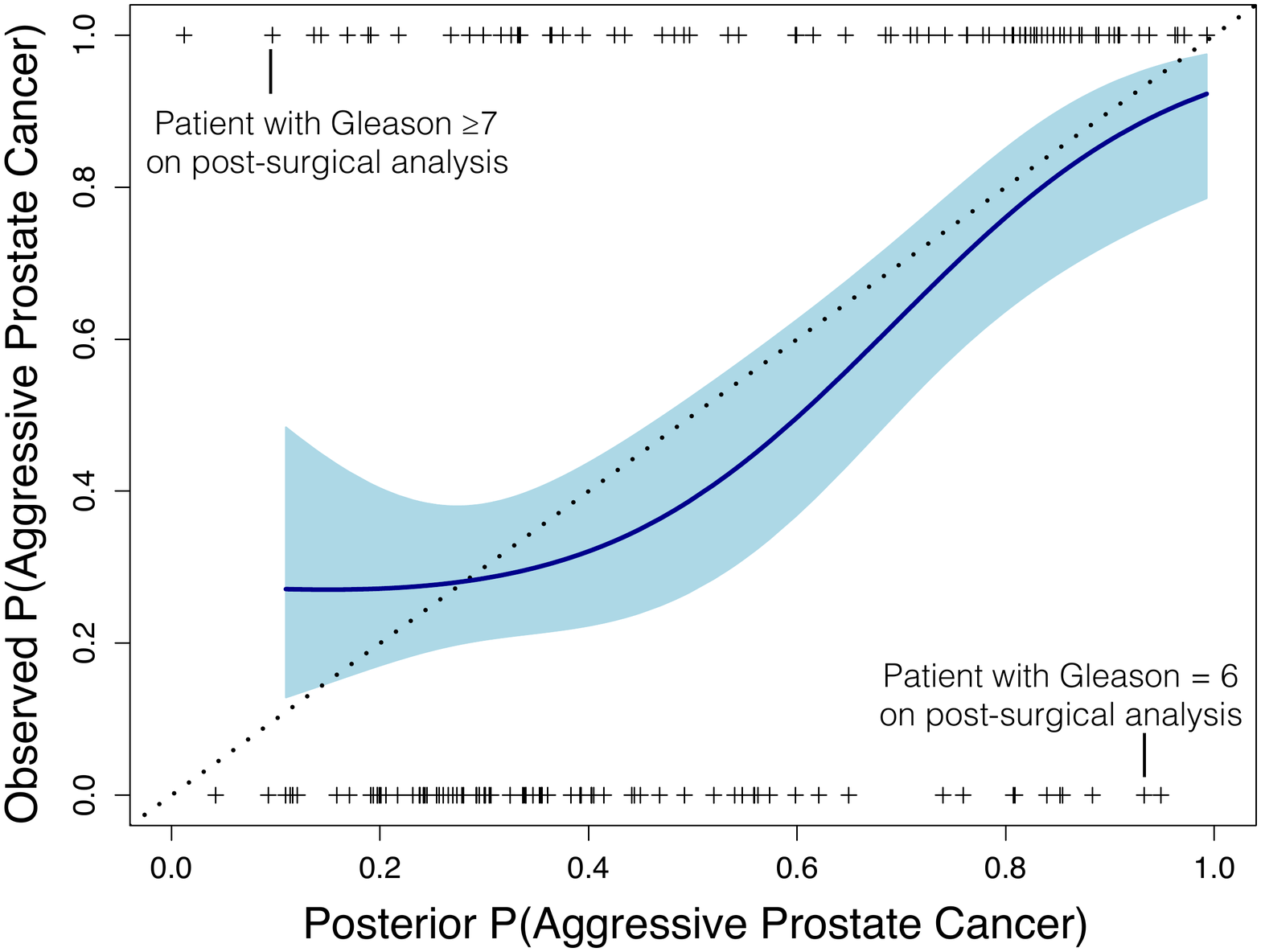}
\caption{Calibration plot: $B,\,S$ IOP predictions}
\label{fig:calibration-eta}
\end{subfigure}
\caption{Predictive accuracy of out-of-sample predictions of $\eta$ among patients with true state observed in the JHAS cohort. In (a), the specificity of predictions from each model is highlighted at the sensitivity of a binary classifier defined by final biopsy result (*). In (b), the dark line shows the empirical rate of observing a true Gleason score of 7 or above (y-axis) given an out-of-sample posterior probability of true state (x-axis) under the model with biopsy and surgery IOP components; shading gives the 95$\%$ point-wise confidence interval. Perfect agreement lies on the x=y axis (dotted line). Hashmarks at y=0 and y=1 correspond to observed cancer states ($\eta=0,1$, respectively) for patients with post-surgery true state observations. Hashmarks are located along the x-axis at each patient's out-of-sample posterior probability of the true state.}
\end{center}
\end{figure}

Posterior predictions of $\eta$ from the proposed model with biopsy and surgery IOP components were more accurate than those from the proposed model with a single or no IOP components or from logistic regression (Figures \ref{fig:jhas-auc} and \ref{fig:roc}). The out-of-sample AUC among patients with observed true cancer state is highest for the biopsy and surgery IOP model (0.75, 95$\%$ bootstrapped interval: 0.67, 0.83), and the MSE from this model was also the lowest (0.201, 95$\%$ Int: 0.17, 0.24; Table 9 in online supplement). While this improvement is slight, it is widely recognized that drastic increases in classification accuracy are rare to achieve \citep{Pepe2004}. The false positive rate (FRP) of predictions from the biopsy and surgery IOP model (0.14, 95$\%$ Int: 0.07, 0.23) is also lower than that of the binary classifier based on final biopsy results (FPR=0.20, 95$\%$ Int: 0.12, 0.29) at a fixed true positive rate of 0.62 (the sensitivity of final biopsies in patients with eventual surgery). The improvement in specificity offered by the biopsy and surgery IOP model corresponds to avoiding, on average, 30$\%$ of unnecessary diagnoses of more aggressive cancer in comparison to a diagnosis based solely on a patient's most recent biopsy (95$\%$ bootstrapped interval of FPR($R$) - FPR($\eta$): -7.5$\%$, 13$\%$). These comparisons are limited because accuracy of posterior predictions can only be assessed among patients with $\eta$ observed. Yet, we expect the predictive accuracy gained by incorporating IOP components to be seen more definitively in patients without true state observations if biopsy and surgery results are indeed MNAR. This is explored further with simulations. 

Posterior predictions of $\eta$ from the IOP model also appear to accurately estimate a patient's risk of having more aggressive cancer. The calibration plot in Figure \ref{fig:calibration-eta} shows that, for patients with known values of $\eta$, the average posterior predicted probability of $\eta=1$ is close to the average observed value of $\eta$, indicating that the model reasonably reproduces the mean of observations. The risks of clinical outcomes (biopsy results) and choices (occurrence of biopsy and surgery) for all patients appear to be accurately estimated by the IOP model as well, as demonstrated by calibration plots in the online supplement (Appendix Figure A3).

Estimation of the time-to-surgery model depends on sufficient evidence among patients with surgery of a relationship between $\eta$ and surgery time. To assess the strength of this evidence, we re-ran the biopsy and surgery IOP model with more informative priors on the coefficients capturing this relationship and found posterior estimates to be robust. Details are given in the online supplement (Appendix Section A3.1.6 and Figure A18).

Through simulation studies, we found that our sampling procedure produced unbiased estimates with nominal coverage (Tables 10-14 in the online supplement) and that the proposed model with biopsy and surgery IOP components outperforms other model variations and logistic regression when it correctly reflects the data-generating mechanism. The AUC for each model among patients with and without post-surgery true state observations are given in Figure \ref{fig:auc-sim}. Differences in predictive accuracy across models are similar in magnitude to those observed in the application (Figure \ref{fig:jhas-auc}).  As expected, we also see that the logistic regression model, which is only estimated with data from patients with true state observations, has the poorest predictive accuracy among patients without surgery. Predictions from models incorporating the surgery IOP component show appropriate calibration while those without overestimate the risk of aggressive prostate cancer (Figure 20, online supplement).


\section{Discussion}
In this paper, we have presented a hierarchical Bayesian model for predicting latent cancer state among low risk prostate cancer patients. Multiple models have been developed to predict biopsy results in this population \citep{Ankerst2015, Truong2013}. However, our model predicts the outcome of chief interest--the true underlying state of an individual's prostate cancer. Focusing on the actual health state, even when latent, is equivalent to subsetting patients into subgroups for which optimal treatments differ. Subsetting is the goal of precision medicine \citep{Saria2015}.

The proposed model integrates four sources of information about whether a tumor is aggressive or indolent: repeated measures of the biomarker PSA; repeated results from tissue biopsies; repeated decisions to have a biopsy; and the time to surgical removal of the prostate. In the subset of patients that have their prostates removed, the true tumor pathology state is observed. This data-integrating method is an example of semi-supervised learning because patients both with and without true state observations are included in model estimation (Chapelle, Sch{\"o}lkopf, and Zien, 2006)\nocite{Chapelle2006}. We adjust for possible informative missingness by modeling the time until surgery depending on the true state. While it is ideal to assess model sensitivity to parametric assumptions embedded in selection models for missing not at random data mechanisms \citep{Daniels2008}, existing methods for re-parameterizing selection models as pattern mixture models do not accommodate event time outcomes with the possibility of censoring. Further research is needed to develop these methods. 

The methods proposed here are tailored to available measurements that address the clinical questions arising from active surveillance of prostate cancer: should I have a biopsy this year; what is the chance my tumor is indolent; should I undertake removal or irradiation of my prostate despite the known serious side effects? The model extends naturally to provide improved answers to these questions as additional data become available. For example, when genetic markers for prostate cancer risk are identified, the probability distribution for latent state ($\rho_i$) could easily be informed by subgroups defined by their expression. Or, when MRI or ultrasound images are commonly used before biopsy, these data will be included in the model as well. In the case that some measurements are not available for all patients, the proposed framework is also able to adjust for informative observation of predictors and outcomes. 

The proposed model can also be modified in response to advancement in scientific understanding about the relationship between clinical measurements and the underlying cancer state. In particular, in the event of new research findings on the rate of progression in this population, the model could be extended to allow an indolent cancer to transition to a lethal one, for example, as a Markov process. Because an individual's true cancer state can only be observed once, the current data contain insufficient information to simultaneously support identifiability of both the rate of biopsy misclassification and the rate of pathological progression in the underlying state. The model currently assumes that an individual's cancer categorization (Gleason score) does not change over the time period under surveillance while allowing for imperfect sensitivity and specificity of biopsies. This assumption reflects the current clinical understanding that biopsy upgrading in AS is more frequently due to misdiagnosis rather than true grade progression \citep{Porten2011}. A more recent analysis by \citet{Inoue2014} suggests a rate of disease progression in the JH AS cohort of 12-14$\%$ within a decade of enrollment, but this estimate is sensitive to prior specification. A dynamic state extension would require strong prior knowledge about the progression rate parameter in order to be identifiable from the current data. The effect of allowing for a state transition would be to give greater weight to more recent PSA and biopsy outcomes when predicting the underlying state rather than giving equal weight to all observations.

The proposed prediction model exemplifies the statistical underpinnings of a learning health care system (Goolsby, Olsen, and McGinni, 2012; Smith et al., 2013)\nocite{IOM2012,Smith2013}, a system with the ability to continuously integrate patient data and medical knowledge to optimize patient care. As more patients enroll in the Johns Hopkins active surveillance cohort, and as more information is collected on existing patients, our ability to predict underlying health states and the likely trajectory of clinical outcomes will improve. Furthermore, importance sampling methods can be used to obtain real-time prediction updates based on the most current information in order to support decision-making in a clinical setting. An example interactive decision-support tool that provides fast predictions of a patient's latent prostate cancer state is demonstrated at \texttt{https://rycoley.shinyapps.io/} \texttt{dynamic-prostate-surveillance}.


\section*{Supplementary Materials}
Simulated data, \texttt{JAGS} scripts, and \texttt{R} code to reproduce the analysis and figures are provided at \texttt{http://github.com/rycoley/prediction-prostate-surveillance}.

\bibliographystyle{biom}
\bibliography{/Users/ryc/Documents/yates-master-bib}


\pagebreak

\setcounter{equation}{0}
\setcounter{figure}{0}
\setcounter{table}{0}
\setcounter{page}{1}
\setcounter{section}{0}
\makeatletter
\renewcommand{\theequation}{A\arabic{equation}}
\renewcommand{\thefigure}{A\arabic{figure}}
\renewcommand{\thetable}{A\arabic{table}}
\renewcommand{\thesection}{A\arabic{section}}
\renewcommand{\bibnumfmt}[1]{[A#1]}
\renewcommand{\citenumfont}[1]{A#1}

\begin{center}
\begin{large}
APPENDIX:\\
A Bayesian hierarchical model for prediction of latent health states from multiple data sources with application to active surveillance of prostate cancer\\
\end{large}
\end{center}


\noindent This appendix contains additional details and results related to the proposed method, its application to the Johns Hopkins Active Surveillance (JHAS) cohort, and simulations based on the JHAS estimates.

\section{Methods}

\subsection{Posterior Estimation}
The joint posterior distribution of the parameters, latent states, and patient-level effects is written as proportional to the likelihood given in Equation (3) and joint prior density of model parameters:
\bea
&& p\Big\{\rho, \bmbeta, \xi,  \sigma^2, \bmnu, \bmgamma, \bmomega; \, (\bmmu_k,\Sigma_k); \, \check{\bmb_i}, i=1,\mydots,n; \,\eta_i, i=1,\mydots,n_{S=0} \, | \nonumber\\ \, 
&& \qquad   \eta_i, i=n_{S=0}+1, \mydots, n; (\bmY_i, \underline{\bmX_i}, \underline{\bmZ_i}), (\mathbf{B}_{i}, \underline{\bmU_{i}}), (\mathbf{R}_{i}, \underline{\bmV_{i}}), (\mathbf{S}_{i}, \underline{\bmW_{i}}), i=1,\mydots,n; \,\bmTheta\Big\} \nonumber\\
&&\qquad  \propto L\Big\{\rho,  \bmbeta, \xi,  \sigma^2, \bmnu, \bmgamma, \bmomega; \, (\bmmu_k, \Sigma_k); \, \check{\bmb_i}, i=1,\mydots,n; \, \eta_i,  i=1,\mydots,n_{S=0}  \, | \nonumber\\ \, 
&& \qquad \qquad \eta_i, i=n_{S=0}+1,\dots,n;\, (\bmY_i,  \underline{\bmX_i}, \underline{\bmZ_i}), (\mathbf{B}_{i}, \underline{\bmU_{i}}), (\mathbf{R}_{i}, \underline{\bmV_{i}}), (\mathbf{S}_{i}, \underline{\bmW_{i}}), i=1,\mydots,n  \Big\} \nonumber\\
\label{eq:post-inf}
&& \qquad \qquad \qquad \qquad \times \bmpi\big\{\rho,\bmbeta, \xi, \sigma^2, \bmnu, \bmgamma, \bmomega; (\bmmu_k, \Sigma_k)\, |\,\bmTheta\big\} \nonumber
\eea
where $\bmpi(\cdot|\bmTheta)$ denotes the joint prior density for model parameters with hyperparameters $\bmTheta$ and indexing on $j$ and $k$ suppressed for clarity in presentation. Patients are indexed such that  $i=1,\dots,n_{S=0}$ refers to patients without surgery ($S=0$) and for whom $\eta_i$ is latent and $i=n_{S=0}+1,\dots,n$ refers to patients with eventual surgery and observation of $\eta_i$. Similar to the notation used in Equation (3),  $f$ and $g$ are multivariate normal densities for the vector of log-transformed PSAs $\bmY_i$ and unscaled random effects $\check{\bmb_i}$, respectively, each with mean and covariance as in Section 2.2 of the accompanying paper. $\underline{\bmX_i}$ denotes the matrix of covariate vectors $[\bmX_{i1},\dots, \bmX_{iM_i}]$; $\underline{\bmZ_i}$, $\underline{\bmU_{i}},\underline{\bmV_{i}}$, and $\underline{\bmW_{i}}$ are similarly defined. $\mathbf{B}_{i}, \mathbf{R}_{i}$, and $\mathbf{S}_{i}$ denote vectors of all biopsy, reclassification, and surgery observations for individual $i$.

\section{Application to Johns Hopkins Active Surveillance Cohort}

\subsection{The Data}

The number of observations and years of follow-up available for analysis are summarized in Table \ref{tab:num_obs}. 318 patients (36$\%$) were censored due to receiving some treatment, 130 (15$\%$) were lost to follow-up, and 19 (2.2$\%$) were censored due to death. (No patients died of prostate cancer.) Loss to follow-up was defined as two years without a PSA or biopsy after the most recent observation. 407 patients (47$\%$) remained active in the program at the time of data collection.

As shown in the CONSORT diagram in Figure 3 of the accompanying paper, grade reclassification was observed in 160 patients (18$\%$ of analysis cohort). Among patients with grade reclassification, 67 patients elected surgical removal of the prostate. An additional 100 patients elected prostatectomy in the absence of grade reclassification. In total, 167 patients (19$\%$ of analysis cohort) underwent surgery,  of which 161 had a definitive post-surgical Gleason score determination. Results of the biopsy-based estimated Gleason score and post-surgical true value are shown in Table \ref{tab:eta-vs-rc}.

\subsection{Model Specification}

\subsubsection{PSA model} Prostate volume is a known source of patient-level variability in PSA and, for this reason, was included as a predictor in the multilevel model for log-PSA. Prostate volume was measured via ultrasound at some biopsies. Since increases in prostate volume due to age and cancer activity are expected to be of a smaller magnitude than the measurement error in ultrasound-guided volume assessment, the average of all available prostate volume observations was used for each patient.  

\subsubsection{Biopsy, reclassification, and surgery models} The JHAS protocol is to perform a biopsy once per year. Hence, biopsy, reclassification, and surgery observations were categorized into annual intervals. A small number (1$\%$) of intervals contained two biopsies. To accommodate this, we redefine the logistic regression model in Equation (1) as the probability of any biopsies during the year. Intervals with two biopsies then contributed two conditionally independent reclassification outcomes (Equation (2)) to the likelihood.

For the biopsy, reclassification, and surgery models, natural spline representations of continuous and discrete predictors (age, time in AS, calendar date, number of previous biopsies, and extent of cancer found in previous biopsies) were included when doing so lowered the AIC. The selected degrees of freedom and location of quantile-based knots for each predictor are identified in Tables \ref{tab:jhas-post-est-bx}, \ref{tab:jhas-post-est-rc}, and \ref{tab:jhas-post-est-surg}.

\subsection{Simulations}

200 simulation datasets were generated using characteristics of the JHAS data and posterior estimates from the proposed model with biopsy and surgery IOP components. For each simulated dataset, the proposed model was estimated with no IOP components, biopsy only IOP, surgery only IOP, and both biopsy and surgery IOP components. The posterior median was recorded for each parameter as well as whether the 95$\%$ quantile-based posterior credible interval contained the true data-generating value. The mean posterior median and coverage were summarized across all simulated datasets.
Data-generating values for model parameters are give in Tables \ref{tab:sim-post-est-rho}-\ref{tab:sim-post-est-surg}. Code for simulating data is provided in the online supplement.

\section{Results}

\subsection{Application to Johns Hopkins Active Surveillance cohort}
\subsubsection{Posterior parameter estimates}
Posterior estimates and 95$\%$ quantile-based credible intervals for model parameters are reported in Tables \ref{tab:jhas-post-est-rho}-\ref{tab:jhas-post-est-surg}. Results are given for four versions of the proposed models: no IOP components (``Unadjusted"), biopsy IOP component only ($B$ IOP), surgery IOP component only ($S$ IOP), and both biopsy and surgery IOP components ($B,\,S$ IOP). 

In Table \ref{tab:jhas-post-est-rho}, we see that the estimated marginal probability of harboring aggressive cancer ($\rho$) is higher in models that include a biopsy IOP component and lower in models with a surgery IOP component. This observation is consistent with posterior estimates of $\nu$ (Table \ref{tab:jhas-post-est-bx}) and $\omega$ (Table \ref{tab:jhas-post-est-surg}). Coefficient estimates in the biopsy model indicate that patients with $\eta=1$ are less likely to receive an annual biopsy (last row in Table \ref{tab:jhas-post-est-bx}). Without adjusting for MNAR biopsy results, the modeling approach is overly optimistic about a patient's true cancer state because it assumes that a patient who skips a biopsy is as likely to have a favorable biopsy as a patient with the same covariate data who does have a biopsy performed. Meanwhile, inclusion of the surgery IOP component identifies evidence that patients with $\eta=1$ are more likely to elect surgical removal of the prostate, particularly if they have also experienced grade reclassification (last two rows in Table \ref{tab:jhas-post-est-surg}). Without accounting for this informative missing data mechanism, we run the risk of overestimating risk in this population.

\subsubsection{Posterior estimates of $\eta$}
Figure \ref{fig:jhas-density-post-eta} shows density plots for posterior predictions of $\eta$ for all patients in the JHAS cohort from four non-IOP/IOP combinations of the proposed model and a logistic regression model (coefficient estimates in Table \ref{tab:jhas-logistic-est}). This diagram reinforces the model comparisons illustrated in the histogram and scatterplot in Figure 4 of the accompanying paper. In particular, predictions of the true cancer state for patients with grade reclassification and no surgery (top right plot) are considerably lower in the models with surgery IOP components. Below, in the simulation study, we see a similar trend when the data is generated according to the estimated dual IOP model: if patients with aggressive cancer are more likely to have surgery sooner (particularly after grade reclassification), models that do not adjust for informative surgery decisions will overestimate patient risk.

\subsubsection{Predictive accuracy}
We provide additional assessment of predictive accuracy of all models considered here. Measures of predictive accuracy of the proposed model among patients with post-surgery true state observation are summarized in Table \ref{tab:jhas-pred-accuracy}. (AUC and FPR estimates correspond to those presented in Figures 5(a) and 6(a) of the accompanying paper.) We see that the AUC, MSE, and FPR at TPR=0.62 are improved by using the proposed model with both biopsy and surgery IOP components. Calibration plots for the proposed model with no, only biopsy, only surgery, and both IOP components are given in Figures \ref{fig:jhas-calibration-unadj}-\ref{fig:jhas-calibration-surg}; a calibration plot for the logistic regression model is given in Figure \ref{fig:jhas-calibration-logistic}. All models appear to produce well-calibrated estimates. 

Calibration plots were also constructed for outcomes observed on all patients. Figure \ref{fig:jhas-calibration-observables} presents a calibration plot for the probability of clinical outcomes (biopsy results) and choices (occurrence of biopsy and surgery) under the proposed model with biopsy and surgery IOP components. Solid lines show, for each saved iteration of the sampling chain, the fitted values of a logistic regression of the observed outcome on the natural spline representation of each person-year's posterior probability of an event. Plotting symbols at y=0 and y=1 correspond to the observed outcome ($B_{ij}$, $R_{ij}$, and $S_{ij}$) and are plotted on the x-axis at the mean posterior probability for that person-year; plotting symbol shape and color indicate eventual observations of the true state. Posterior probabilities and observed rates are generally similar to each other, with closer agreement occurring in ranges with more data.

\subsubsection{Individual-level effect estimates in PSA model}
Posterior estimates from the multilevel model for PSA are displayed in Figure \ref{fig:psa-random-effects}. In this plot, each plotting circle represents the scaled patient-level intercept (x-axis) and slope (y-axis) estimates for a single patient. Filled circles represent patients for whom the true cancer state was observed, with red indicating an aggressive cancer found after surgery and green indicating a determination of indolent cancer. The color of open circles reflects the posterior probability of aggressive cancer, ranging from 0-25$\%$ (green) to 76-100$\%$ (red), among patients for whom true state was not observed. Finally, credible ellipses show the posterior mean and covariance of patient-level coefficients in each partially latent class. We see that there is a fair amount of overlap in these intervals, indicating that PSA level and trajectory does not provide strong evidence of the true state for many patients. PSA  is more informative for patients with particularly high or low levels and trajectories or for patients with shorter follow-up. We also note that the Hopkins cohort has PSA requirements for enrollment; PSA data may be more informative in a cohort with less strict enrollment criteria.

\subsubsection{MCMC settings and convergence diagnostics}
For all IOP combinations, five independent posterior sampling chains were run for 50,000 iterations. The first 25,000 iterations were discarded as burn-in and posterior samples were saved at every twentieth iteration thereafter. Convergence of the posterior sampling algorithm was assessed with cumulative density and trace plots; these are given for the model with biopsy and surgery IOP components in Figures \ref{fig:jhas-post-rho}-\ref{fig:jhas-post-omega-4} and exhibit appropriate convergence. Trace plots (left) show sampled values for each chain (indicated by color). Cumulative quantile plots (middle) show running posterior quantiles for the median (solid line) and lower 2.5 and upper 97.5 percentiles (dotted lines) for one sampling chain. Plots in the rightmost column show posterior densities for each sampling chain (indicated by color) alongside the prior probability (dotted lines).

\subsubsection{Robustness of IOP estimates}
The posterior distributions of IOP coefficients, i.e., the effect of $\eta$ on biopsy occurrence and surgery, indicate that the data contain evidence of informative missingness, as shown in Figure \ref{fig:post-coef-iop} (black, solid lines in each plot). 95$\%$ quantile-based credible intervals of the log-odds ratio (OR) for the effect of $\eta=1$ on the probability of having a biopsy in an interval (lefthand plot) and the log-OR for the interaction between $\eta=1$ and prior grade reclassification (righthand plot) exclude zero (vertical line).

An important question is whether estimation of the additional parameters in the IOP model, especially those associated with observation of the true cancer state, is supported by evidence in the data or, instead, only identifiable by the likelihood construction and prior specification. To assess robustness of posterior predictions to prior specification, we refit the IOP model with multiple informative priors on both the log-OR of surgery given true state and the log-OR of surgery given an interaction between true state and prior biopsy results. Specifically, we considered all combinations of normal priors with a variance of one and mean OR of one-half, one, and two for the association of $\eta_i$ and $\eta_i\times\mathbf{1}_{[R_{i\bar{j}}=1]}$ with the probability of surgery for patient $i$ in year $j$ (where $\mathbf{1}_{[R_{i\bar{j}}=1]}$ is an indicator of grade reclassification for patient $i$ during or prior to year $j$). The resulting posterior distributions,  shown in Figure \ref{fig:post-coef-iop}, demonstrate relative robustness to prior specification and affirm confidence in posterior predictions from the IOP model with vague priors. The primary effects of specifying these more informative priors appear to be a reduction in the variability of posterior distributions and an attenuation of the estimated effect of the interaction of $\eta=1$ and prior grade reclassification on the risk of surgery. Posterior predictions of $\eta$ and the model's predictive accuracy were not changed by specifying informative priors on IOP components (not shown). It appears that repeated contributions to the likelihood of the probability of not having surgery ($P(S_{ij}=0$)) in intervals prior to the decision to have surgery provide appropriate evidence about the relationship between the true cancer state and its eventual observation.

\subsection{Simulations}

\subsubsection{Posterior parameter estimates}
Posterior estimates and coverage for all models considered are given in Tables \ref{tab:sim-post-est-rho}-\ref{tab:sim-post-est-surg}. Estimation appears unbiased for the model with biopsy and surgery IOP components (which was used to generate data), and credible intervals from that model have nominal or slightly conservative coverage. Biased estimation of models without both biopsy and surgery IOP components is most prominent in coefficients related to the true cancer state. For example, the log odds-ratio for the association between true cancer state $\eta=1$ and risk of reclassification was overestimated in the unadjusted and biopsy IOP only models (last row, Table \ref{tab:sim-post-est-rc}).

\subsubsection{Posterior estimates of $\eta$}
\begin{figure}
\begin{center}
\includegraphics[width=\textwidth]{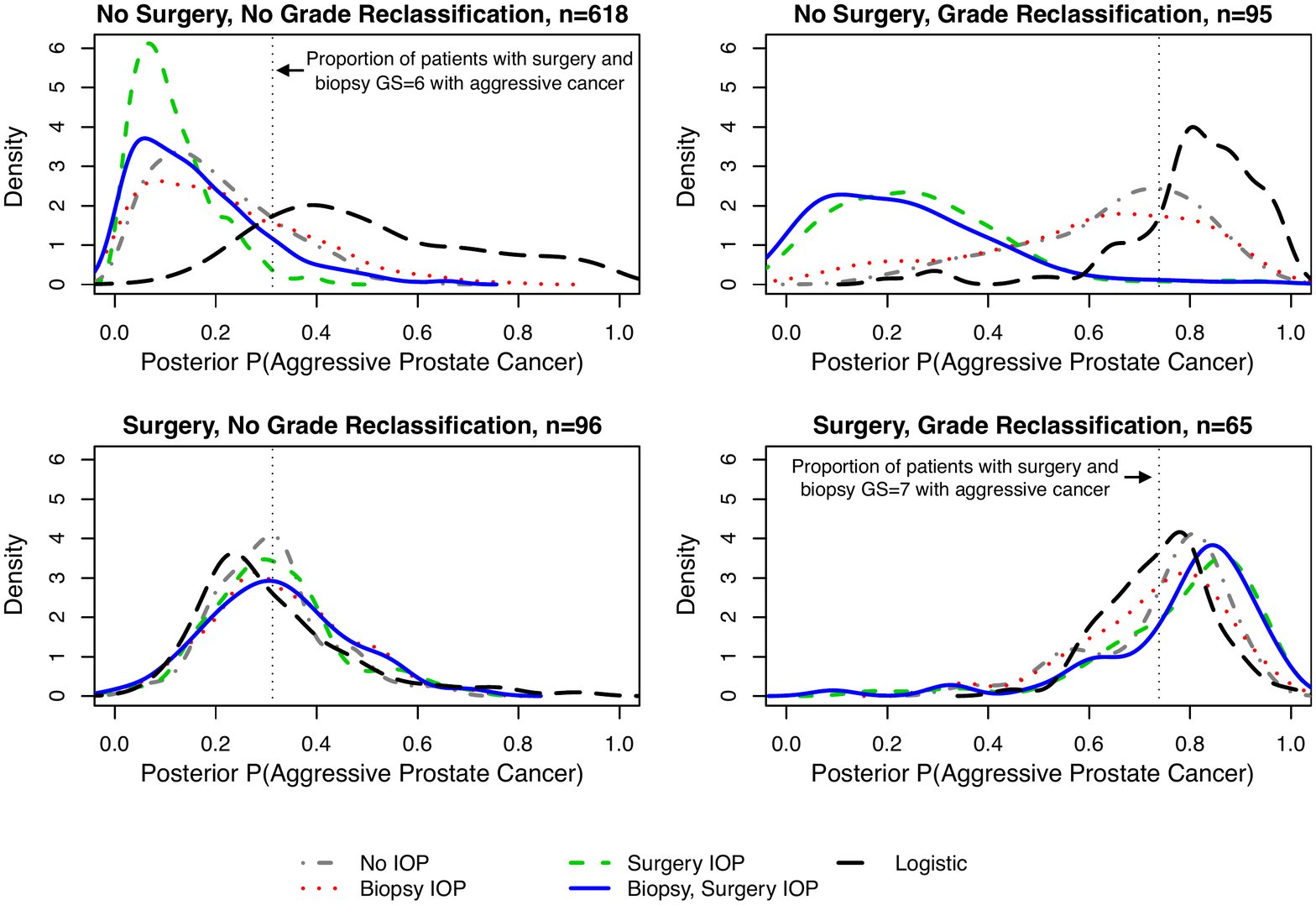}
\caption{Density plots of posterior predictions of $\eta$, stratified by biopsy results and the decision to have surgery, from the JHAS analysis. Line types correspond to different statistical models, as indicated by the legend at the bottom. The vertical dotted line represents the proportion of surgery patients with no grade reclassification (left) and reclassification (right) on biopsy who had higher grade prostate cancer on the full prostate examination.}
\label{fig:jhas-density-post-eta}
\end{center}
\end{figure}

Density estimates for the posterior predictions of $\eta$ from a single simulated dataset are shown in Figure \ref{fig:sim-density-post-eta}. These plots show similar trends in posterior predictions across model options to those observed in the application to JHAS cohort data (Figure \ref{fig:jhas-density-post-eta}), which indicates that the differences in posterior predictions across models (particularly those seen in the subgroup with grade reclassification and no surgery) would be expected if the dual biopsy and surgery IOP data generating mechanism was correct.

\subsubsection{Predictive accuracy}
Table \ref{tab:sim-pred-accuracy} gives the average AUC and MSE among patients with $\eta$ observed and unobserved from 200 simulation studies. We see that the AUC is highest for both groups of patients in the proposed model with biopsy and surgery IOP components. MSE is actually higher among patients without $\eta$ observed for all versions of the proposed model, likely due to a calibration accuracy similar to patients with $\eta$ observed and a greater sample size. MSE of predictions from the logistic regression model increased among patients without $\eta$ observed. This increase is expected since the logistic model was estimated using only data from patients with post-surgery $\eta$ observations, instead of the semi-supervised approach of the proposed model.

\begin{figure}
\begin{center}

\begin{subfigure}[b]{0.45\textwidth}
\includegraphics[width=\textwidth]{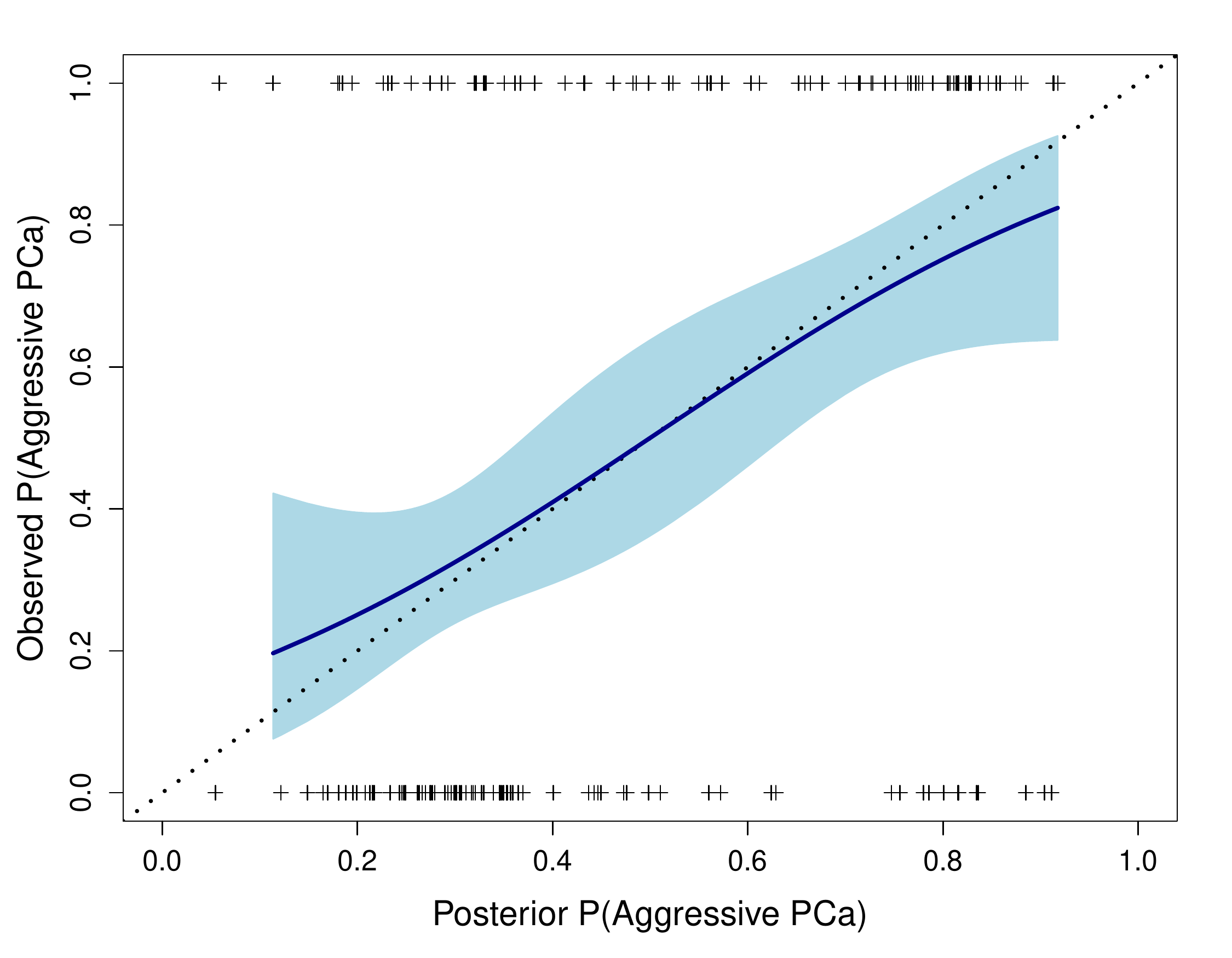}
\caption{Proposed model, unadjusted}
\label{fig:jhas-calibration-unadj}
\end{subfigure}
\begin{subfigure}[b]{0.45\textwidth}
\includegraphics[width=\textwidth]{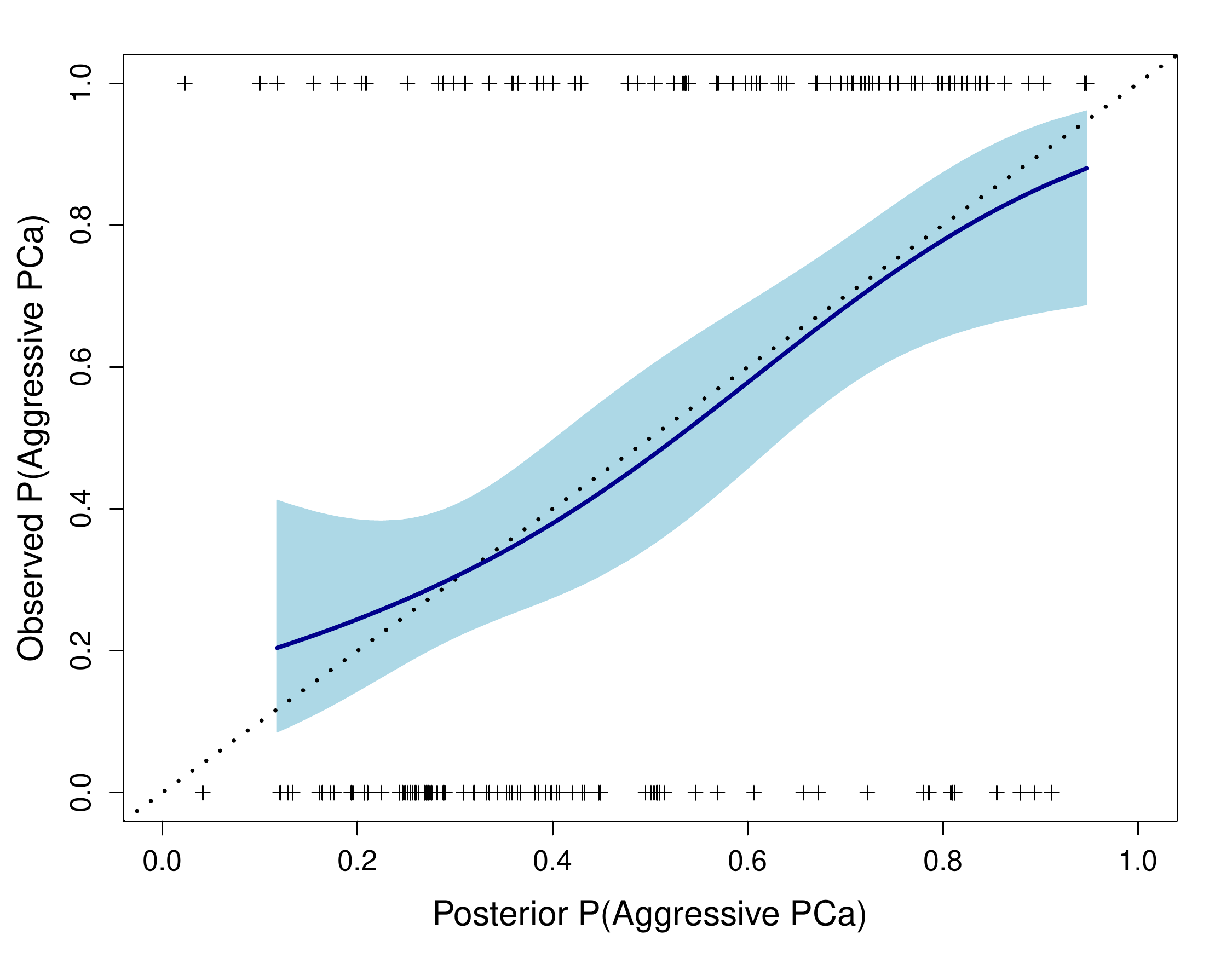}
\caption{Proposed model, biopsy IOP}
\label{fig:jhas-calibration-bx}
\end{subfigure}

\begin{subfigure}[b]{0.45\textwidth}
\includegraphics[width=\textwidth]{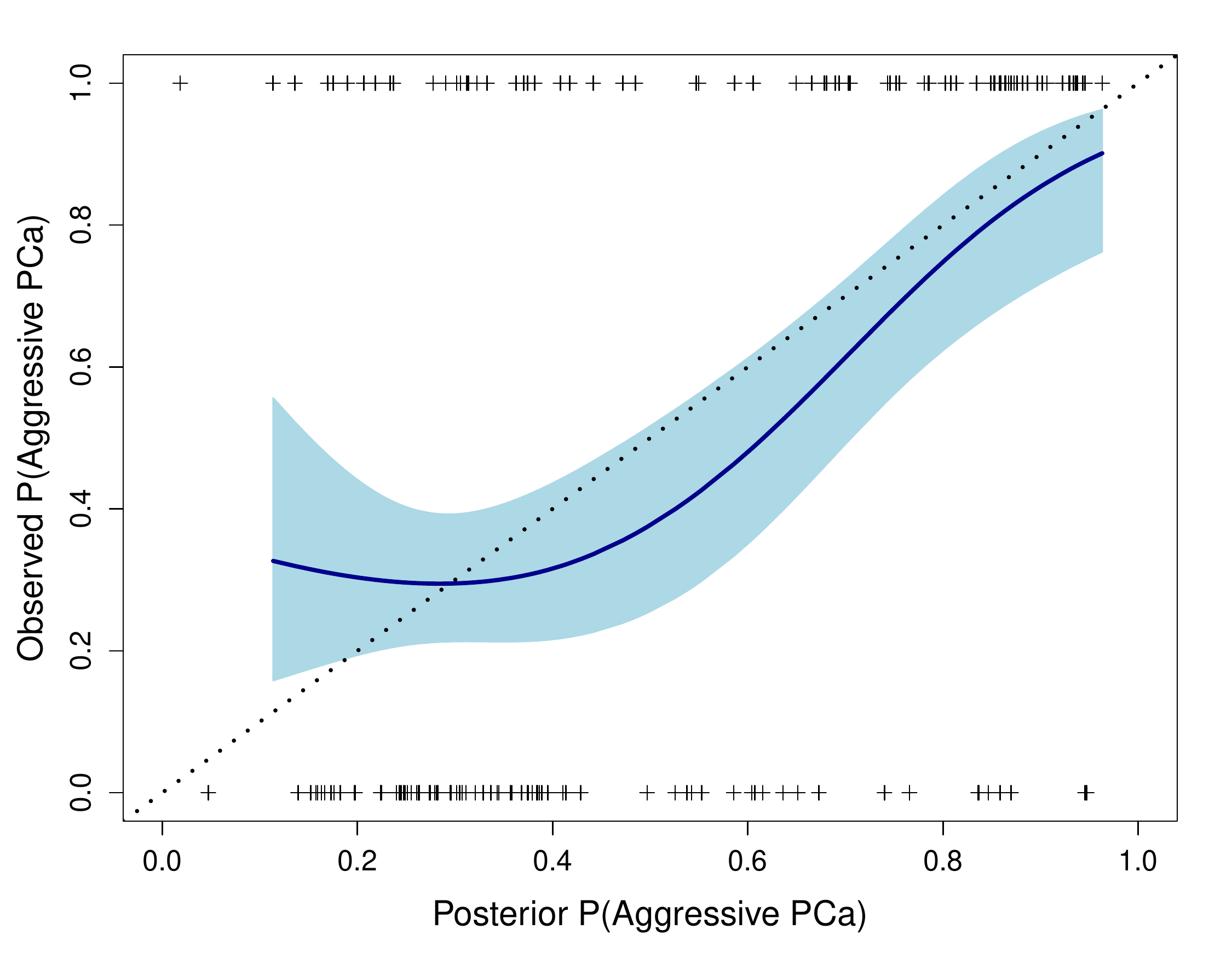}
\caption{Proposed model, surgery IOP}
\label{fig:jhas-calibration-surg}
\end{subfigure}
\begin{subfigure}[b]{0.45\textwidth}
\includegraphics[width=\textwidth]{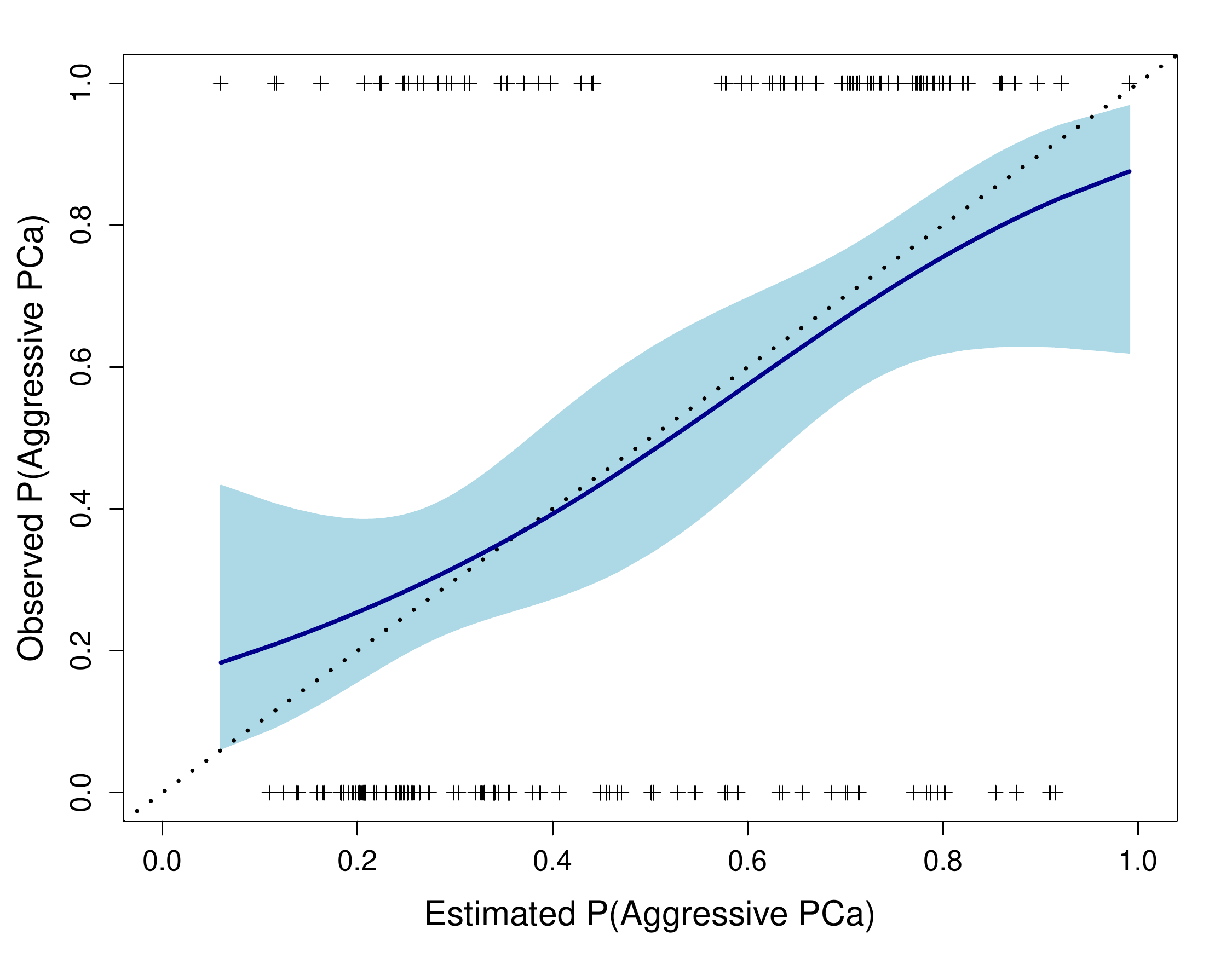}
\caption{Logistic regression}
\label{fig:jhas-calibration-logistic}
\end{subfigure}
\caption{Calibration plots for predictions of true cancer state in JHAS analysis. (Calibration plot for predictions from model with both biopsy and surgery IOP components given in Figure 6(b) of accompanying paper.)}
\label{fig:jhas-calibration}
\end{center}
\end{figure}

\begin{figure}
\begin{center}
\includegraphics[width=\textwidth]{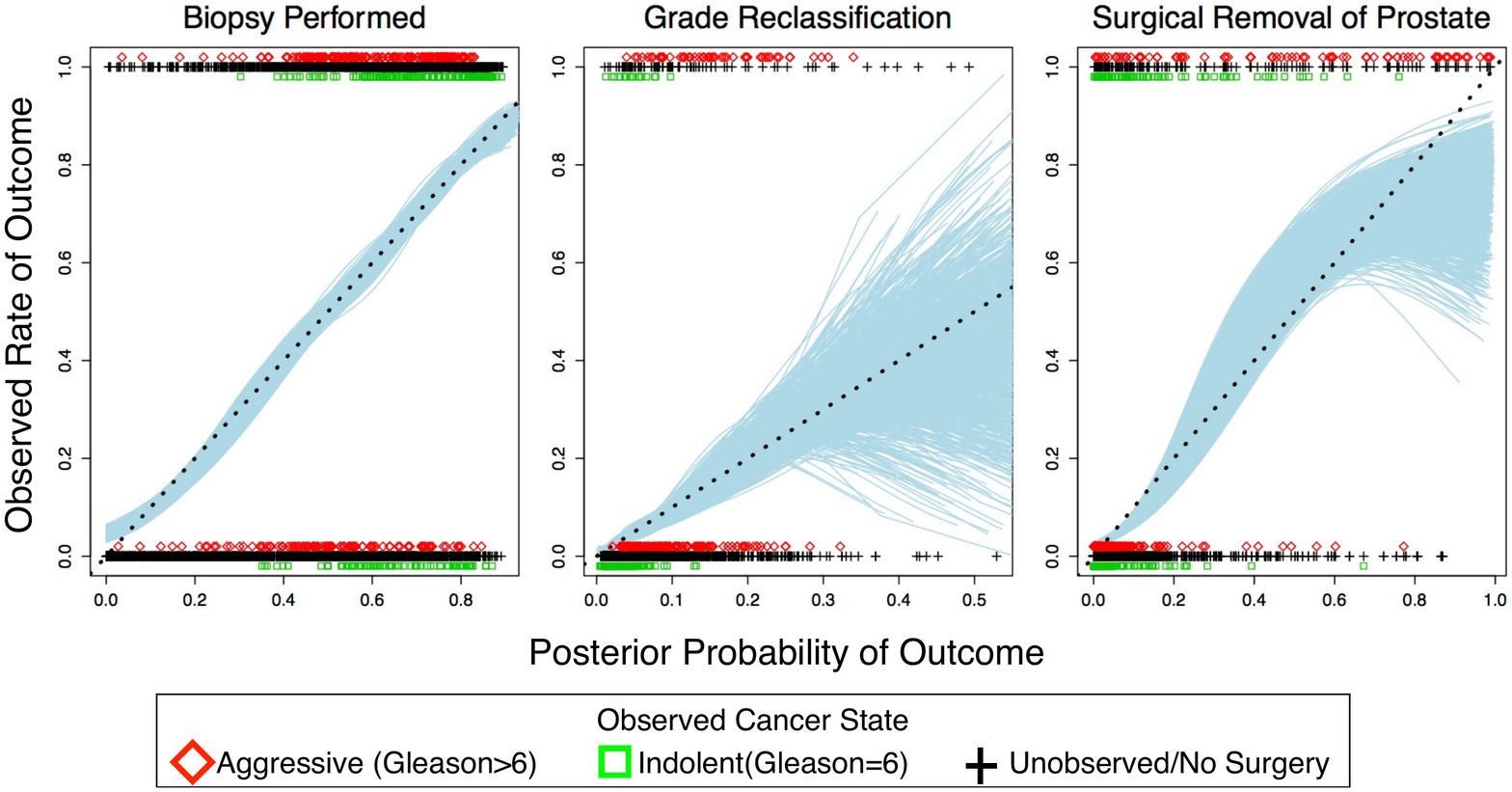}
\caption{Calibration plots for predictions of the occurrence of a biopsy (left), grade reclassification on biopsy (center), and surgery (right) at annual intervals for all patients in the JHAS cohort. Each solid line represents agreement between the posterior probability and observed event rate for a single iteration of the sampling algorithm for the proposed model with surgery and biopsy IOP components.}
\label{fig:jhas-calibration-observables}
\end{center}
\end{figure}

\begin{figure}
\begin{center}
\includegraphics[width=\textwidth]{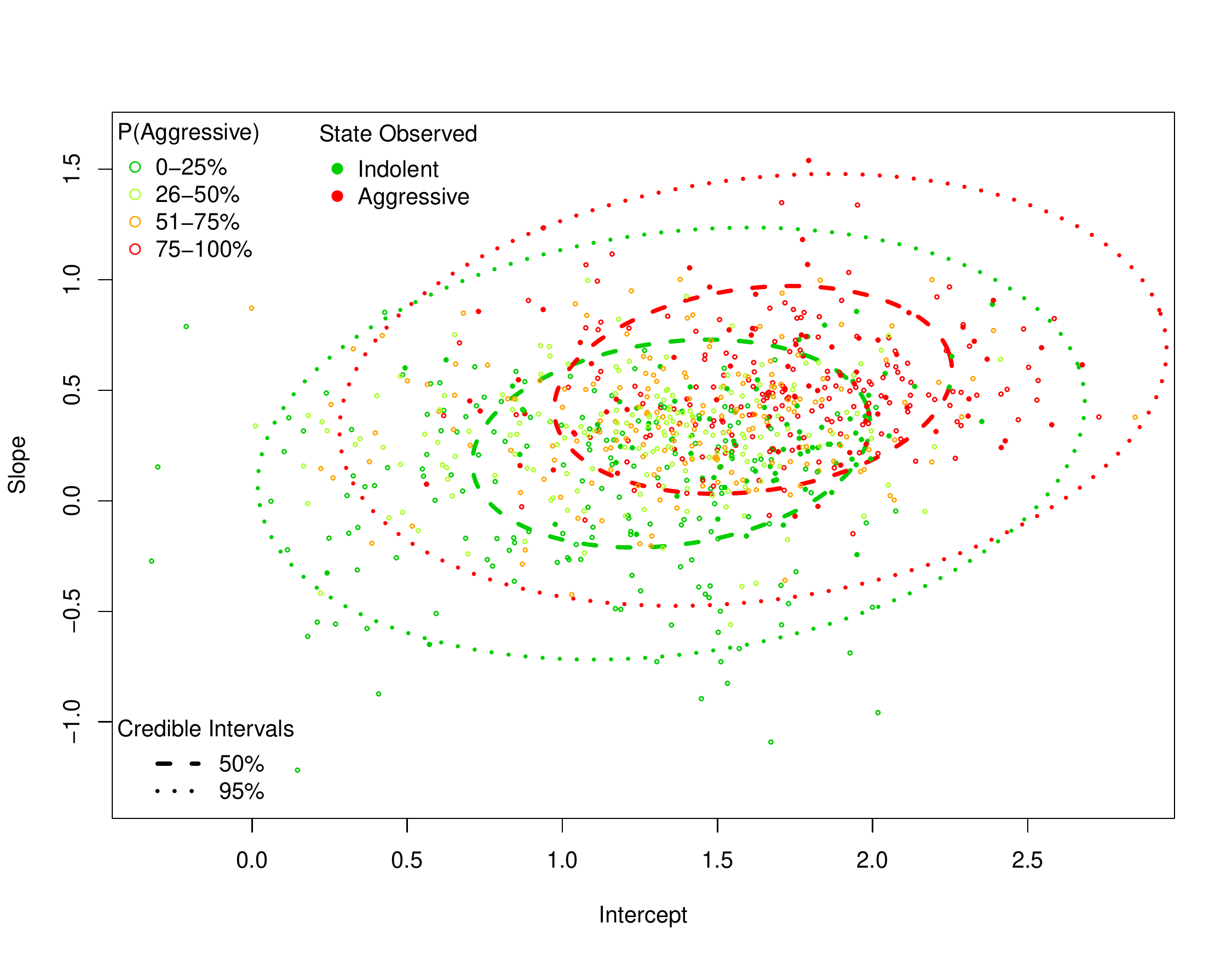}
\caption{Patient-level intercept (x-axis) and slope (y-axis) estimates from the multilevel PSA model in the proposed model with biopsy and surgery IOP components. }
\label{fig:psa-random-effects}
\end{center}
\end{figure}

\begin{figure}
\begin{center}
\includegraphics[width=\textwidth]{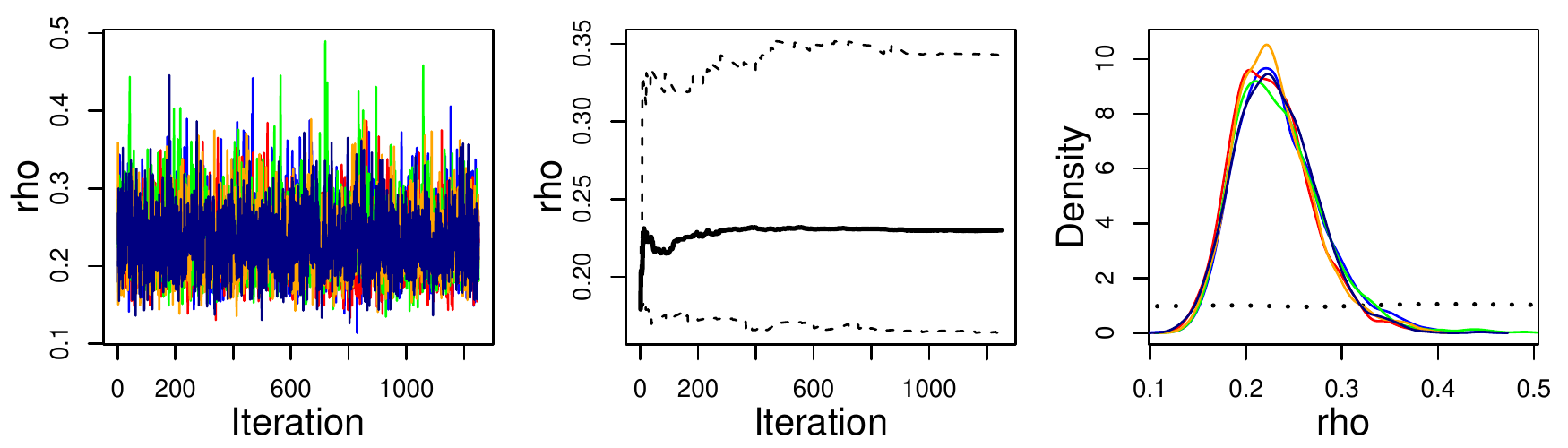}
\caption{Plots to assess convergence for $\rho$, the marginal probability of $\eta=1$, for the proposed model with biopsy and surgery IOP components applied to JHAS data: trace plots for five sampling chains, indicated by color (left); cumulative quantile plot for a representative sampling chain (center); and (right) plot comparing prior (dotted line) to posterior density (solid lines).}
\label{fig:jhas-post-rho}
\end{center}
\end{figure}

\begin{figure}
\begin{center}
\includegraphics[width=\textwidth]{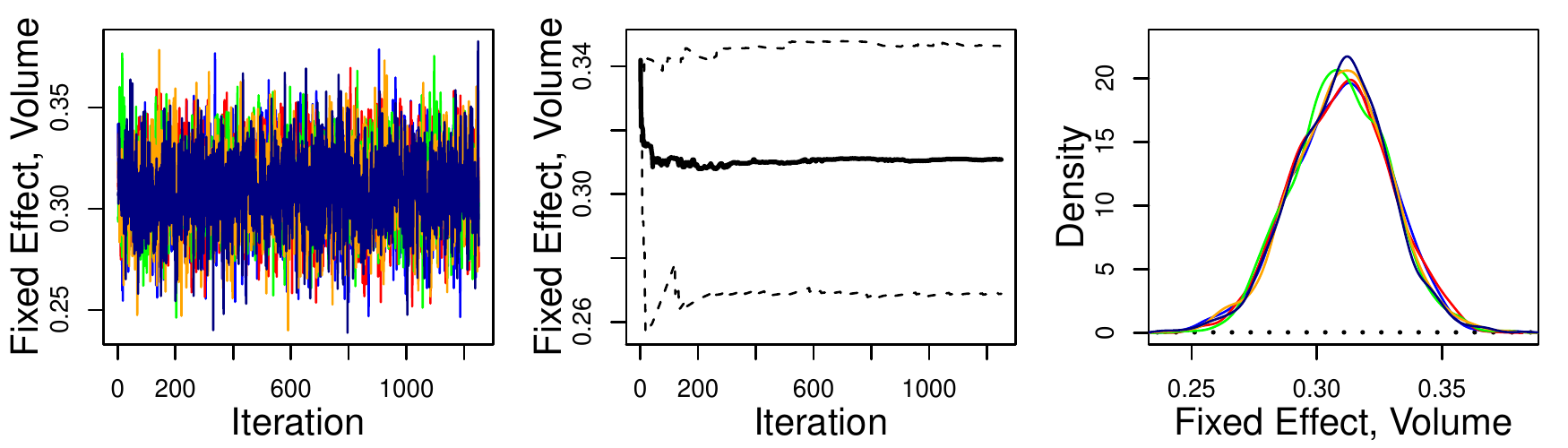}
\caption{Plots to assess convergence for $\beta$, the population-level coefficient for prostate volume in the multilevel regression model for PSA. Plots are from the joint posterior of the proposed model with biopsy and surgery IOP components applied to JHAS data: trace plots for five sampling chains, indicated by color (left); cumulative quantile plot for a representative sampling chain (center); and (right) plot comparing prior (dotted line) to posterior density (solid lines).}
\label{fig:jhas-post-bet}
\end{center}
\end{figure}

\begin{figure}
\begin{center}
\includegraphics[width=\textwidth]{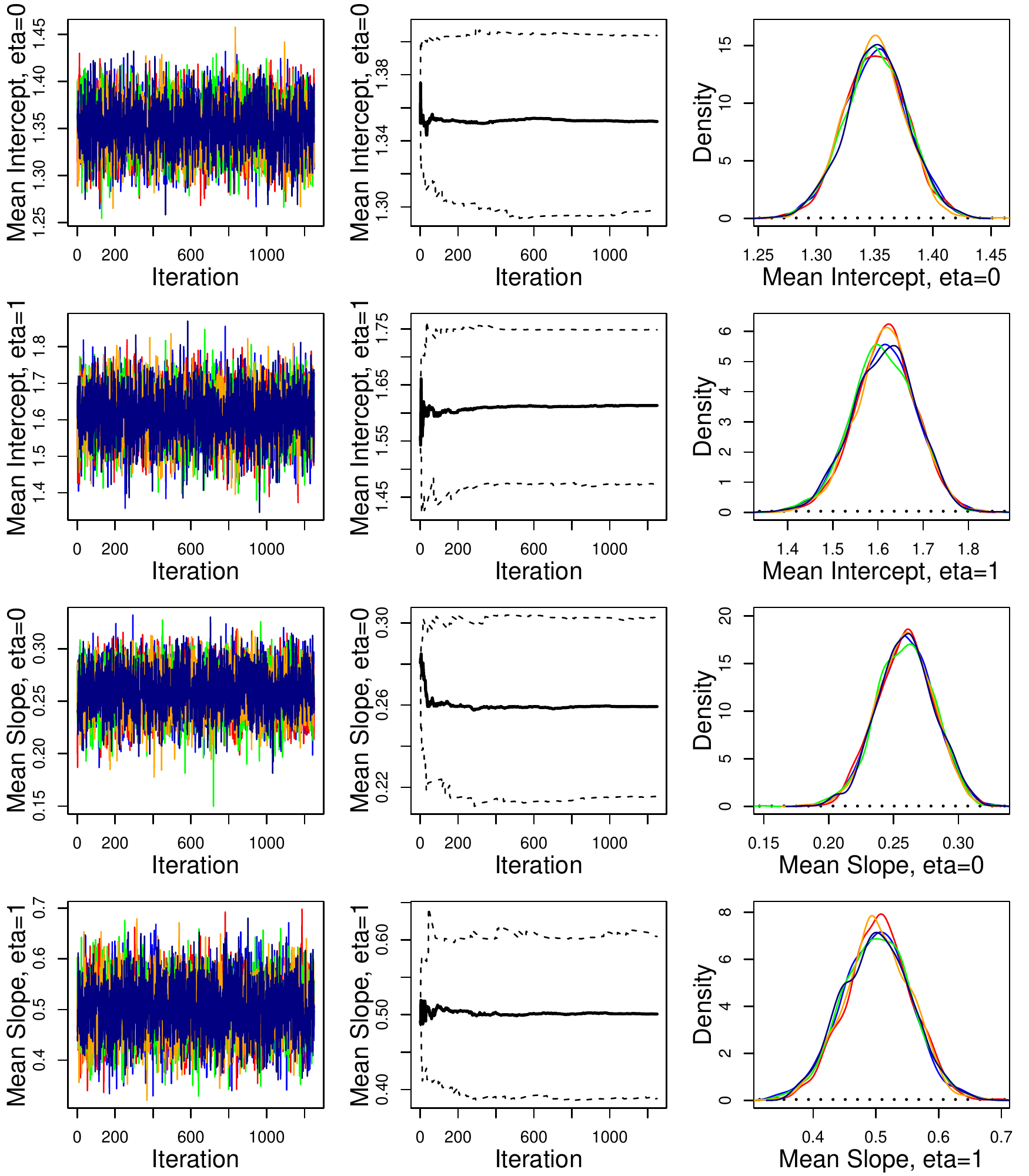}
\caption{Plots to assess convergence for $\bmmu$, the mean intercept and slope for latent classes in the multilevel regression model for PSA. Plots are from the joint posterior of the proposed model with biopsy and surgery IOP components applied to JHAS data: trace plots for five sampling chains, indicated by color (left); cumulative quantile plot for a representative sampling chain (center); and (right) plot comparing prior (dotted line) to posterior density (solid lines).}
\label{fig:jhas-post-mu}
\end{center}
\end{figure}

\begin{figure}
\begin{center}
\includegraphics[width=\textwidth]{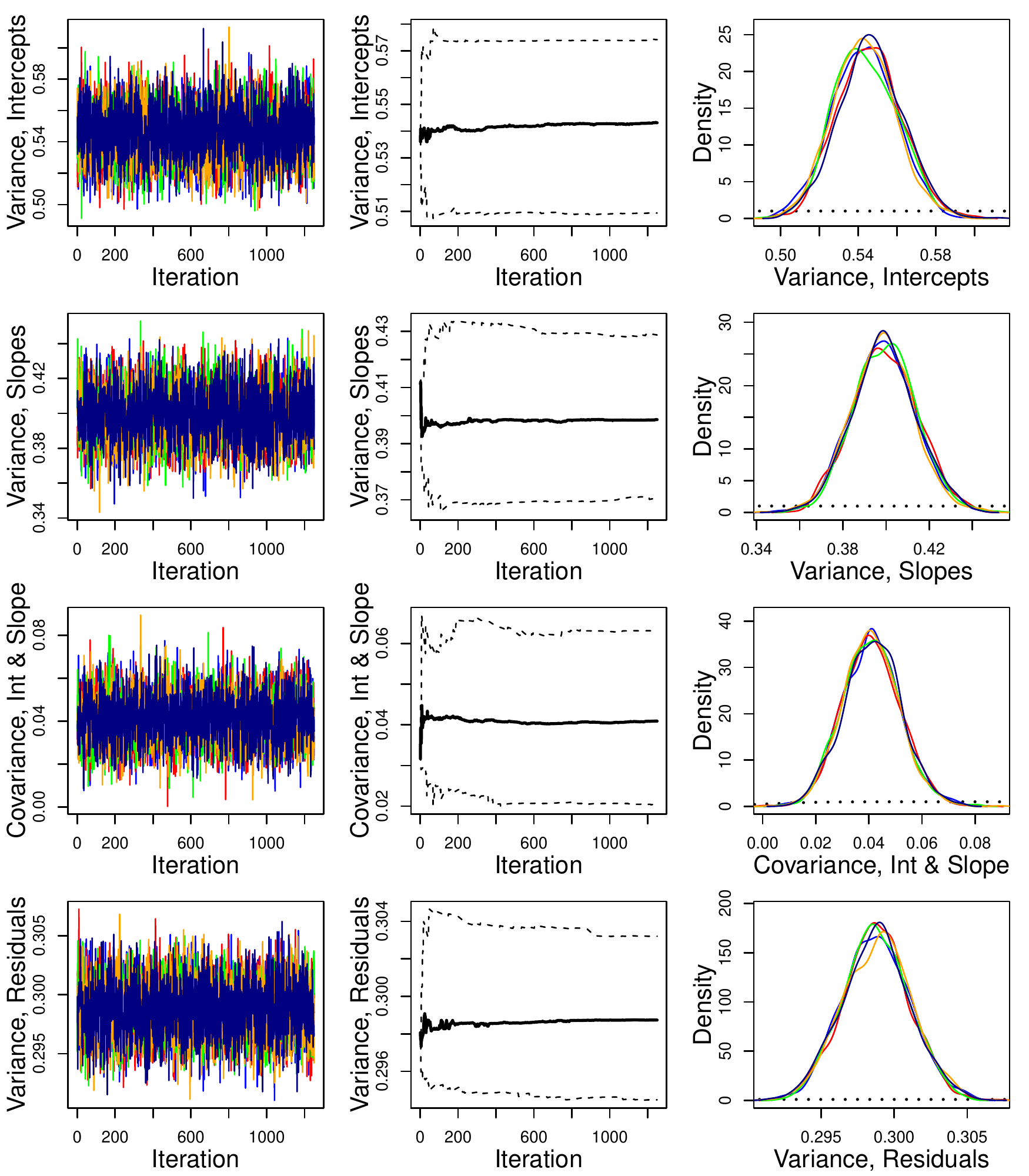}
\caption{Plots to assess convergence for the variance of patient-level intercepts and slopes, the covariance between patient-level intercepts and slopes, and the residual variance for the multilevel PSA regression model. Plots are from the joint posterior of the proposed model with biopsy and surgery IOP components applied to JHAS data: trace plots for five sampling chains, indicated by color (left); cumulative quantile plot for a representative sampling chain (center); and (right) plot comparing prior (dotted line) to posterior density (solid lines).}
\label{fig:jhas-post-sigma}
\end{center}
\end{figure}

\begin{figure}
\begin{center}
\includegraphics[width=0.8\textwidth]{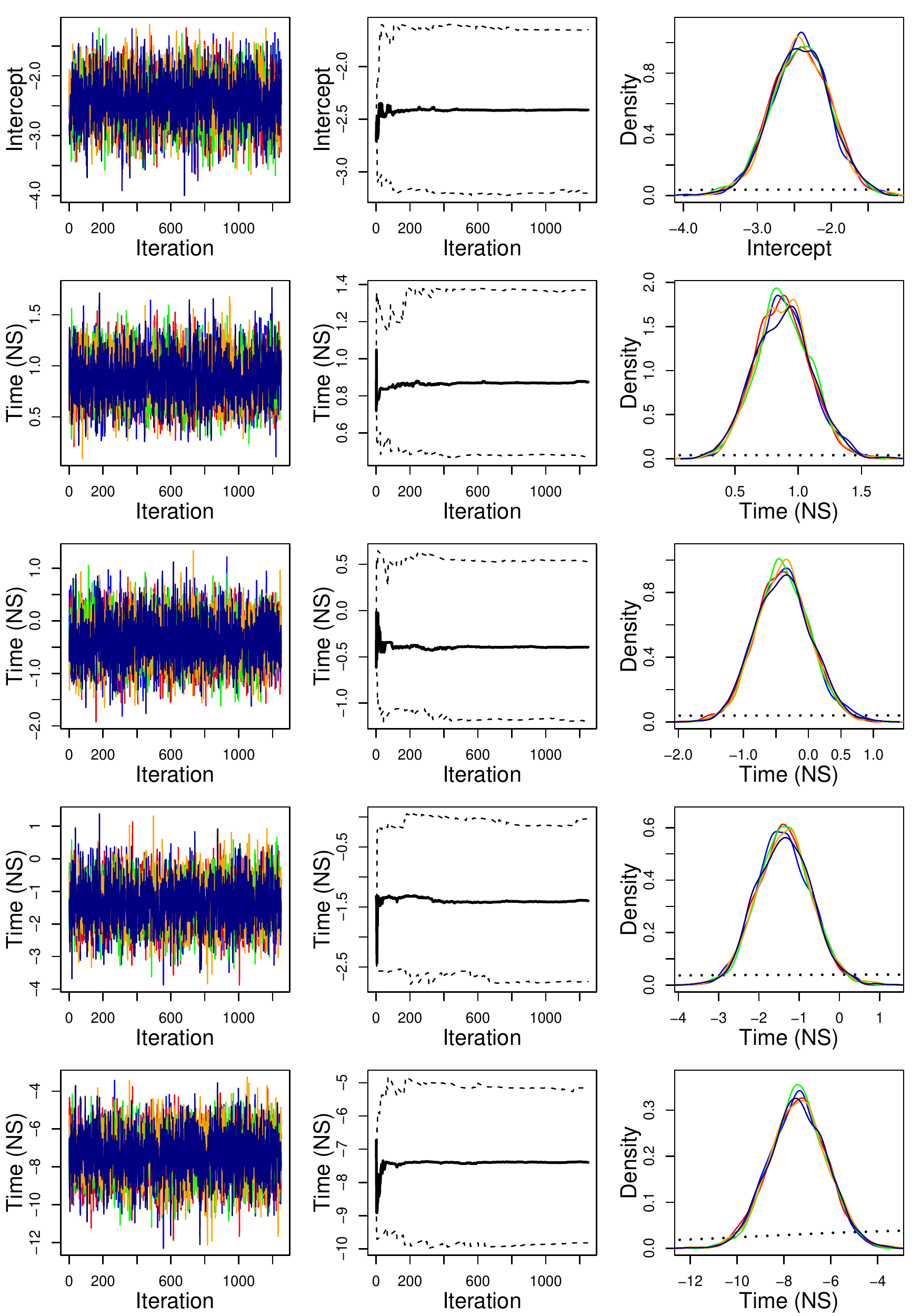}
\caption{Plots to assess convergence for $\bmnu$, coefficients in the logistic regression for patient decision to receive an annual biopsy; associated covariates are indicated on the far left y-axis. Plots are from the joint posterior of the proposed model with biopsy and surgery IOP components applied to JHAS data: trace plots for five sampling chains, indicated by color (left); cumulative quantile plot for a representative sampling chain (center); and (right) plot comparing prior (dotted line) to posterior density (solid lines).}
\label{fig:jhas-post-nu-1}
\end{center}
\end{figure}

\begin{figure}
\begin{center}
\includegraphics[width=\textwidth]{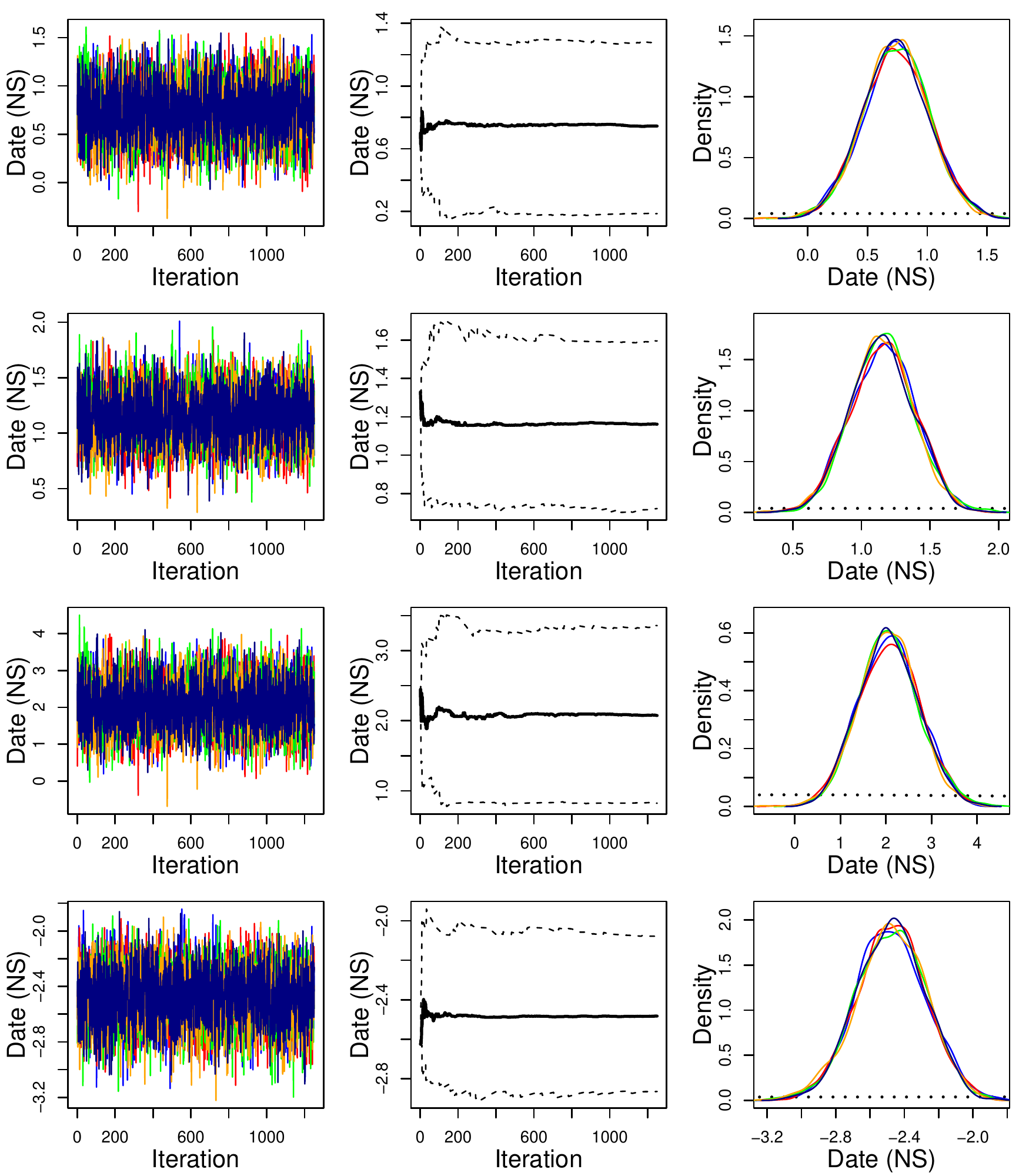}
\caption{Plots to assess convergence for $\bmnu$, coefficients in the logistic regression for patient decision to receive an annual biopsy; associated covariates are indicated on the far left y-axis. Plots are from the joint posterior of the proposed model with biopsy and surgery IOP components applied to JHAS data: trace plots for five sampling chains, indicated by color (left); cumulative quantile plot for a representative sampling chain (center); and (right) plot comparing prior (dotted line) to posterior density (solid lines).}
\label{fig:jhas-post-nu-2}
\end{center}
\end{figure}

\begin{figure}
\begin{center}
\includegraphics[width=0.8\textwidth]{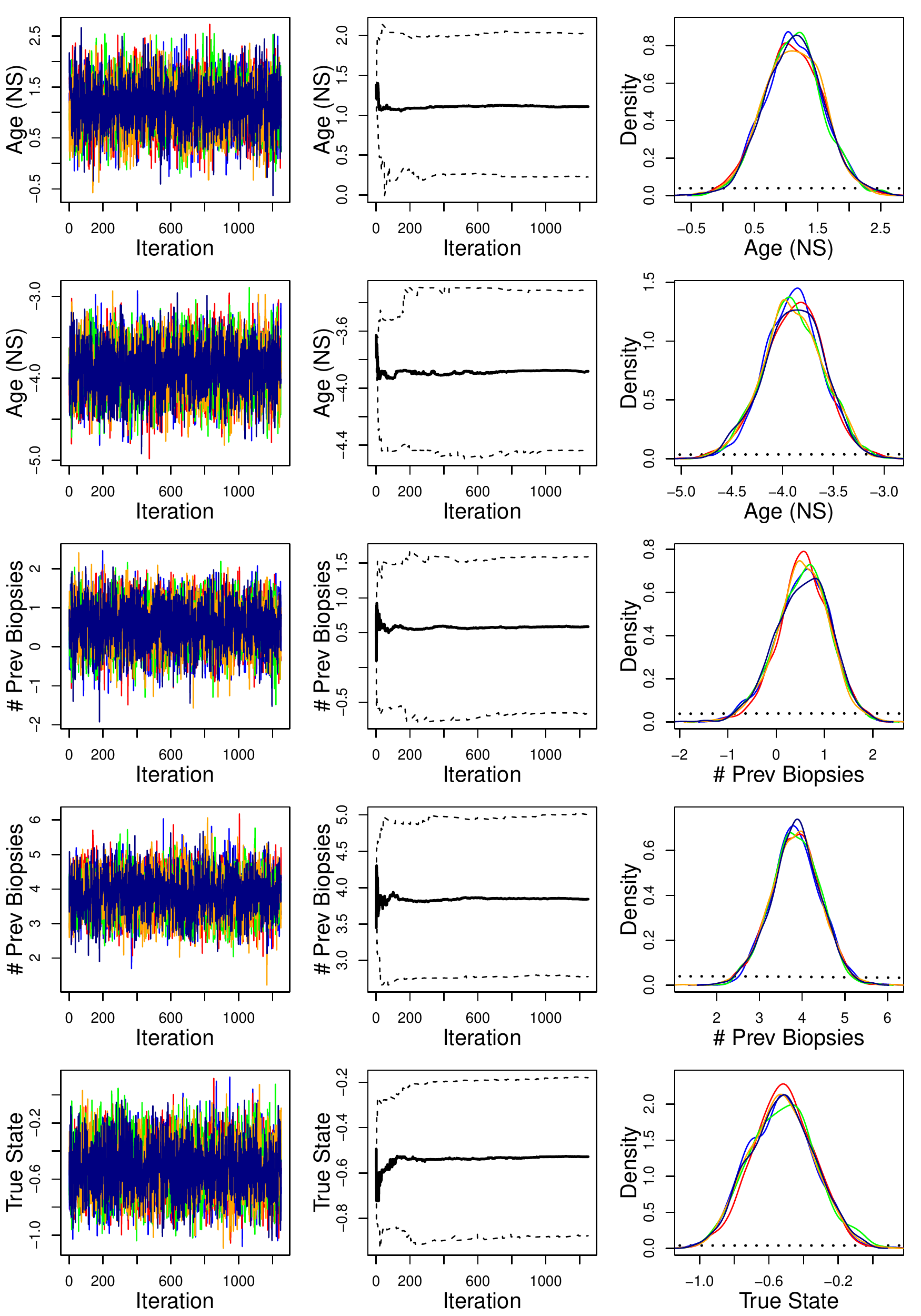}
\caption{Plots to assess convergence for $\bmnu$, coefficients in the logistic regression for patient decision to receive an annual biopsy; associated covariates are indicated on the far left y-axis. Plots are from the joint posterior of the proposed model with biopsy and surgery IOP components applied to JHAS data: trace plots for five sampling chains, indicated by color (left); cumulative quantile plot for a representative sampling chain (center); and (right) plot comparing prior (dotted line) to posterior density (solid lines).}
\label{fig:jhas-post-nu-3}
\end{center}
\end{figure}

\begin{figure}
\begin{center}
\includegraphics[width=\textwidth]{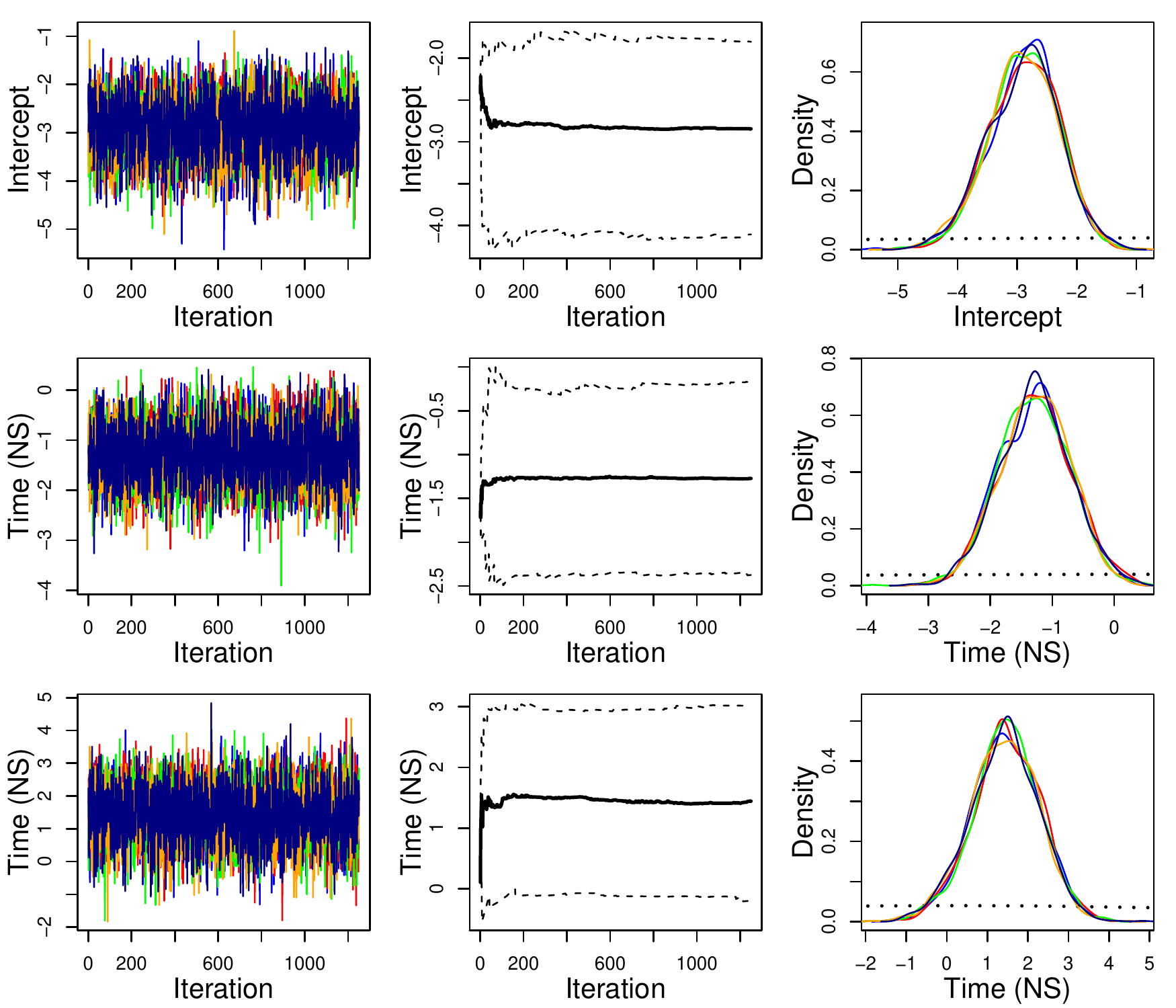}
\caption{Plots to assess convergence for $\bmgamma$, coefficients in the logistic regression for grade reclassification on biopsy; associated covariates are indicated on the far left y-axis. Plots are from the joint posterior of the proposed model with biopsy and surgery IOP components applied to JHAS data: trace plots for five sampling chains, indicated by color (left); cumulative quantile plot for a representative sampling chain (center); and (right) plot comparing prior (dotted line) to posterior density (solid lines).}
\label{fig:jhas-post-gamma-1}
\end{center}
\end{figure}

\begin{figure}
\begin{center}
\includegraphics[width=\textwidth]{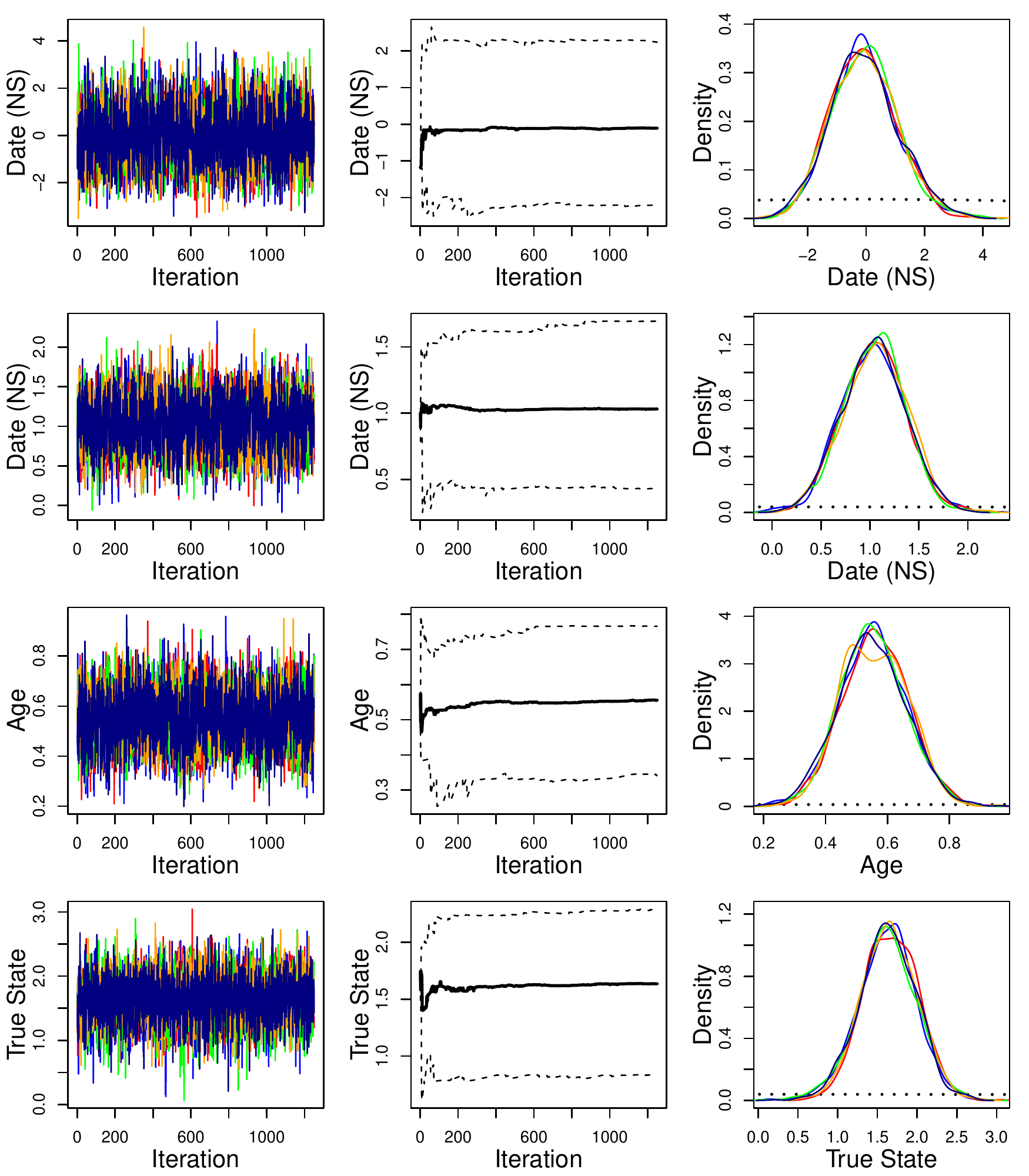}
\caption{Plots to assess convergence for $\bmgamma$, coefficients in the logistic regression for grade reclassification on biopsy; associated covariates are indicated on the far left y-axis. Plots are from the joint posterior of the proposed model with biopsy and surgery IOP components applied to JHAS data: trace plots for five sampling chains, indicated by color (left); cumulative quantile plot for a representative sampling chain (center); and (right) plot comparing prior (dotted line) to posterior density (solid lines).}
\label{fig:jhas-post-gamma-2}
\end{center}
\end{figure}

\begin{figure}
\begin{center}
\includegraphics[width=0.8\textwidth]{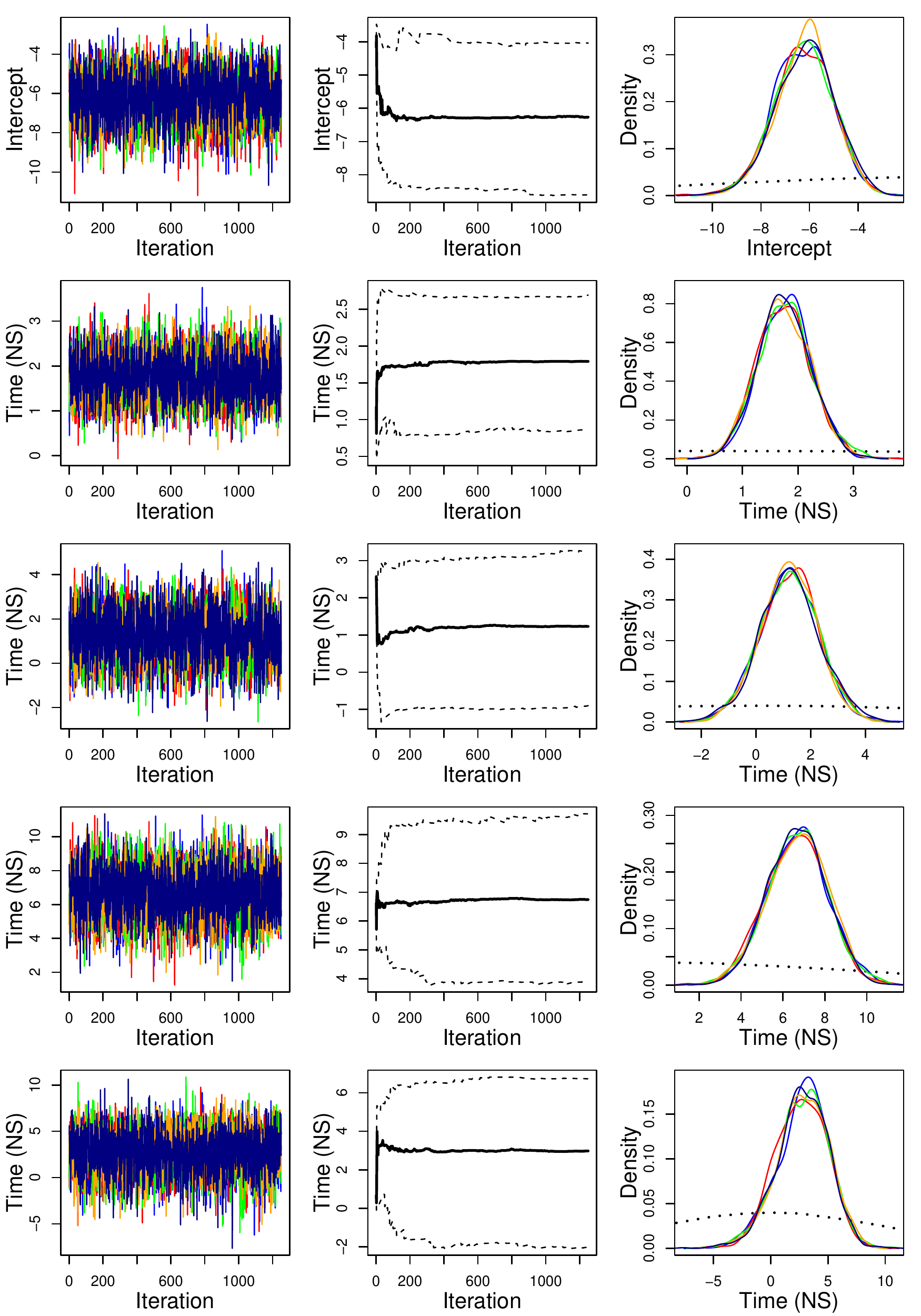}
\caption{Plots to assess convergence for $\bmomega$, coefficients in the logistic regression for patient decision to undergo surgical removal of the prostate; associated covariates are indicated on the far left y-axis. Plots are from the joint posterior of the proposed model with biopsy and surgery IOP components applied to JHAS data: trace plots for five sampling chains, indicated by color (left); cumulative quantile plot for a representative sampling chain (center); and (right) plot comparing prior (dotted line) to posterior density (solid lines).}
\label{fig:jhas-post-omega-1}
\end{center}
\end{figure}

\begin{figure}
\begin{center}
\includegraphics[width=0.8\textwidth]{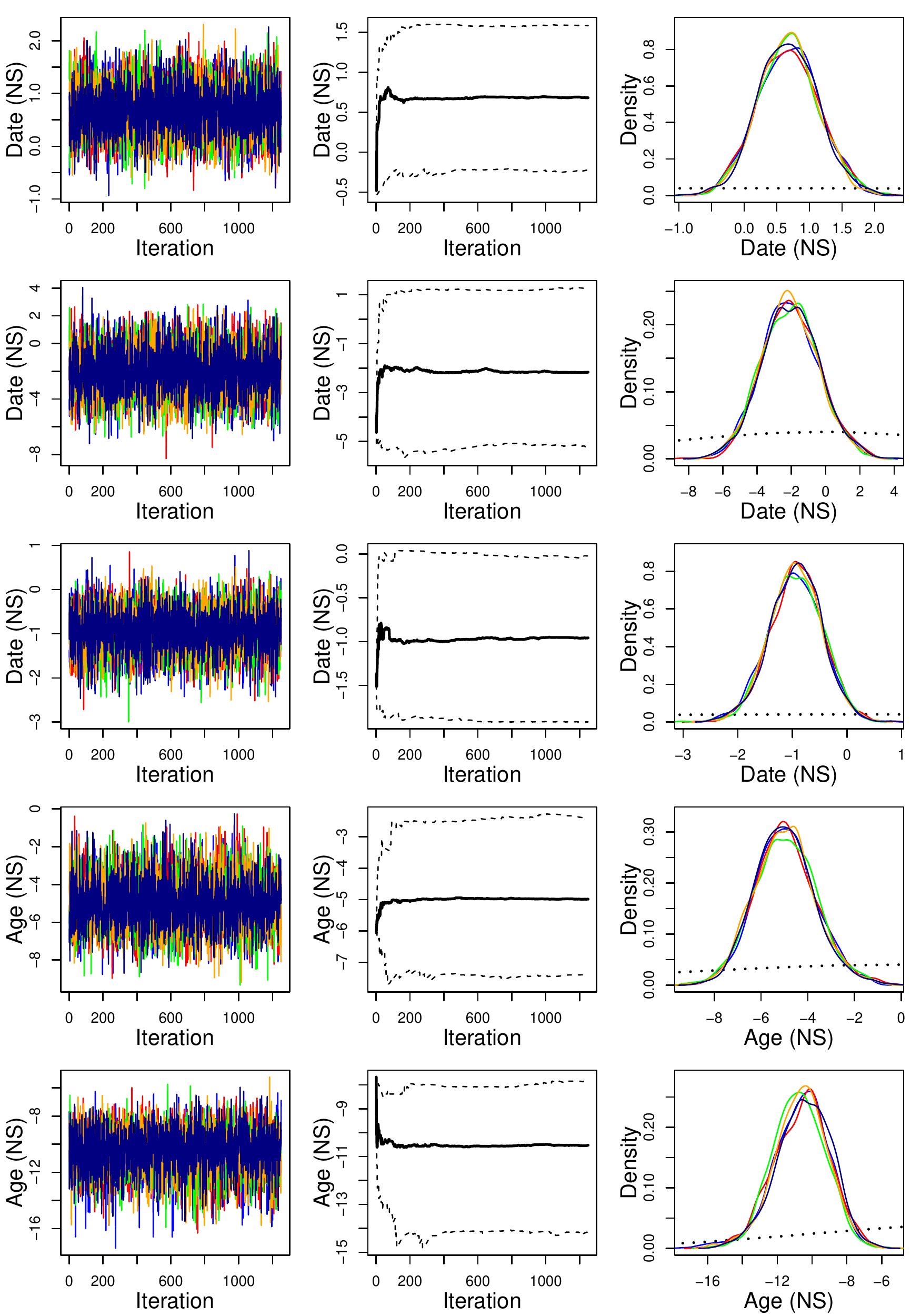}
\caption{Plots to assess convergence for $\bmomega$, coefficients in the logistic regression for patient decision to undergo surgical removal of the prostate; associated covariates are indicated on the far left y-axis. Plots are from the joint posterior of the proposed model with biopsy and surgery IOP components applied to JHAS data: trace plots for five sampling chains, indicated by color (left); cumulative quantile plot for a representative sampling chain (center); and (right) plot comparing prior (dotted line) to posterior density (solid lines).}
\label{fig:jhas-post-omega-2}
\end{center}
\end{figure}

\begin{figure}
\begin{center}
\includegraphics[width=\textwidth]{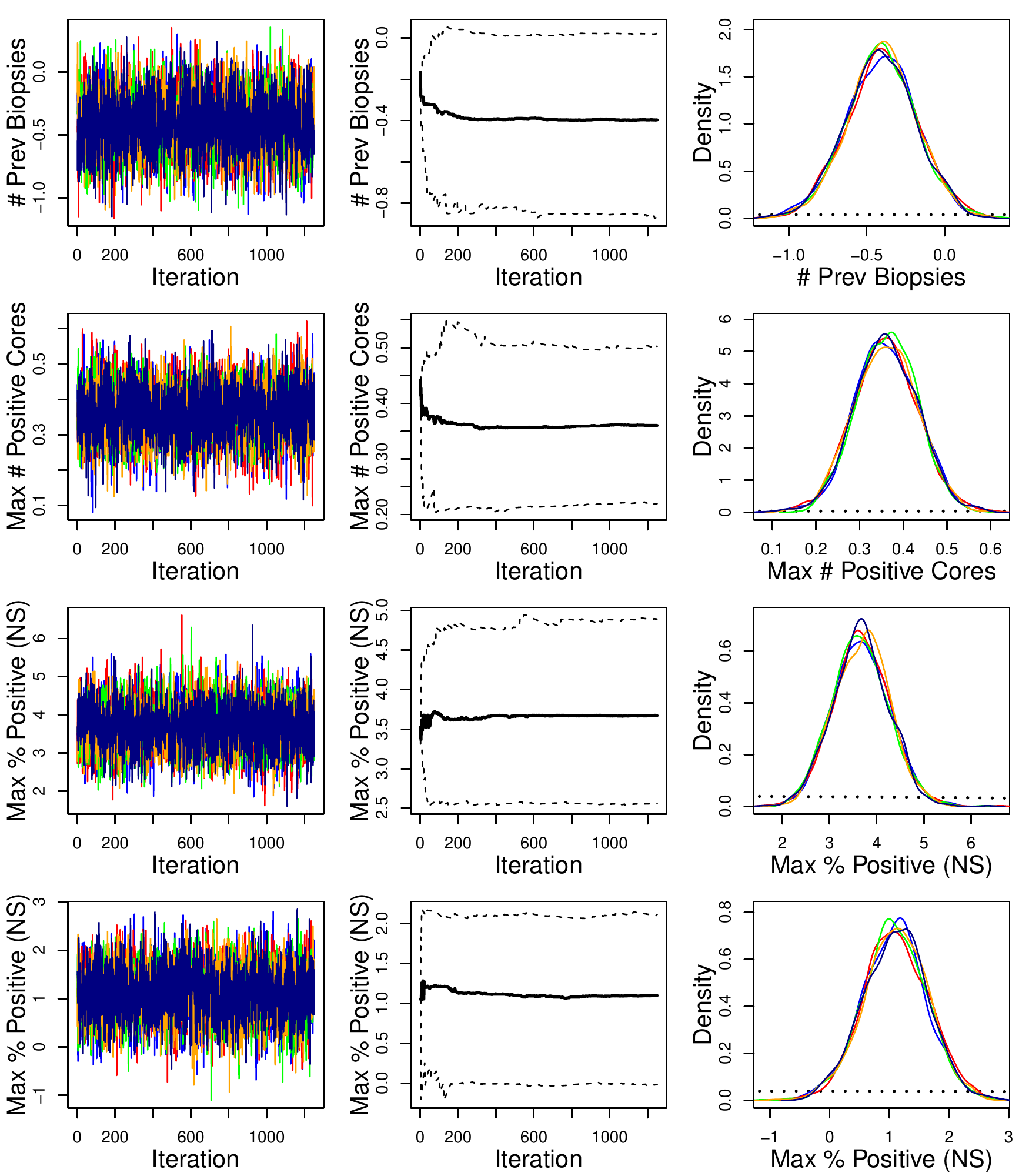}
\caption{Plots to assess convergence for $\bmomega$, coefficients in the logistic regression for patient decision to undergo surgical removal of the prostate; associated covariates are indicated on the far left y-axis. Plots are from the joint posterior of the proposed model with biopsy and surgery IOP components applied to JHAS data: trace plots for five sampling chains, indicated by color (left); cumulative quantile plot for a representative sampling chain (center); and (right) plot comparing prior (dotted line) to posterior density (solid lines).}
\label{fig:jhas-post-omega-3}
\end{center}
\end{figure}

\begin{figure}
\begin{center}
\includegraphics[width=\textwidth]{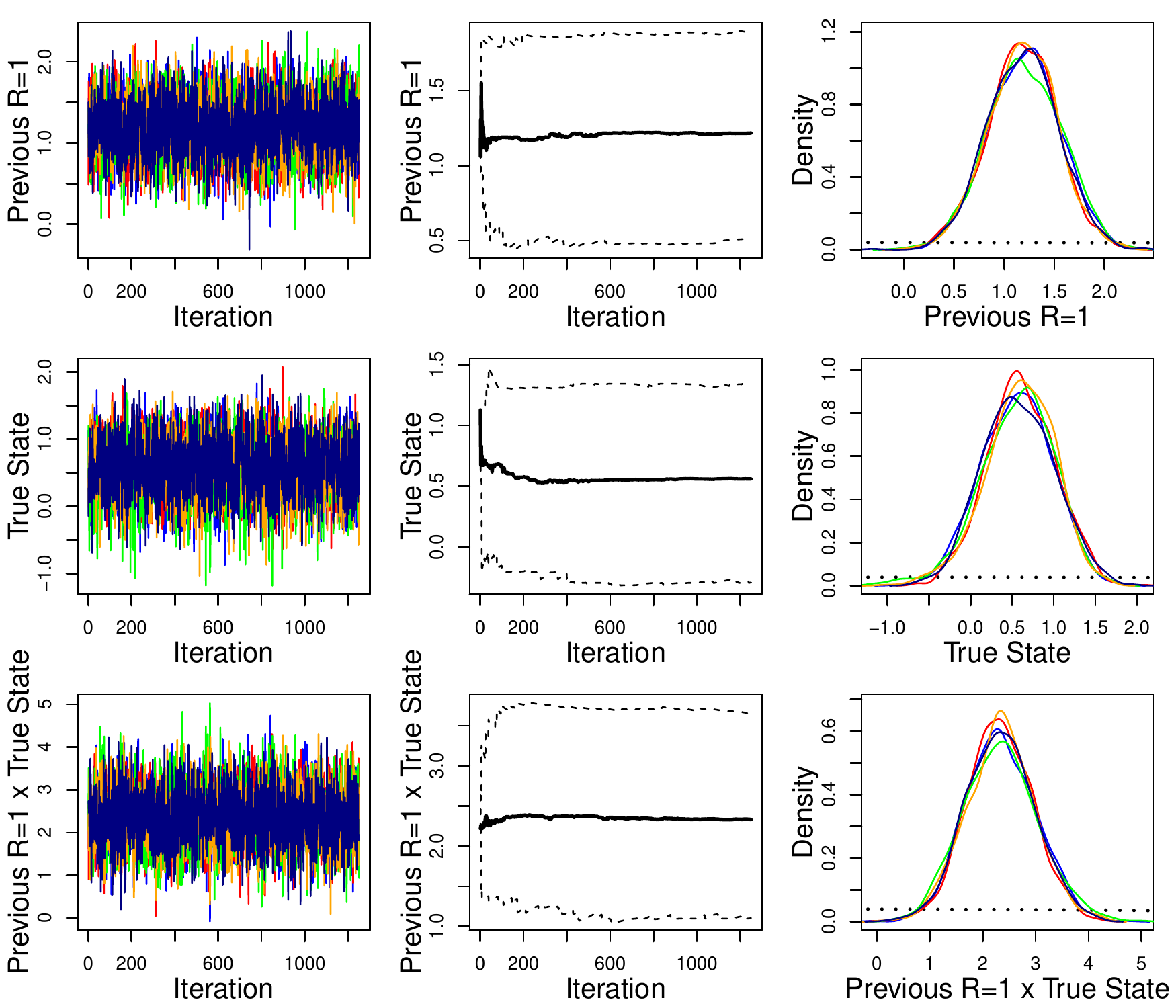}
\caption{Plots to assess convergence for $\bmomega$, coefficients in the logistic regression for patient decision to undergo surgical removal of the prostate; associated covariates are indicated on the far left y-axis. Plots are from the joint posterior of the proposed model with biopsy and surgery IOP components applied to JHAS data: trace plots for five sampling chains, indicated by color (left); cumulative quantile plot for a representative sampling chain (center); and (right) plot comparing prior (dotted line) to posterior density (solid lines).}
\label{fig:jhas-post-omega-4}
\end{center}
\end{figure}

\begin{figure}
\begin{center}
\includegraphics[width=\textwidth]{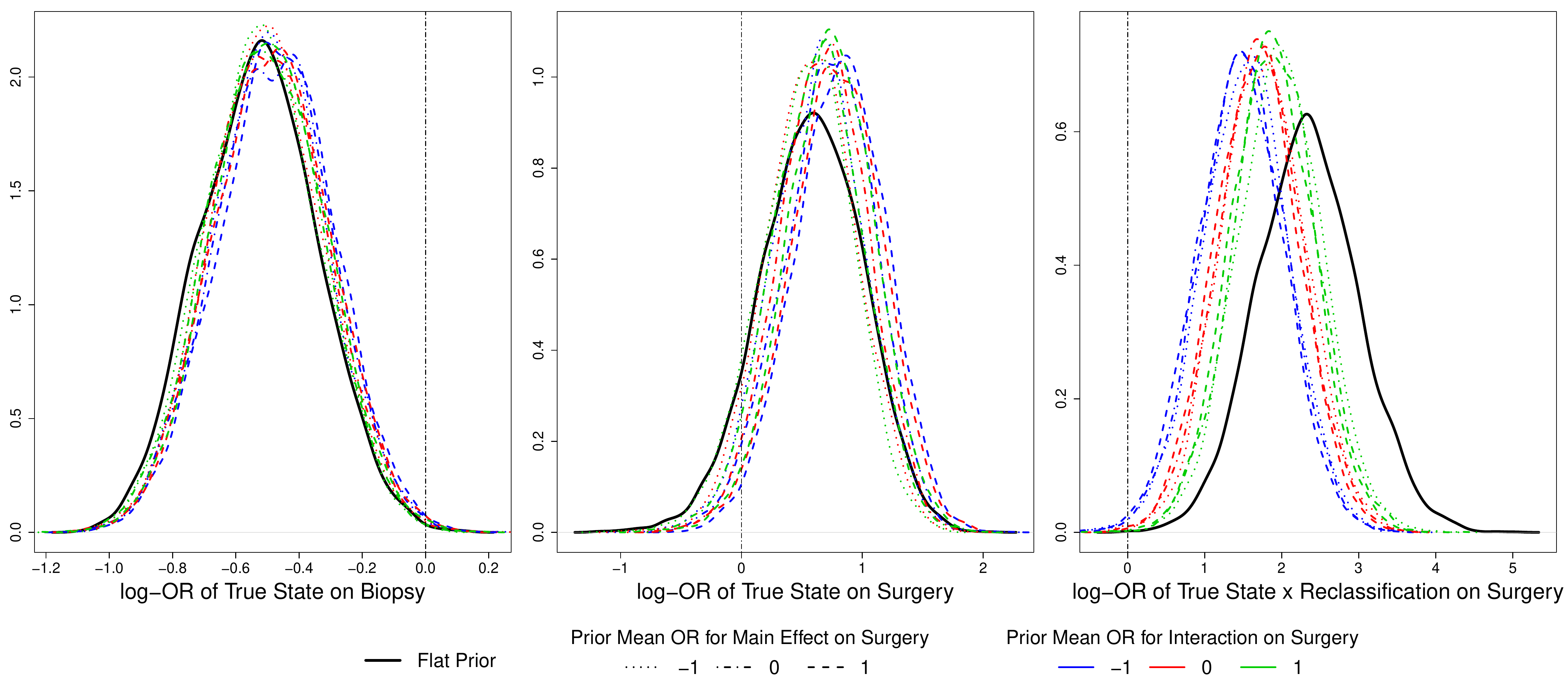}
\caption{Posterior distributions for biopsy and surgery IOP coefficients under vague and informative priors. Vertical line drawn at log-OR=0.}
\label{fig:post-coef-iop}
\end{center}
\end{figure}

\begin{figure}
\begin{center}
\includegraphics[width=\textwidth]{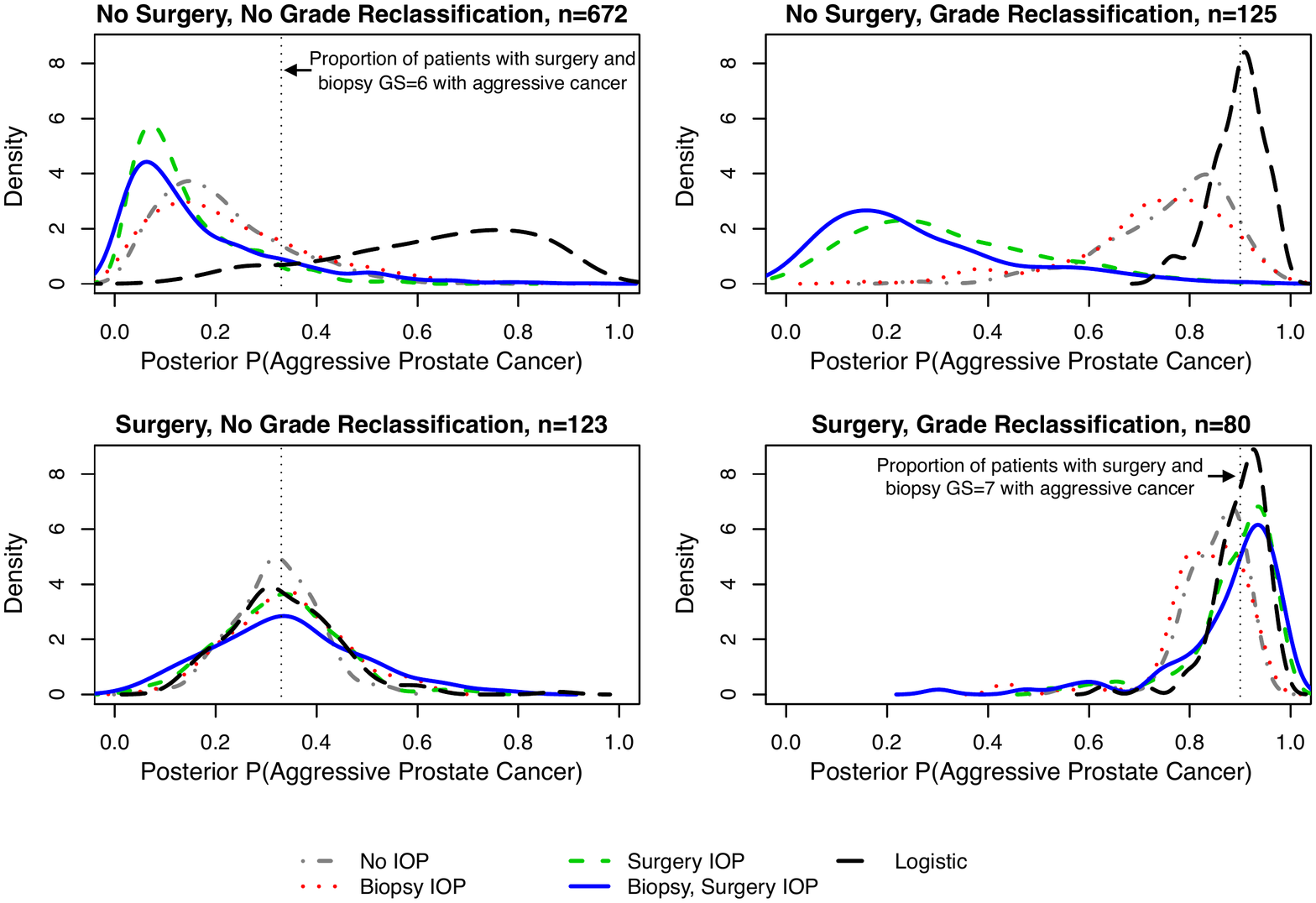}
\caption{Density plots of posterior predictions of $\eta$, stratified by biopsy results and the decision to have surgery, from a single simulated dataset. Line types correspond to different statistical models, as indicated by the legend at the bottom. The vertical dotted line represents the proportion of surgery patients with no grade reclassification (left) and reclassification (right) on biopsy who had higher grade prostate cancer on the full prostate examination.}
\label{fig:sim-density-post-eta}
\end{center}
\end{figure}

\begin{figure}
\begin{center}

\begin{subfigure}[b]{0.45\textwidth}
\includegraphics[width=\textwidth]{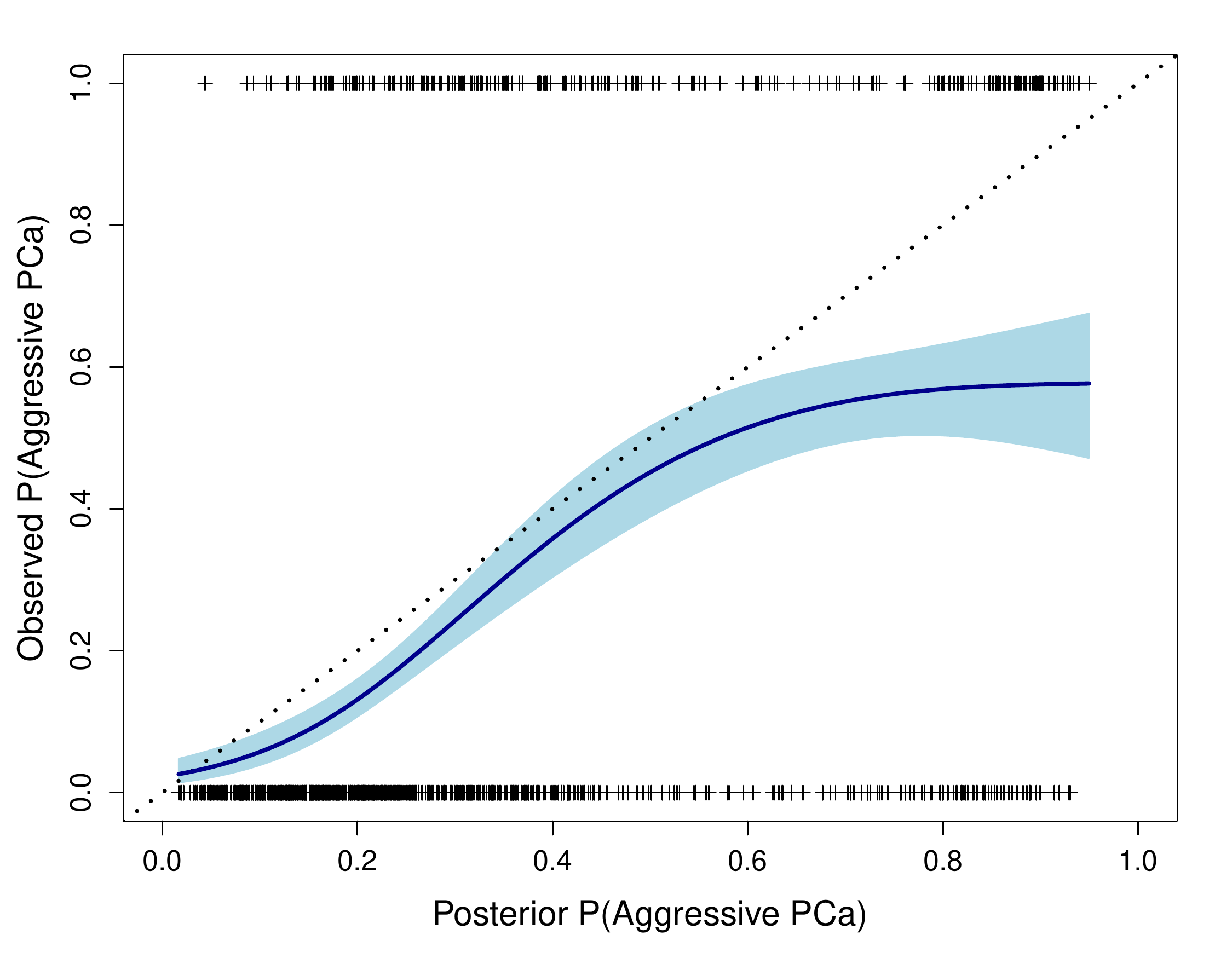}
\caption{Proposed model, unadjusted}
\label{fig:sim-calibration-unadj}
\end{subfigure}
\begin{subfigure}[b]{0.45\textwidth}
\includegraphics[width=\textwidth]{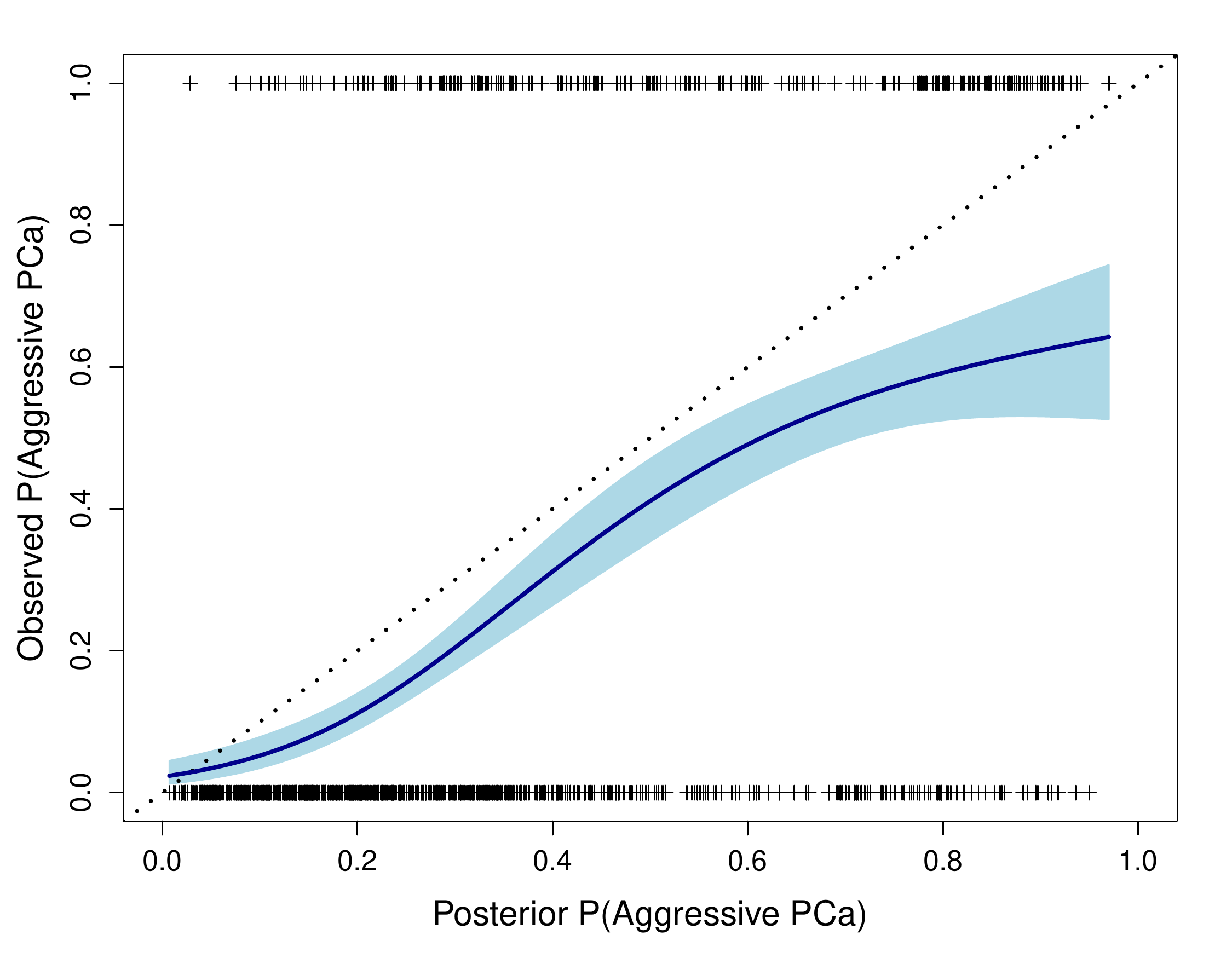}
\caption{Proposed model, biopsy IOP}
\label{fig:sim-calibration-bx}
\end{subfigure}

\begin{subfigure}[b]{0.45\textwidth}
\includegraphics[width=\textwidth]{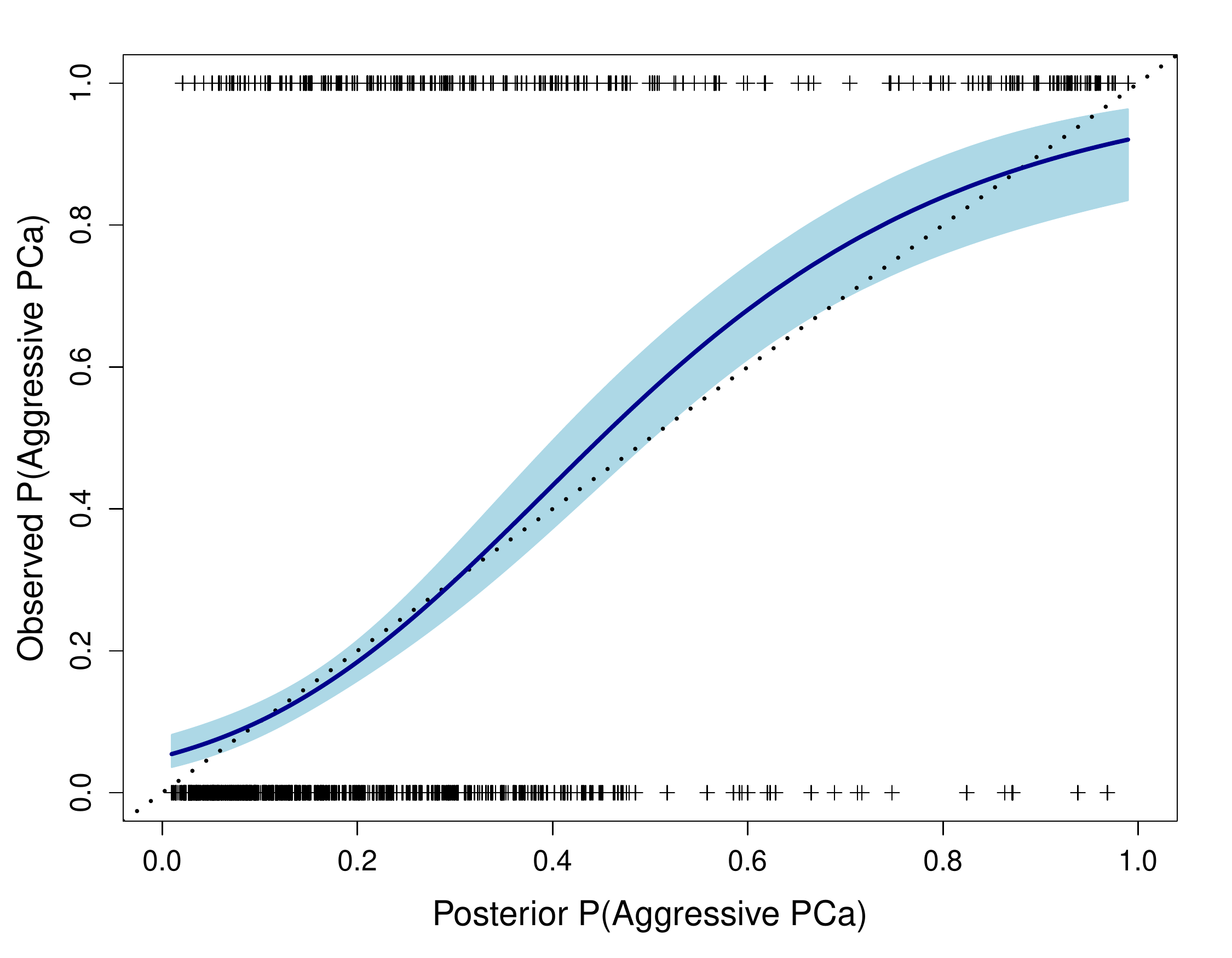}
\caption{Proposed model, surgery IOP}
\label{fig:sim-calibration-surg}
\end{subfigure}
\begin{subfigure}[b]{0.45\textwidth}
\includegraphics[width=\textwidth]{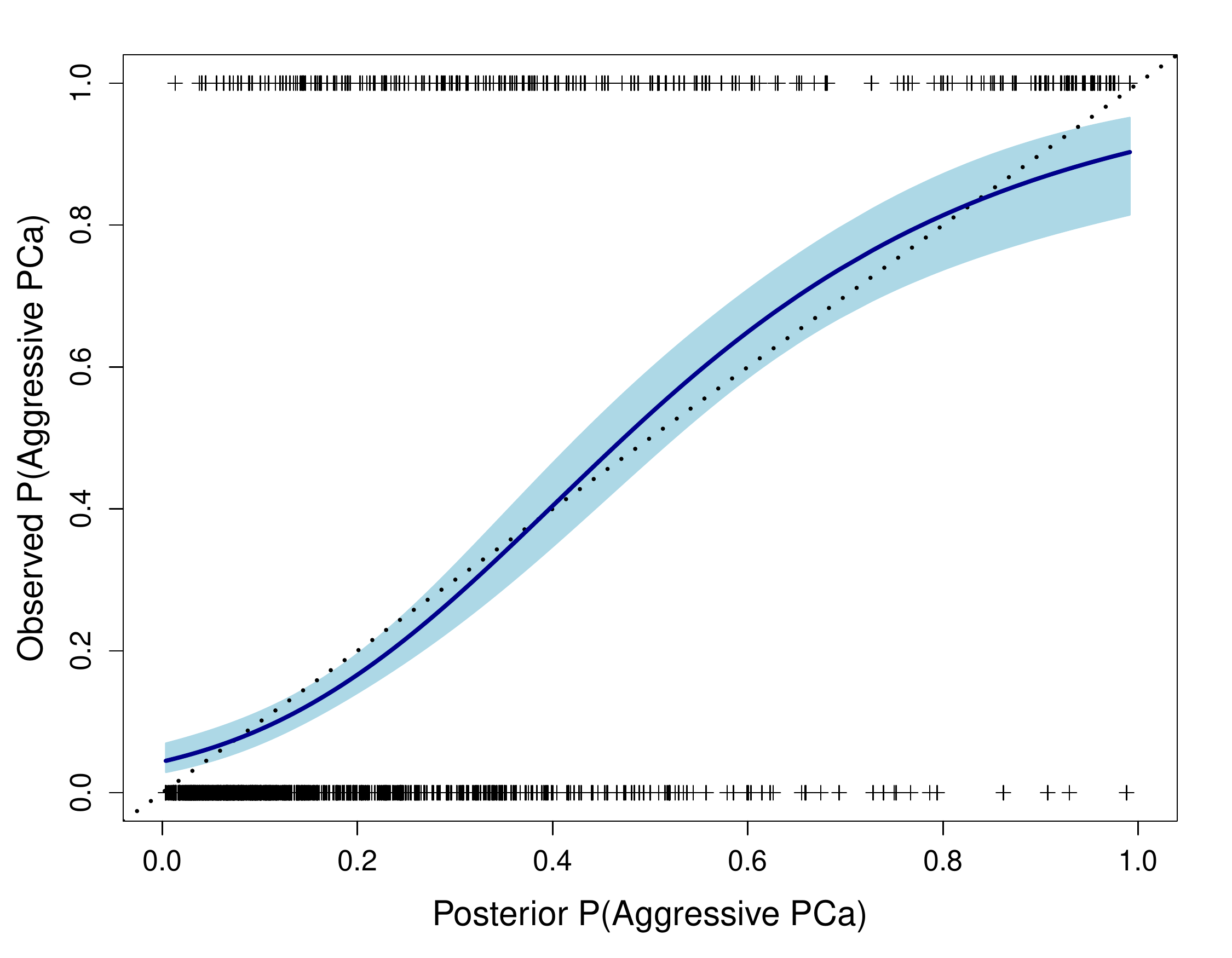}
\caption{Proposed model, biopsy and surgery IOP}
\label{fig:sim-calibration-iop}
\end{subfigure}

\begin{subfigure}[b]{0.45\textwidth}
\includegraphics[width=\textwidth]{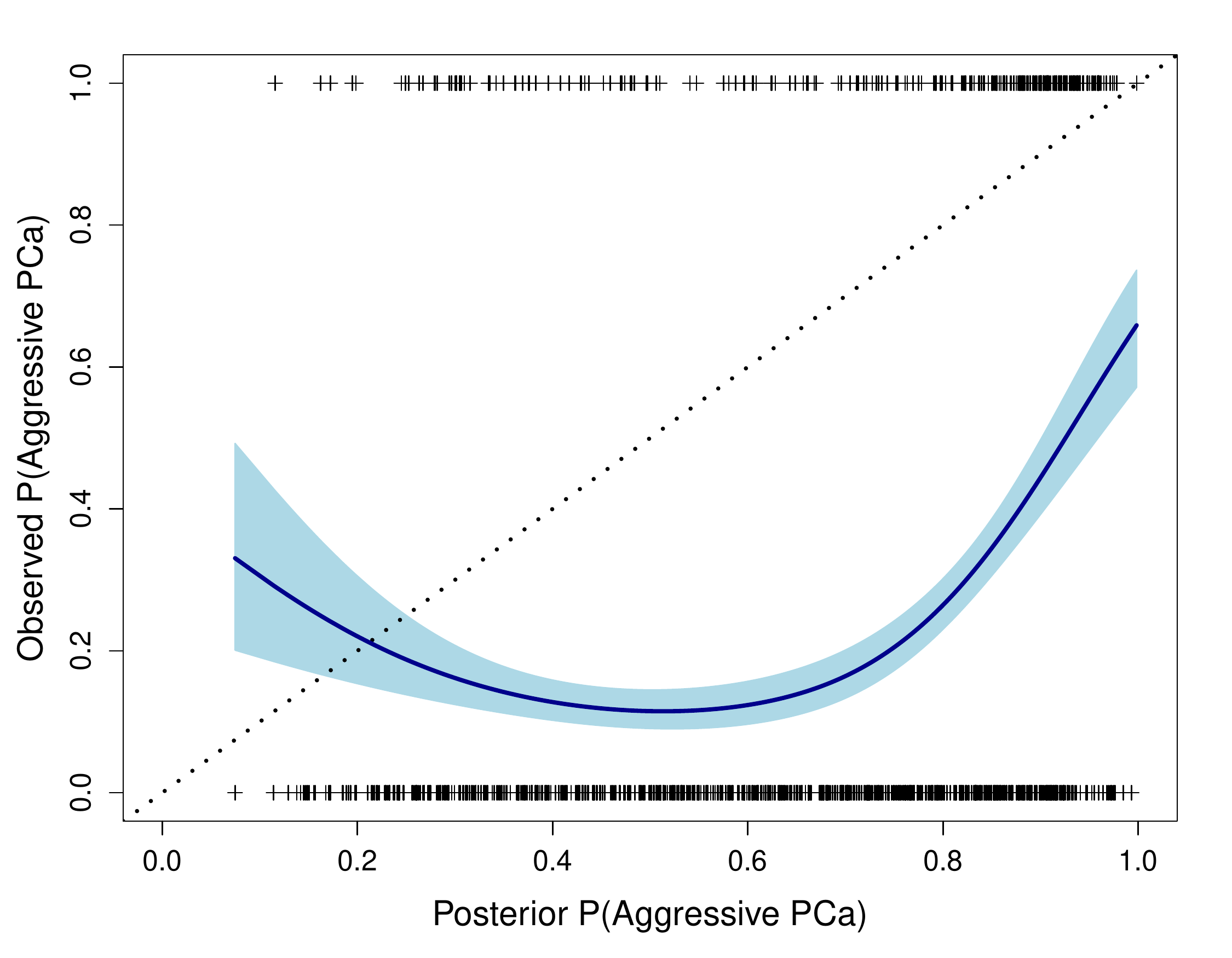}
\caption{Proposed model, logistic regression}
\label{fig:sim-calibration-logistic}
\end{subfigure}

\caption{Calibration plots \textbf{among all patients} for predictions of true cancer state in one simulated dataset}
\label{fig:sim-calibration}
\end{center}
\end{figure}

Figure \ref{fig:sim-calibration} gives calibration plots for predictions from each of these models. Across all models the pointwise confidence intervals for calibration plots are much more narrow than those for the JHAS application (Figure \ref{fig:jhas-calibration}) because these plots contain predictions on all patients, not just the smaller subset with surgery. We see in the top row of plots that predictions from the model with no IOP and only biopsy IOP components tend to overestimate the probability of harboring aggressive prostate cancer; this observation is consistent with higher estimates of $\rho$ seen in both the application and simulation results and with our discussion above (Section 3.1.1) regarding the influence of surgery IOP components on risk estimates. We also see that predictions from the logistic regression model are poorly calibrated \ref{fig:sim-calibration-logistic}. In comparison, predictions from the logistic regression model in the JHAS cohort analysis (Figure \ref{fig:jhas-calibration-logistic}) seem well calibrated when only patients with $\eta$ observed are considered.

\begin{figure}
\begin{center}

\begin{subfigure}[b]{0.45\textwidth}
\includegraphics[width=\textwidth]{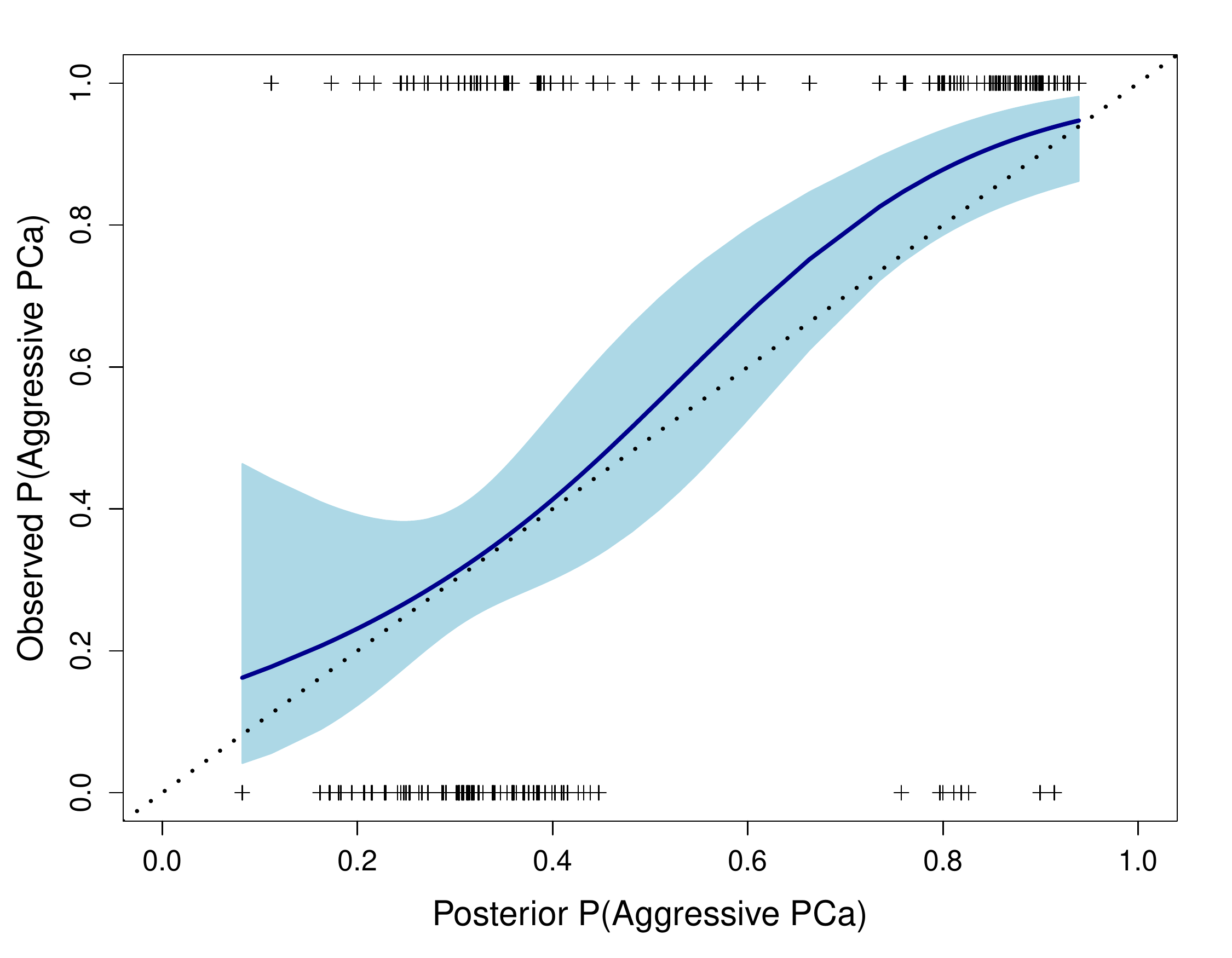}
\caption{Proposed model, unadjusted}
\label{fig:sim-calibration-unadj-ek}
\end{subfigure}
\begin{subfigure}[b]{0.45\textwidth}
\includegraphics[width=\textwidth]{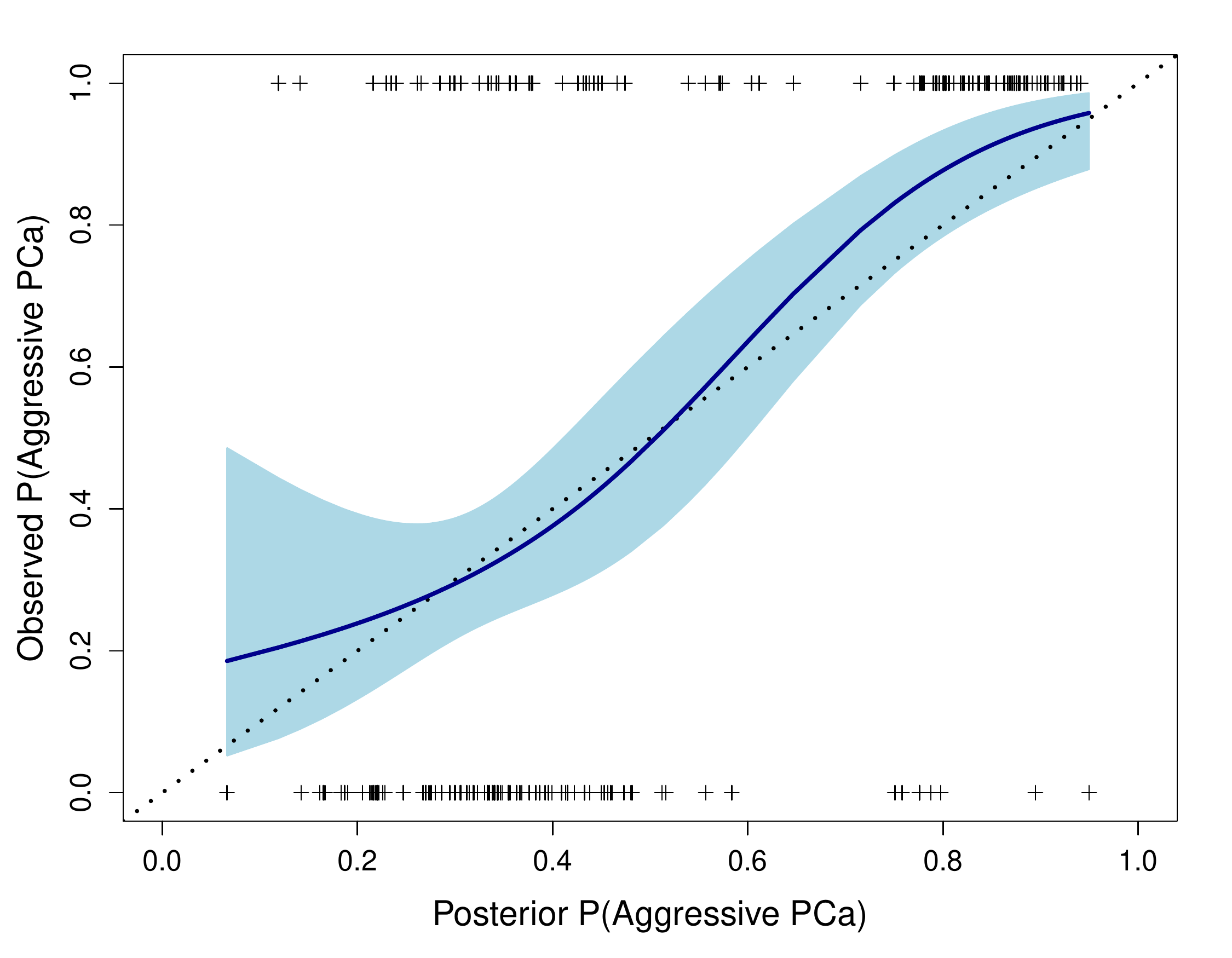}
\caption{Proposed model, biopsy IOP}
\label{fig:sim-calibration-bx-ek}
\end{subfigure}

\begin{subfigure}[b]{0.45\textwidth}
\includegraphics[width=\textwidth]{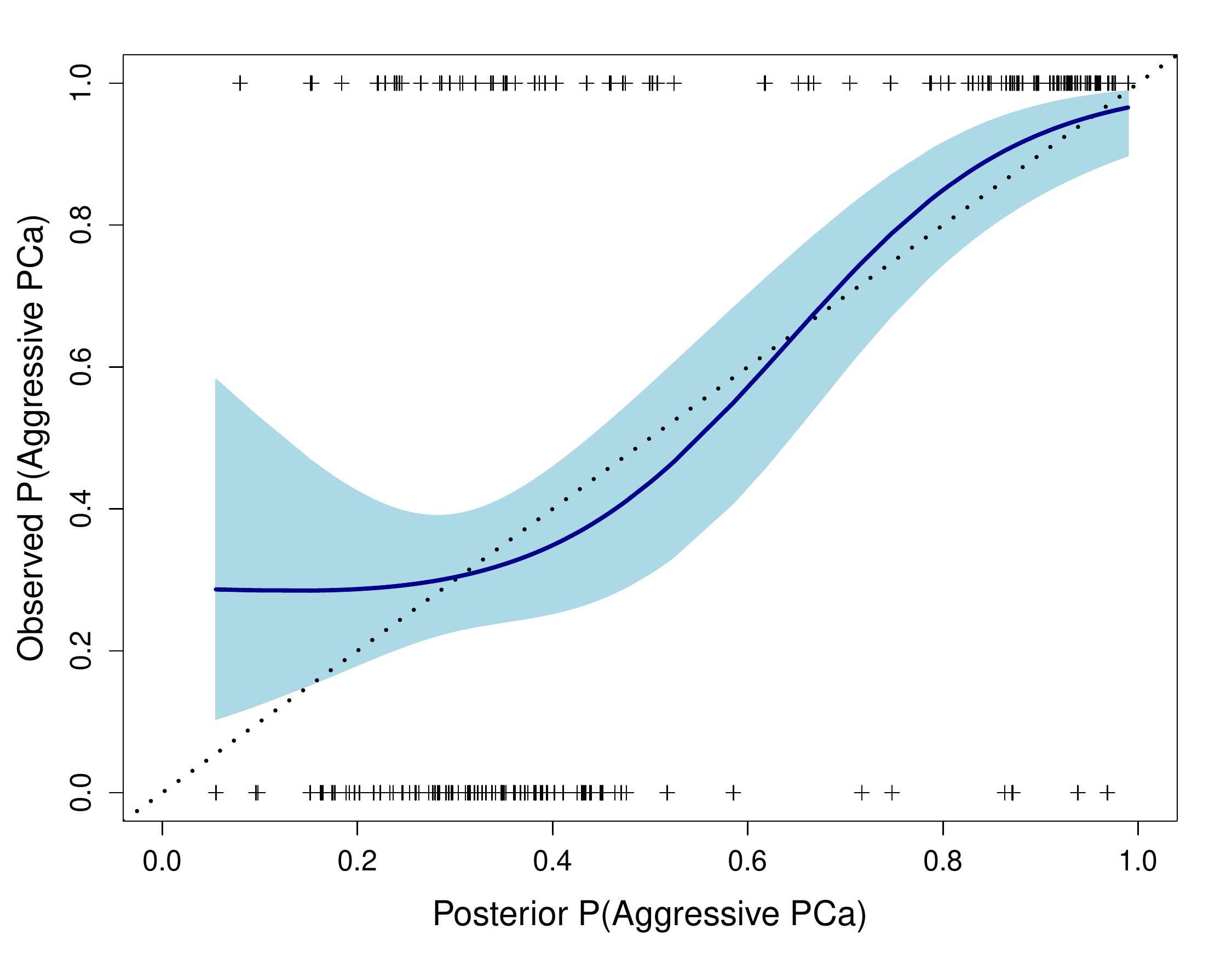}
\caption{Proposed model, surgery IOP}
\label{fig:sim-calibration-surg-ek}
\end{subfigure}
\begin{subfigure}[b]{0.45\textwidth}
\includegraphics[width=\textwidth]{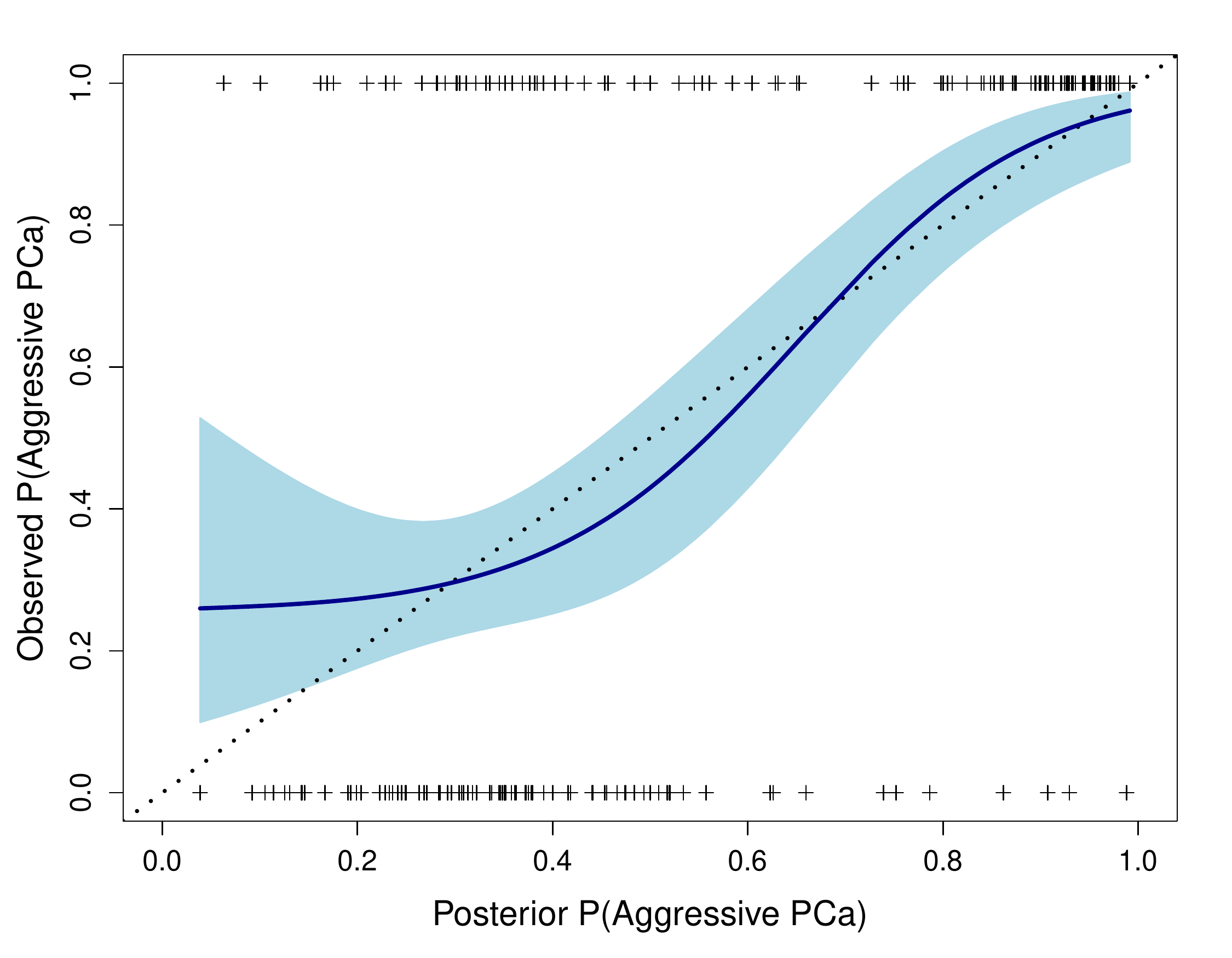}
\caption{Proposed model, biopsy and surgery IOP}
\label{fig:sim-calibration-iop-ek}
\end{subfigure}

\begin{subfigure}[b]{0.45\textwidth}
\includegraphics[width=\textwidth]{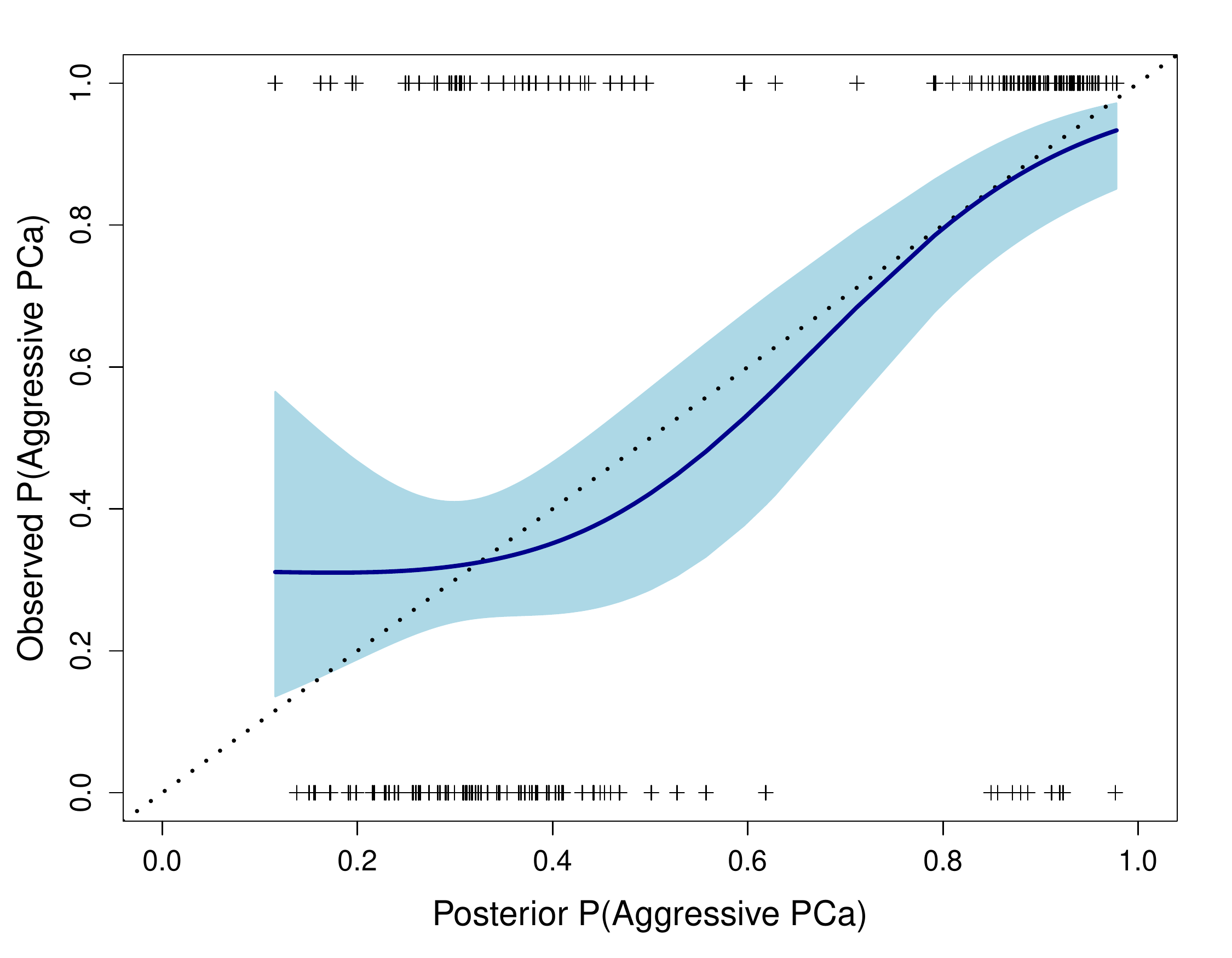}
\caption{Logistic regression}
\label{fig:sim-calibration-logistic}
\end{subfigure}

\caption{Calibration plots \textbf{among patients with true state observations} ($\eta$ known) for predictions of true cancer state in one simulated dataset}
\label{fig:sim-calibration-ek}
\end{center}
\end{figure}

Figure \ref{fig:sim-calibration-ek} shows calibration plots limited to patients with true state observations. These plots are comparable to those for the JHAS application (Figure 6(b) in the accompanying paper and Figure \ref{fig:jhas-calibration} in the appendix). We see here that the proposed model appears to underestimate the risk of more aggressive disease for those at lowest risk (posterior $P(\eta=1) <0.2$) when only patients with surgery are considered, though plots in Figure \ref{fig:sim-calibration} showed accurate calibration. This suggests that patients with and without surgery are not exchangeable at given levels of posterior risk. Perhaps a larger number of patients receiving surgery or a stronger signal for the association between $\eta$ and surgery is needed to reduce this apparent bias in predictions.

\subsection{Individualized predictions}

\begin{figure}
\begin{center}
\includegraphics[width=0.9\textwidth]{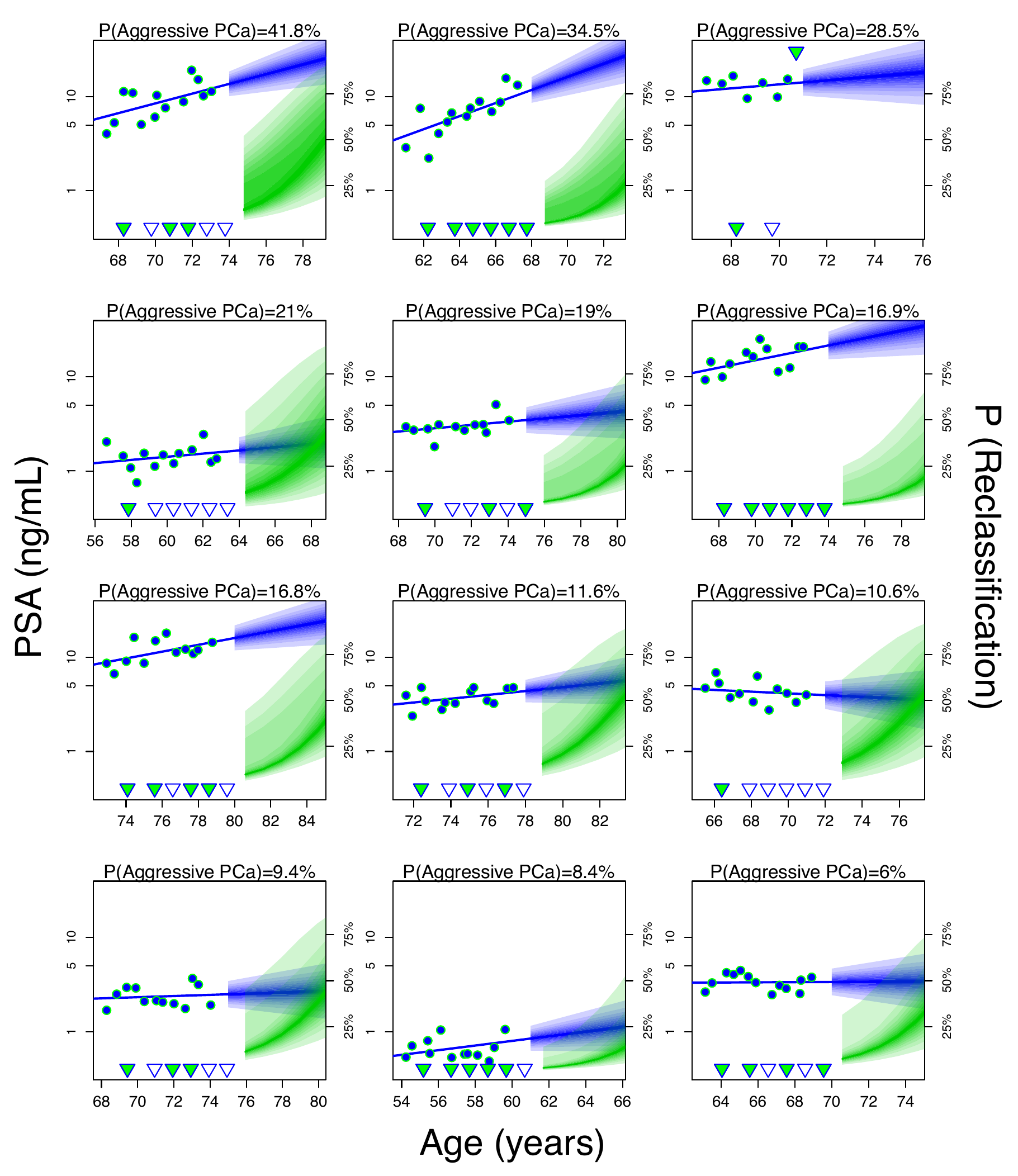}
\caption{Simulated PSA (circles) and reclassification (triangle) data for a dozen patients. Vertical position of filled triangles indicate results of biopsies received-- triangles at the bottom represent Gleason 6 observations while those on top represent Gleason 7 or above; open triangles indicate missed biopsies. Posterior probabilities of having aggressive prostate cancer (PCa) are shown above each patient's data. Shaded intervals show pointwise posterior credible intervals around projected PSA and reclassification trajectories with shading gradations indicating deciles of the interval and darkest shading occurring at the posterior median.}
\label{fig:singles}
\end{center}
\end{figure}

The goal of this modeling approach is to provide individual patients with predictions of their true cancer state in order to support clinical decision making. Plots in Figure \ref{fig:singles} show posterior predictions of $\eta$ from the biopsy and surgery IOP model as well as predictions of future PSA and biopsy values for a dozen simulated patients. For each patient, plotting circles represent simulated PSA observations, with the scale given on the lefthand y-axis. Triangles represent simulated biopsies, with open triangles indicating no biopsy in an annual interval (and, thus, no reclassification observed) and filled triangles indicating biopsy results: triangles at the bottom of the plot represent a Gleason score of 6 while those at the top represent a Gleason score of 7 or higher on biopsy. Posterior predictions of each patient's $\eta$ value are given above the plot. Shaded 95$\%$ credible intervals indicate the likely PSA trajectory and risk of reclassification ($P(R=1|\text{data})$, scale on righthand axis) for each patient. The dark green line represents the predicted risk of reclassification on a future biopsy (age indicated on the x-axis) given data observed up until this time. Interpretation is similar for PSA predictions: the dark blue line shows the expected PSA value at a future age given currently observed data. The darkest shading for biopsy and PSA predictions occur at the center of the posterior distribution (47.5-52.5 percentile) and progressively lighter shading is used at every posterior decile (42.5-57.5, $\dots$, 2.5-97.5 percentiles). Note that no reclassification projection is given for the patient with a biopsy Gleason score of 7 or higher. Since future biopsy outcomes are censored at the time of grade reclassification, post-reclassification biopsy predictions are not supported by this model.


\newpage 


{\renewcommand{\arraystretch}{1.5}
\begin{table}
\begin{center}
\begin{tabular}{|l|c|c|}
\hline
 &Total $\#$ observations & Median $\#$ per patient (IQR) \\
\hline
PSA & 10,425 & 10 (6,16)\\
Biopsy & 2,741 & 3 (1,4)\\
Years of follow-up & \multirow{2}{*}{4,980} & \multirow{2}{*}{5 (3,8)}\\
(prior to reclassification) & & \\
\hline
\end{tabular}
\vspace{20pt}
\caption{Summary of observations and follow-up time for $n$=874 patients included in JHAS analysis.}
\label{tab:num_obs}
\end{center}
\end{table}
}

{\renewcommand{\arraystretch}{1.5}
\begin{table}
\begin{center}
\begin{tabular}{|cc|cc|c|}
\hline
& &\multicolumn{2}{c|}{Biopsy Gleason Score} & \\
 & &6& $\geq 7$ & Total\\
 \hline
\multirow{2}{*}{Post-surgical True Value} &Indolent, $\eta=0$ & 66  (69$\%$) & 17 (26$\%$) & 83 \\
& Aggressive, $\eta=1$ & 30 (31$\%$) & 48 (74$\%$) & 78\\
 \hline 
 \multicolumn{2}{|c|}{Total} & 96 & 65 & 161\\
\hline
\end{tabular}
\vspace{20pt}
\caption{Summary of post-surgical cancer state determination ($\eta$) compared to final biopsy-based Gleason score (with column percentages) in JHAS analysis.}
\label{tab:eta-vs-rc}
\end{center}
\end{table}
}

{\renewcommand{\arraystretch}{1.5}
\begin{table}
\begin{center}
\caption{JHAS results: Latent class distribution parameter $\rho$,  marginal probability of more aggressive cancer ($\eta=1$)}
\label{tab:jhas-post-est-rho}
\begin{tabular}{|l|c|}
\hline
Proposed Model Variation & $\qquad$ Estimate (95$\%$ CI) $\qquad$ \\
\hline
Unadjusted (no IOP components) & 0.30 (0.23, 0.38) \\
Biopsy IOP component only & 0.31 (0.24, 0.39) \\
Surgery IOP component only & 0.20 (0.14, 0.28) \\
Biopsy and surgery IOP components & 0.23 (0.16, 0.33)\\
\hline
\end{tabular}
\end{center}
\end{table}
}

\begin{landscape}

{\renewcommand{\arraystretch}{1.5}
\begin{table}
\begin{center}
\caption{JHAS results: Stratified multilevel regression for outcome PSA, $Y$}
\label{tab:jhas-post-est-psa}
\resizebox{\textwidth}{!}{%
\begin{tabular}{|c|c|c|c|c|c|c|}
\hline
\multirow{2}{*}{Parameter} & \multirow{2}{*}{Interpretation} & Coefficient & \multicolumn{4}{c|}{\underline{Estimate (95$\%$ CI)}}  \\
& & Transformation & Unadjusted & $B$ IOP & $S$ IOP & $B,\,S$ IOP\\
\hline
\multirow{5}{*}{$\bmmu$} & Mean intercept, $\eta=0$ & & 1.33 (1.27, 1.38) & 1.33 (1.27, 1.38) &  1.36 (1.30, 1.40) & 1.35 (1.30, 1.40) \\
& Mean intercept, $\eta=1$& & 1.6 (1.5, 1.7) & 1.6 (1.5, 1.7) & 1.6 (1.5, 1.8) &  1.6 (1.5, 1.7)\\
& \multirow{2}{*}{Mean slope (age), $\eta=0$} & Standardized & \multirow{2}{*}{0.24 (0.19, 0.28)} & \multirow{2}{*}{0.24 (0.19, 0.29)} & \multirow{2}{*}{0.26 (0.22, 0.42)} & \multirow{2}{*}{0.26 (0.22, 0.30)}\\ 
&\multirow{2}{*}{Mean slope (age), $\eta=1$} & mean = 67.1 & \multirow{2}{*}{0.50 (0.41, 0.58) } & \multirow{2}{*}{0.48 (0.40, 0.57)} & \multirow{2}{*}{0.53 (0.42, 0.63)} &  \multirow{2}{*}{0.50 (0.39, 0.60)}\\
&  & sd= 6.8 & & & &  \\ 
\hline
\multirow{3}{*}{$\bmSigma$} & Standard deviation, intercepts & & 0.54 (0.51, 0.57) & 0.54 (0.51, 0.57) & 0.54 (0.51, 0.58) & 0.54 (0.51, 0.58) \\
& Standard deviation, slopes & & 0.39 (0.37, 0.42)  & 0.40 (0.37, 0.43) & 0.40 (0.37, 0.43) & 0.40 (0.37, 0.43) \\
& Covariance & & 0.036 (0.016, 0.057) & 0.038 (0.017, 0.059) & 0.040 (0.020, 0.061) & 0.041 (0.020, 0.063)\\
\hline
\multirow{3}{*}{$\bmbeta$} & \multirow{3}{*}{Fixed effect, prostate volume} & Standardized & \multirow{3}{*}{0.31 (0.27, 0.35)} & \multirow{3}{*}{0.31 (0.27, 0.35)} & \multirow{3}{*}{0.31 (0.27, 0.35)} &\multirow{3}{*}{0.31 (0.27, 0.35)}\\
&&mean = 57.5 & & & & \\
&& sd = 24.9 & & & & \\
\hline
$\sigma^2$ & Residual variance & & 0.299 (0.294, 0.303) & 0.299 (0.294, 0.303) & 0.299 (0.294, 0.303) & 0.299 (0.294, 0.303)\\
\hline
\end{tabular} }
\end{center}
\end{table}
}
\end{landscape}

{\renewcommand{\arraystretch}{1.5}
\begin{table}
\begin{center}
\caption{JHAS results: Logistic regression for whether biopsy was performed, $B$; parameter: $\bmnu$}
\label{tab:jhas-post-est-bx}
\resizebox{\textwidth}{!}{%
\begin{tabular}{|c|c|c|c|c|}
\hline
 \multirow{2}{*}{Covariate} & \multirow{2}{*}{Transformation} & \multicolumn{2}{c|}{\underline{Estimate (95$\%$ CI)}}\\
& & $B$ IOP & $B,\,S$ IOP \\
\hline
Intercept & & -2.4 (-3.2, -1.6) & -3.2 (-2.4, -1.7)\\
\hline
\multirow{4}{*}{Time since diagnosis} & Natural splines, $df$=4  & 0.86 (0.45, 1.3) & 0.88 (0.46, 1.3)\\
 & knots = (2, 4, 6)  &-0.41 (-1.2, 0.41)  &-0.39 (-1.2, 0.47)\\
& boundary = (1, 20)  & -1.4 (-2.7, -0.14) & -1.4 (-2.7, -0.036)\\
& & -7.5 (-9.9, -5.2) & -7.4 (-9.8, -5.0)\\
\hline
 \multirow{4}{*}{Date} & Natural splines, $df$=4  & 0.74 (0.21, 1.3) & 0.74 (0.19, 1.3)\\
 & knots = (4/4/07, 7/11/10, 1/28/13)  & 1.2 (0.71, 1.6) &1.2 (0.71, 1.6)\\
& boundary =  (8/17/95, 9/30/15) &2.1 (0.78, 3.4) & 2.1 (0.79, 3.4)\\
&& -2.5 (-2.9, -2.1) & -2.5 (-2.9, -2.1)\\
\hline
 \multirow{4}{*}{Age} & Natural splines, $df$=2  &  \multirow{2}{*}{1.1 (0.18, 2.1)} & \multirow{2}{*}{1.1 (0.21, 2.0)}\\
 & knots = 69.8  & \multirow{2}{*}{-3.9 (-4.5,-3.3)} & \multirow{2}{*}{-3.9 (-4.5,-3.3)}\\
& boundary =  (46.8, 89.5) & &\\
\hline
 \multirow{4}{*}{$\#$ previous biopsies} & Natural splines, $df$=2 & \multirow{2}{*}{0.61 (-0.49, 1.6)} & \multirow{2}{*}{0.58 (-0.63, 1.6)}\\
 & knots = 3 & \multirow{2}{*}{3.9 (2.8, 5.0)} & \multirow{2}{*}{3.9 (2.7, 5.0)}\\
& boundary =  (1, 13)  & &\\
\hline
 $\eta=1$ & & -0.47 (-0.78, -0.12) & -0.52 (-0.88, -0.17)\\
\hline
\end{tabular}}
\end{center}
\end{table}
}

\begin{landscape}
{\renewcommand{\arraystretch}{1.5}
\begin{table}
\begin{center}
\caption{JHAS results: Logistic regression for grade reclassification, $R$; parameter: $\bmgamma$}
\label{tab:jhas-post-est-rc}
\resizebox{\textwidth}{!}{%
\begin{tabular}{|c|c|c|c|c|c|c|}
\hline
 \multirow{2}{*}{Covariate} & \multirow{2}{*}{Transformation} & \multicolumn{4}{c|}{\underline{Estimate (95$\%$ CI)}} \\
 & & Unadjusted & $B$ IOP & $S$ IOP & $B,\,S$ IOP\\
\hline
Intercept & & -3.4 (-4.8, -2.2) & -3.4 (-4.7, -2.2) &-2.9 (-4.2, -1.8)  & -2.9 (-4.1, -1.8)\\
\hline
\multirow{2}{*}{Time since diagnosis} & Natural splines, $df$=2  & \multirow{2}{*}{-1.3 (-2.6, -0.10)} & \multirow{2}{*}{-1.0 (-2.2, 0.21)} & \multirow{2}{*}{-1.4 (-2.5, -0.25)} &\multirow{2}{*}{-1.3 (-2.4, -0.17)} \\
 & knots = 2.3 & \multirow{2}{*}{1.6 (-0.27, 3.4) }& \multirow{2}{*}{1.7 (-0.085, 3.3)} & \multirow{2}{*}{1.4 (-0.29, 3.0)} &\multirow{2}{*}{1.5 (-0.30, 3.0)} \\
& boundary = (0.08, 15.9) & & & & \\
\hline
 \multirow{2}{*}{Date} & Natural splines, $df$=2  & \multirow{2}{*}{0.007 (-2.4, 2.6)} & \multirow{2}{*}{-0.046 (-2.3, 2.5)} & \multirow{2}{*}{-0.037 (-2.2, 2.3)} & \multirow{2}{*}{-0.10 (-2.2, 2.3)}\\
 & knots =  1/7/09 & \multirow{2}{*}{1.1 (0.44, 1.9)} & \multirow{2}{*}{1.1 (0.43, 1.8)} & \multirow{2}{*}{1.1 (0.40, 1.7)}  &\multirow{2}{*}{1.0 (0.40, 1.7)} \\
& boundary =  (10/25/95, 6/19/14) && & &\\
\hline
 \multirow{3}{*}{Age} & Standardized  & & & &  \\
 & mean = 67.7 & 0.61 (0.38, 0.86) & 0.61 (0.39, 0.86) & 0.56 (0.36, 0.78) & 0.55 (0.35, 0.77)\\
& sd = 5.5 && & &\\
\hline
 $\eta=1$ & & 2.1 (1.5, 2.7) & 2.1 (1.5, 2.7) &1.8 (1.1, 2.5)  & 1.6 (0.92, 2.3) \\
\hline
\end{tabular}}
\end{center}
\end{table}
}
\end{landscape}

{\renewcommand{\arraystretch}{1.5}
\begin{table}
\begin{center}
\caption{JHAS results: Logistic regression for whether surgery was performed, $S$; parameter: $\bmomega$}
\label{tab:jhas-post-est-surg}
\resizebox{\textwidth}{!}{%
\begin{tabular}{|c|c|c|c|c|}
\hline
\multirow{2}{*}{Covariate} & \multirow{2}{*}{Transformation} & \multicolumn{2}{c|}{\underline{Estimate (95$\%$ CI)}}\\
& & $S$ IOP & $B,\,S$ IOP\\
\hline
Intercept & & -6.4 (-8.9, -4.2) & -6.2 (-8.6, -4.0)\\
\hline
\multirow{3}{*}{Time since diagnosis} & Natural splines, $df$=3  & 1.8 (0.90, 2.8) & 1.7 (0.82, 2.7)\\
 & knots = (2, 4, 6) & 1.3 (-0.84, 3.3) & 1.2 (-0.93, 3.3)\\
& boundary = (1, 20) & 6.7 (3.9, 9.5) & 6.7 (3.8, 9.4) \\
& & 2.9 (-2.0, 6.8) & 2.9 (-2.0, 6.8)\\
\hline
 \multirow{3}{*}{Date} & Natural splines, $df$=3  & 0.72 (-0.17, 1.7)& 0.67 (-0.22, 1.6)\\
 & knots = (6/18/08, 4/15/12) & -2.1 (-5.2, 1.3) & -2.1 (-5.3, 1.3)\\
& boundary =  (8/17/95, 9/30/15)& -0.91 (-1.9, -0.003) & -0.93 (-1.9, -0.020) \\
\hline
 \multirow{4}{*}{Age} & Natural splines, $df$=2  & \multirow{2}{*}{-4.9 (-7.3, -2.2)} & \multirow{2}{*}{-5.0 (-7.4, -2.3)}\\
 & knots = 69.8 & \multirow{2}{*}{-11 (-14, -7.8)} & \multirow{2}{*}{ -11 (-14, -7.7)}\\
& boundary =  (46.8, 89.6) & & \\
\hline
 \multirow{3}{*}{$\#$ previous biopsies} &  Standardized & & \\
 & mean = 3.8 & -0.45 (-0.87, -0.015) & -0.40 (-0.85, 0.037)\\
& sd = 2.3 & &\\
\hline
 \multirow{3}{*}{max. previous $\#$ positive cores} & Standardized  & & \\
 & mean = 1.6  & 0.36 (0.22, 0.51) & 0.36 (0.22, 0.50)\\
& sd =  0.9 & & \\
\hline
 \multirow{3}{*}{max. previous max $\%$ positive} & Natural splines, $df$=2 & \multirow{2}{*}{3.7 (2.5, 4.9)} & \multirow{2}{*}{3.7 (2.5, 4.9)}\\
 & knots = 15 & \multirow{2}{*}{1.1 (0.051, 2.1)} & \multirow{2}{*}{1.1 (0.015, 2.1)}\\
& boundary = (1, 100) & &\\
\hline
previous $R=1$ & & 1.3 (0.57, 2.0) & 1.2 (0.50, 1.9)\\
\hline
 $\eta=1$ & &0.91 (0.13, 1.6)  & 0.59 (-0.29, 1.4)\\
\hline
previous $R=1 \, \times \, \eta=1$ & & 2.0 (0.70, 3.2) & 2.3 (1.1, 3.6)\\
\hline
\end{tabular}}
\end{center}
\end{table}
}

{\renewcommand{\arraystretch}{1.5}
\begin{table}
\begin{center}
\caption{JHAS results: Estimated odds ratios for aggressive prostate cancer from logistic regression analysis of JHAS cohort patients with post-surgery observations of $\eta$}
\label{tab:jhas-logistic-est}
\begin{tabular}{|l|c|}
\hline
Covariate & Odds Ratio (95$\%$ CI) \\
\hline
Reclassification on biopsy & 4.7 (2.0, 11) \\
Age (years) & 1.1 (0.98, 1.2) \\
$\#$ previous biopsies without reclassification & 0.68 (0.41, 0.1.2) \\
Years in AS & 1.5 (1.0, 2.2)\\
PSA density ($\times 10$) & 1.7 (0.61, 4.9) \\
Slope log-PSA ($\times 10$) & 1.3 (0.98, 1.6)\\
\hline
\end{tabular}
\end{center}
\end{table}
}

{\renewcommand{\arraystretch}{1.5}
\begin{table}[h]
\begin{center}
\caption{JHAS results: Predictive accuracy among patients with post-surgery $\eta$ observations. 95$\%$ quantile-based bootstrapped intervals are given in parentheses.}
\label{tab:jhas-pred-accuracy}
\begin{tabular}{|cl|c|c|c|}
\hline
\multicolumn{2}{|c|}{Estimation Method} & AUC & MSE  & FPR at TPR=0.62\\
\hline
\multirow{2}{*}{Proposed}  & Biopsy and Surgery IOP & 0.75 (0.67, 0.83) & 0.201 (0.17, 0.24) & 0.14  (0.07, 0.30) \\
 & Biopsy only IOP & 0.74 (0.66, 0.82) & 0.205 (0.17, 0.24) & 0.17 (0.09, 0.26) \\
\multirow{2}{*}{Model} & Surgery only IOP & 0.74 (0.65, 0.81) & 0.207 (0.17, 0.24) & 0.19 (0.08, 0.32)\\
 & No IOP components & 0.72 (0.64, 0.80) & 0.210 (0.18, 0.25) & 0.19 (0.11, 0.29)\\
 \hline
\multicolumn{2}{|c|}{Logistic Regression} & 0.74 (0.66, 0.81) & 0.209 (0.18, 0.24) & 0.19 (0.10, 0.29)\\
\hline
 \multicolumn{2}{|c|}{Grade Reclassification on Final Biopsy} & n/a & 0.292  (0.22, 0.37) & 0.20 (0.12, 0.29) \\
 \hline
\end{tabular}
\end{center}
\end{table}
}

{\renewcommand{\arraystretch}{1.5}
\begin{table}
\begin{center}
\caption{Simulation results: Latent class distribution parameter $\rho$,  marginal probability of more aggressive cancer ($\eta=1$). Data generating value was $\mathbf{\rho=0.23}$.}
\label{tab:sim-post-est-rho}
\begin{tabular}{|l|cc|}
\hline
Proposed Model Variation & Estimate & Coverage\\
\hline
Unadjusted (no IOP components) & 0.34 & 8.5$\%$ \\
Biopsy IOP component only &  0.34 & 4.5$\%$ \\
Surgery IOP component only &  0.24 & 94$\%$ \\
Biopsy and surgery IOP components & 0.25 & 96$\%$ \\
\hline
\end{tabular}
\end{center}
\end{table}
}

\newpage
\begin{landscape}

{\renewcommand{\arraystretch}{1.5}
\begin{table}
\begin{center}
\caption{Simulation results: Stratified multilevel regression for outcome PSA, $Y$}
\label{tab:sim-post-est-psa}
\resizebox{\textwidth}{!}{%
\begin{tabular}{|c|c|c|c|c|c|c|c|} 
\hline
\multirow{2}{*}{Parameter} & \multirow{2}{*}{Interpretation} & Generating & Coefficient & \multicolumn{4}{c|}{\underline{Estimate (Coverage of 95$\%$ Interval)}}  \\
& & Value & Transformation & Unadjusted & $B$ IOP & $S$ IOP & $B,\,S$ IOP\\
\hline
\multirow{5}{*}{$\bmmu$} &  Mean intercept, $\eta=0$ & 1.4 & &  1.3 (95$\%$) & 1.3 (94$\%$) & 1.3 (97$\%$) & 1.3 (96$\%$) \\
& Mean intercept, $\eta=1$ &  1.6 & & 1.5 (71$\%$) & 1.6 (74$\%$) & 1.6 (96$\%$) & 1.6 (97$\%$) \\ 
& \multirow{2}{*}{Mean slope (age), $\eta=0$} & \multirow{2}{*}{0.26} & Standardized & \multirow{2}{*}{0.25 (95$\%$)} & \multirow{2}{*}{0.25 (94$\%$)} & \multirow{2}{*}{0.26 (97$\%$)} & \multirow{2}{*}{0.25 (97$\%$)} \\ 
&\multirow{2}{*}{Mean slope (age), $\eta=1$} & \multirow{2}{*}{0.50} & mean = 67.1 & \multirow{2}{*}{0.44 (65$\%$)} & \multirow{2}{*}{0.44 (66$\%$)} & \multirow{2}{*}{0.49 (93$\%$)} & \multirow{2}{*}{0.48 (92$\%$)} \\
&  & &  sd= 6.8 & & & &  \\ 
\hline
\multirow{3}{*}{$\bmSigma$} & Standard deviation, intercepts & 0.54 & & 0.55 (93$\%$) & 0.55 (92$\%$) & 0.54 (93$\%$) & 0.54 (92$\%$) \\
& Standard deviation, slopes & 0.40 & &   0.40 (95$\%$) & 0.40 (94$\%$) & 0.40 (95$\%$) & 0.40 (94$\%$) \\
& Covariance & 0.041 &  &   0.042 (94$\%$) & 0.041 (95$\%$) & 0.040 (96$\%$) & 0.040 (94$\%$) \\
\hline
\multirow{3}{*}{$\bmbeta$} & \multirow{3}{*}{Fixed effect, prostate volume} & \multirow{3}{*}{0.31} & Standardized & \multirow{3}{*}{0.31 (95$\%$)} & \multirow{3}{*}{0.31 (95$\%$)} & \multirow{3}{*}{0.31 (95$\%$)} & \multirow{3}{*}{0.31 (95$\%$)}\\
& & & mean = 57.5 & & & & \\
& & & sd = 24.9 & & & & \\
\hline
$\sigma^2$ & Residual variance & 0.30 &  &   0.30 (95$\%$) & 0.30 (96$\%$) & 0.30 (95$\%$) & 0.30 (95$\%$) \\
\hline
\end{tabular}}
\end{center}
\end{table}
}

\end{landscape}

{\renewcommand{\arraystretch}{1.5}
\begin{table}
\begin{center}
\caption{Simulation results: Logistic regression for whether biopsy was performed, $B$; parameter: $\bmnu$}
\label{tab:sim-post-est-bx}
\resizebox{\textwidth}{!}{%
\begin{tabular}{|c|c|c|c|c|} 
\hline
 \multirow{2}{*}{Covariate} & Generating &  \multirow{2}{*}{Transformation} & \multicolumn{2}{c|}{\underline{Estimate (95$\%$ CI)}}\\
& Value & & $B$ IOP & $B,\,S$ IOP \\
\hline
Intercept & -2.4 & & -2.4 (91$\%$) & -2.4 (96$\%$)\\
\hline
\multirow{4}{*}{Time since diagnosis} &  0.88 & Natural splines, $df$=4  & 0.91 (99$\%$) &  0.94 (97$\%$)\\
& -0.39 & knots = (2, 4, 6)  & -0.42 (96$\%$) &  0.36 (96$\%$)\\
& -1.4 & boundary = (1, 20)  & -1.1 (96$\%$) &  -1.1 (96$\%$)\\
& -7.4 &  & -6.9 (98$\%$) & -6.9 (98$\%$)\\
\hline
 \multirow{4}{*}{Date} & 0.74 & Natural splines, $df$=4  & 0.72 (91$\%$) &  0.73 (96$\%$)\\
 & 1.2 & knots = (4/4/07, 7/11/10, 1/28/13)  & 1.2 (66$\%$) &  1.2 (97$\%$)\\
& 2.1 & boundary =  (8/17/95, 9/30/15) & 2.0 (91$\%$) &  2.1 (96$\%$)\\
& -2.5 & & -2.5 (99$\%$) &  -2.5 (97$\%$)\\
\hline
 \multirow{4}{*}{Age} &  \multirow{2}{*}{1.1} & Natural splines, $df$=2 &  \multirow{2}{*}{1.1 (96$\%$)} & \multirow{2}{*}{1.1 (96$\%$)}\\
 & \multirow{2}{*}{-3.9} & knots = 69.8  &  \multirow{2}{*}{-4.0 (96$\%$)} & \multirow{2}{*}{-4.0 (96$\%$)}\\
& & boundary =  (46.8, 89.5) & &\\
\hline
 \multirow{4}{*}{$\#$ previous biopsies} & \multirow{2}{*}{0.58} & Natural splines, $df$=2 &\multirow{2}{*}{0.55 (98$\%$)} & \multirow{2}{*}{0.44 (98$\%$)}\\
 & \multirow{2}{*}{3.9} & knots = 3 &\multirow{2}{*}{3.8 (91$\%$)} & \multirow{2}{*}{3.8 (96$\%$)}\\
& & boundary =  (1, 13)  & &\\
\hline
 $\eta=1$ & -0.52 & & -0.40 (66$\%$) & -0.53 (97$\%$) \\
\hline
\end{tabular} }
\end{center}
\end{table}
}

\begin{landscape}

{\renewcommand{\arraystretch}{1.5}
\begin{table}
\begin{center}
\caption{Simulation results: Logistic regression for grade reclassification, $R$; parameter: $\bmgamma$}
\label{tab:sim-post-est-rc}
\resizebox{\textwidth}{!}{%
\begin{tabular}{|c|c|c|c|c|c|c|c|}
\hline
 \multirow{2}{*}{Covariate} & Generating &\multirow{2}{*}{Transformation} & \multicolumn{4}{c|}{\underline{Estimate (95$\%$ CI)}} \\
 & Value &  & Unadjusted & $B$ IOP & $S$ IOP & $B,\,S$ IOP\\
\hline
Intercept & -2.9 & & -3.6 (90$\%$) & -3.5 (91$\%$) & -3.0 (95$\%$)  & -2.9 (96$\%$) \\
\hline
\multirow{2}{*}{Time since diagnosis} & \multirow{2}{*}{-1.3} & Natural splines, $df$=2  & \multirow{2}{*}{-1.3 (98$\%$)} & \multirow{2}{*}{-1.2 (99$\%$)} & \multirow{2}{*}{-1.3 (97$\%$)} &\multirow{2}{*}{-1.4 (97$\%$)} \\
&  \multirow{2}{*}{1.4} & knots = 2.3 & \multirow{2}{*}{1.4 (95$\%$) }& \multirow{2}{*}{1.6 (96$\%$)} & \multirow{2}{*}{1.2 (96$\%$)} &\multirow{2}{*}{1.3 (96$\%$)} \\
& & boundary = (0.08, 15.9) & & & & \\
\hline
 \multirow{2}{*}{Date} & \multirow{2}{*}{-0.07} & Natural splines, $df$=2  & \multirow{2}{*}{0.018 (96$\%$)} & \multirow{2}{*}{-0.031 (96$\%$)} & \multirow{2}{*}{0.098 (95$\%$)} & \multirow{2}{*}{0.065 (96$\%$)}\\
& \multirow{2}{*}{1.0} & knots =  1/7/09 & \multirow{2}{*}{1.1 (98$\%$)} & \multirow{2}{*}{1.1 (98$\%$)} & \multirow{2}{*}{1.1 (97$\%$)}  &\multirow{2}{*}{1.1 (98$\%$)} \\
&  & boundary =  (10/25/95, 6/19/14) && & &\\
\hline
 \multirow{3}{*}{Age} & & Standardized  & & & &  \\
&  0.55 & mean = 67.7 & 0.61 (91$\%$) & 0.61 (91$\%$) & 0.56 (96$\%$)  & 0.56 (96$\%$) \\
& & sd = 5.5 && & &\\
\hline
 $\eta=1$ & 1.6 & & 2.2 (67$\%$) & 2.1 (66$\%$) & 1.6 (92$\%$)  & 1.5 (97$\%$) \\
\hline
\end{tabular} }
\end{center}
\end{table}
}
\end{landscape}

{\renewcommand{\arraystretch}{1.5}
\begin{table}
\begin{center}
\caption{Simulation results: Logistic regression for whether surgery was performed, $S$; parameter: $\bmomega$}
\label{tab:sim-post-est-surg}
\resizebox{\textwidth}{!}{%
\begin{tabular}{|c|c|c|c|c|c|}
\hline
\multirow{2}{*}{Covariate} & Generating & \multirow{2}{*}{Transformation} & \multicolumn{2}{c|}{\underline{Estimate (95$\%$ CI)}}\\
& Value & & $S$ IOP & $B,\,S$ IOP\\
\hline
Intercept &  -5.0 & &  -5.5 (95$\%$) & -5.3 (96$\%$) \\
\hline
\multirow{3}{*}{Time since diagnosis} & 1.8 & Natural splines, $df$=3  & 1.9 (97$\%$) & 1.8 (97$\%$) \\
 & 1.2 & knots = (2, 4, 6) & 1.6 (96$\%$) & 1.4 (96$\%$) \\
& 6.7 &  boundary = (1, 20) & 6.0 (95$\%$) & 5.8 (96$\%$) \\
& 2.8 & & 0.57 (97$\%$) & 0.49 (98$\%$) \\
\hline
 \multirow{3}{*}{Date} & 0.67 &  Natural splines, $df$=3  & 0.81 (96$\%$) & 0.80 (96$\%$) \\
 & -2.1 & knots = (6/18/08, 4/15/12) & -1.8 (92$\%$) & -1.7 (97$\%$) \\
& -0.93 & boundary =  (8/17/95, 9/30/15)& -0.99 (95$\%$) & -0.95 (96$\%$) \\
\hline
 \multirow{4}{*}{Age} & \multirow{2}{*}{-5.0} & Natural splines, $df$=2  & \multirow{2}{*}{-4.9 (97$\%$)} & \multirow{2}{*}{-4.9 (97$\%$)}\\
 & \multirow{2}{*}{-11} & knots = 69.8 & \multirow{2}{*}{-11 (96$\%$)} & \multirow{2}{*}{-11 (96$\%$)}\\
& & boundary =  (46.8, 89.6) & & \\
\hline
 \multirow{3}{*}{$\#$ previous biopsies} &  & Standardized & & \\
 & -0.40&  mean = 3.8 & -0.46 (95$\%$) & -0.39 (96$\%$) \\
& &  sd = 2.3 & &\\
\hline
previous $R=1$ & 1.2&  & 1.3 (97$\%$) & 1.2 (98$\%$) \\
\hline
 $\eta=1$ & 0.59 &  & 0.64 (96$\%$) & 0.46(96$\%$) \\
\hline
previous $R=1 \, \times \, \eta=1$ & 2.3 & & 2.2 (92$\%$) & 2.5 (97$\%$) \\
\hline
\end{tabular} }
\end{center}
\end{table}
}

{\renewcommand{\arraystretch}{1.5}
\begin{table}
\begin{center}
\caption{Simulation accuracy: Predictive accuracy in simulation studies with data generated according to biopsy and surgery IOP model. The mean AUC and MSE across 200 simulations are given for patients with and without $\eta$ observed. 95$\%$ quantile-based intervals of estimated AUC and MSE from all sims are in parentheses.}
\label{tab:sim-pred-accuracy}
\begin{tabular}{|l|cc|cc|}
\hline
\multirow{2}{*}{Estimation Method} &\multicolumn{2}{c|}{\underline{$\eta$ observed}} & \multicolumn{2}{c|}{\underline{$\eta$ unobserved}}\\
& AUC (95$\%$ Int) &  MSE (95$\%$ Int) & AUC (95$\%$ Int) & MSE (95$\%$ Int)\\
\hline 
Biopsy and Surgery IOP & 0.83 (0.77, 0.88) & 0.17 (0.14, 0.19) & 0.77 (0.72, 0.81) & 0.12 (0.11, 0.16) \\
Biopsy IOP only & 0.81 (0.76, 0.86) & 0.18 (0.15, 0.20) & 0.74 (0.69, 0.78) & 0.16 (0.14, 0.18) \\
Surgery IOP only & 0.81 (0.76, 0.86) & 0.17 (0.15, 0.20) & 0.73 (0.67, 0.78) & 0.13 (0.11, 0.18) \\
Unadjusted (No IOP) & 0.80 (0.74, 0.85) & 0.18 (0.16, 0.20) & 0.71 (0.66, 0.75) & 0.17 (0.14, 0.19) \\
\hline
Logistic Regression & 0.77 (0.70, 0.83) & 0.19 (0.16, 0.21) & 0.68 (0.63, 0.73) & 0.26 (0.18, 0.56) \\
\hline
\end{tabular}
\end{center}
\end{table}
}

\clearpage

\end{document}